\documentclass{aa}
\usepackage[varg]{txfonts}
\usepackage{graphicx}
\bibpunct{(}{)}{;}{a}{}{,} % to follow the A&A style

\begin{document}
%%%%%%%%%%%%%%%%%%%%%%%%%%%%%%%%%%%%%%%%%%%%%%%%%%%%%%%%%%%%%%%%%%%%
\title{Jiamusi pulsar observations:
   III. Nulling of 20 pulsars}

\author{P.~F. Wang\inst{1,2,3} \thanks{E-mail: pfwang@nao.cas.cn}
        \and J.~L. Han\inst{1,2,3} \thanks{E-mail: hjl@nao.cas.cn}
        \and L. Han\inst{4,5}
        \and B. Y. Cai\inst{4,5}
        \and C. Wang\inst{1,2,3}
        \and T. Wang\inst{1,3}
        \and X. Chen\inst{1,3}
        \and \\D. J. Zhou\inst{1,3}
        \and Y. Z. Yu\inst{6}
        \and J. Han\inst{1}
        \and J. Xu\inst{1,2}
        \and X. Y. Gao\inst{1,2,3}
        \and T. Hong\inst{1,2}
        \and L. G. Hou\inst{1,2}
        \and B. Dong\inst{1,2}}
\institute{National Astronomical Observatories, Chinese Academy of Sciences,
         Jia-20 Datun Road, ChaoYang District, Beijing 100101, PR China
         \and
         CAS Key Laboratory of FAST, NAOC, Chinese Academy of Sciences,
         Beijing 100101, PR China
         \and
         School of Astronomy, University of Chinese Academy of Sciences,
         Beijing 100049, PR China
         \and
         The State Key Laboratory of Astronautic Dynamics, Xi'an,
         Shaanxi 710043, PR China
         \and
         Jiamusi Deep Space Station, China Xi'an Satellite Control Center,
         Jiamusi, Heilongjiang 154002, PR China
         \and
         Qiannan Normal University for Nationalities,
         Duyun, Guizhou 558000, PR China
       }

\date{Received YYY/ Accepted XXX}

\abstract
    {}
    {Most of pulsar nulling observations were conducted at frequencies
      lower than 1400~MHz. We aim to understand the nulling behaviors
      of pulsars at relatively high frequency, and to check if nulling
      is caused by a global change of pulsar magnetosphere. }
    {20 bright pulsars are observed at 2250~MHz with unprecedented
      lengths of time by using Jiamusi 66m telescope. Nulling
      fractions of these pulsars are estimated, and the null and
      emission states of pulses are identified. Nulling degrees and
      scales of the emission-null pairs are calculated to
      describe the distributions of emission and null
      lengths. }
    {Three pulsars, PSRs J0248+6021, J0543+2329 and J1844+00, are
      found to null for the first time. The details of
      null-to-emission and emission-to-null transitions within pulse
      window are first observed for PSR J1509+5531, which is a small
      probability event. A complete cycle of long nulls for hours is
      observed for PSR J1709$-$1640. For most of these pulsars, the
      K-S tests of nulling degrees and nulling scales reject the
      hypothesis that null and emission are of random processes at
      high significance levels. Emission-null sequences of some
      pulsars exhibit quasi-periodic, low-frequency or featureless
      modulations, which might be related to different origins. During
      transitions between emission and null states, pulse intensities
      have diverse tendencies for variations. Significant correlations
      are found for nulling fraction, nulling cadence and nulling
      scales with the energy loss rate of the pulsars. Combined with
      the nulling fractions reported in literatures for 146 nulling
      pulsars, we found that statistically large nulling fractions are
      more tightly related to pulsar period than to characteristic age
      or energy loss rate.}
    % Conclusions heading (optional)
    {}


\maketitle
\titlerunning{Nulling of 20 pulsars at S-band}
\authorrunning{P. F. Wang, et al.}

\section{INTRODUCTION}

%% what is nulling
Pulse profiles obtained by integrating tens of thousands of individual
pulses are generally stable and represent the unique feature of each
pulsar. However, individual pulses vary a lot. For some pulsars, the
pulse-by-pulse emission is suddenly ceased and no radio emission can
be detected, but after some periods (one to a few thousands) the
emission is restored and the pulsar can be detected again. This
phenomenon is called nulling, as was first reported by
\citet{bac70}. To date, 214 pulsars have been reported to null
\citep[e.g.][]{rit76,big92,wmj07,gjk12,bmm17}, less than 10 percent of
the known pulsars. The actual nulling pulsars may be more than knowns
since observations are restricted by sensitivity and the length of
observation sessions.

%% null length and explanization
Pulsars may null over a wide range of timescales from one rotation to
many hours even as long as months. The most prevalent nulls last for
just a single period, e.g. PSR B2021+51, which is generally considered
as a stochastic process within the pulsar magnetosphere
\citep[e.g.][]{bmm17}. Long nulls of PSR B1706$-$16 last for 2-5 hours
\citep{njmk18}, which might be related to changes of plasma processes
within a pulsar magnetosphere. Nulling for much longer time of days to
months has been found for intermittent pulsars, such as PSR B1931+24
\citep{klo+06}, which is closely related to the spin-down energy loss.

%%%%%%%%%%%%%%%%%%%%%%%%%
\begin{table*}
  \centering     
  \caption{Observational parameters of 20 pulsars.}
  \label{table:obs}
  \tabcolsep 2.0mm
  %\small
  %\footnotesize
  %\scriptsize
  %\tiny
  \begin{tabular}{ccrlcrrrrl}
    \hline  
    \hline
JName & Name & Period & \multicolumn{1}{c}{Date} & $N_{\rm ch}$ & $T_{\rm obs}$ & $N_{\rm sub}$ & $N_{\rm int}$ & \multicolumn{1}{c}{$N_{\rm blk}$} & Plot\\
      &       &  (s)  &                          &     & (min)       &             &  (P)        &                 &   \\
 (1)  & (2)   &  (3)  & \multicolumn{1}{c}{(4)}  & (5) & (6)         &  (7)        & (8) & \multicolumn{1}{c}{(9)} & (10) \\
\hline
%\endhead
%\hline
%\endfoot 
J0034$-$0721& B0031$-$07& 0.943 & 2015-12-12& 128& 191.1&  380 & 32 &  2& Fig.\ref{fig:J0034_20151212}\\
J0248$+$6021&           & 0.217 & 2015-06-16& 256&  32.4&  280 & 32 &  1& Fig.\ref{fig:J0248_20150616-20160811}\\
            &           &       & 2016-08-11& 256& 119.4& 1032 & 32 &  1& Fig.\ref{fig:J0248_20150616-20160811}\\
J0304$+$1932& B0301$+$19& 1.387 & 2015-12-14& 256&  55.5&  300 &  8 &  1& Fig.\ref{fig:J0304_20151214}\\
J0332$+$5434& B0329$+$54& 0.714 & 2016-02-21& 256& 178.5&15000 &  1 &  2& Fig.\ref{fig:J0332_20160221}\\
            &           &       & 2016-02-24& 256&  83.3& 7000 &  1 &  1& Fig.\ref{fig:J0332_20160224}\\
J0528$+$2200& B0525$+$21& 3.745 & 2015-06-19& 256&  28.7&  115 &  4 &  1& Fig.\ref{fig:J0528_20150618-20160127}\\
            &           &       & 2016-01-27& 128&  32.0&  128 &  4 &  1& Fig.\ref{fig:J0528_20150618-20160127}\\
J0543$+$2329& B0540$+$23& 0.245 & 2015-06-25& 256&  49.3&  754 & 16 &  1& Fig.\ref{fig:J0543_20150625}\\
            &           &       & 2017-11-01& 256& 250.1& 3828 & 16 &  2& Fig.\ref{fig:J0543_20171101}\\
J0826$+$2637& B0823$+$26& 0.530 & 2015-11-12& 128&  66.1& 1870 &  4 &  2& Fig.\ref{fig:J0826_20151112}\\
            &           &       & 2015-11-14& 256&  53.0& 1500 &  4 &  2& Fig.\ref{fig:J0826_20151114}\\
            &           &       & 2015-11-15& 256&  48.6& 1375 &  4 &  2& Fig.\ref{fig:J0826_20151115}\\
J0908$-$1739& B0906$-$17& 0.401 & 2016-05-23& 256&  37.4&  175 & 32 &  1& Fig.\ref{fig:J0908_20160523}\\
J0922$+$0638& B0919$+$06& 0.430 & 2015-07-14& 256&  57.3& 4000 &  2 &  1& Fig.\ref{fig:J0922_20150714}\\
            &           &       & 2015-08-17& 128&  45.9& 3200 &  2 &  1& Fig.\ref{fig:J0922_20150817-20150818}\\
            &           &       & 2015-08-18& 256&  34.4& 2400 &  2 &  1& Fig.\ref{fig:J0922_20150817-20150818} \\
J0953$+$0755& B0950$+$08& 0.253 & 2015-12-12& 128& 197.3& 5850 &  8 &  1& Fig.\ref{fig:J0953_20151212}\\
J1136$+$1551& B1133$+$16& 1.187 & 2015-09-19& 256&  10.1&  128 &  4 &  1& Fig.\ref{fig:J1136_20150919-20180916} \\
            &           &       & 2017-10-25& 256& 130.6& 1650 &  4 &  4& Fig.\ref{fig:J1136_20171025AB}, Fig.\ref{fig:J1136_20171025CD}\\
            &           &       & 2018-09-16& 128&  23.7&  300 &  4 &  1& Fig.\ref{fig:J1136_20150919-20180916}\\
J1239$+$2453& B1237$+$25& 1.382 & 2015-12-12& 128&  23.0&  500 &  2 &  2& Fig.\ref{fig:J1239_20151212-20151216}\\
            &           &       & 2015-12-16& 128&  46.0& 1000 &  2 &  2& Fig.\ref{fig:J1239_20151212-20151216}\\
            &           &       & 2015-12-19& 128& 276.4& 6000 &  2 &  1& Fig.\ref{fig:J1239_20151219-20160218}\\
            &           &       & 2016-02-18& 256&  69.1& 1500 &  2 &  1& Fig.\ref{fig:J1239_20151219-20160218} \\
J1509$+$5531& B1508$+$55& 0.739 & 2017-10-30& 256&  56.7& 4600 &  1 &  2& Fig.\ref{fig:J1509_20171030AB}\\
J1709$-$1640& B1706$-$16& 0.653 & 2017-10-27& 256& 217.7& 1250 & 16 &  1& Fig.\ref{fig:J1709phase-t}\\
            &           &       & 2017-10-27$\rm \_A$& 256&  76.2& 3500 &  2 &  1& Fig.\ref{fig:J1709_20171027-20171103}\\
J1844$+$00  &           & 0.460 & 2018-08-09& 128& 180.1&  734 & 32 &  1& Fig.\ref{fig:J1844_20180809}\\
J1932$+$1059& B1929$+$10& 0.226 & 2016-02-21& 256& 274.9&73000 &  1 &  4& Fig.\ref{fig:J1932_20160221AB}, Fig.\ref{fig:J1932_20160221CD}\\
J2022$+$5154& B2021$+$51& 0.529 & 2015-06-16& 256&  30.9& 3500 &  1 &  1& Fig.\ref{fig:J2022_20150615-20171101}\\
            &           &       & 2017-11-01& 256&  88.2&10000 &  1 &  1& Fig.\ref{fig:J2022_20150615-20171101}\\
            &           &       & 2017-11-06& 256&  97.0&11000 &  1 &  1& Fig.\ref{fig:J2022_20171106}\\
J2048$-$1616& B2045$-$16& 1.961 & 2015-06-18& 256&  29.3&  448 &  2 &  1& Fig.\ref{fig:J2048_20150618-20160808}\\
            &           &       & 2016-08-08& 256&   3.8&   58 &  2 &  1& Fig.\ref{fig:J2048_20150618-20160808} \\
J2313$+$4253& B2310$+$42& 0.349 & 2017-10-25& 256&  63.7& 8000 &  1 &  2& Fig.\ref{fig:J2313_20171025AB} \\
J2321$+$6024& B2319$+$60& 2.256 & 2015-12-17& 256& 298.3& 1984 &  4 &  1& Fig.\ref{fig:J2321_20151217}\\
            &           &       & 2017-10-30& 256& 400.8& 2672 &  4 &  4& Fig.\ref{fig:J2321_20171030AB}, Fig.\ref{fig:J2321_20171030CD}\\
\hline
\end{tabular}  
\end{table*}
%    }
%    \clearpage
%    \twocolumn

% nulling fraction, null-emission interaction
To quantify the degree of nulling of a pulsar, nulling fraction,
$f_{\rm n}$, has been used to describe the percentage of periods with
no detectable emission \citep{rit76}, which varies from about zero to
more than 90\%, \citep[e.g.][]{njmk18}. Nulling pulsars with a similar
nulling fraction may have quite different nulling lengths
\citep{gjk12}. It is not clear if nulling is a random
phenomenon. Non-randomness of nulls suggests that the emission or null
of a pulse is not independent on the state of the preceding pulse. The
non-randomness assumption can be tested either by the Waled-Wolframite
statistics or the Kolmogorov-Smirnov (K-S) test
\citep[e.g.][]{rr09}. \citet{gjk12} argued that the occurrence of
nulls is not random from the statistics of the distributions of null
lengths, but the duration of which is random from speculation. In
fact, both the occurrence and duration should be tested for the
randomness. If the nulling of a pulsar is not random, could it be
modulated by periodic processes? Some attempts were made in previous
studies and the periodic nulling was identified for pulsars like PSR
B1133+16 \citep{hr07,bmm17}. Nulling degree and nulling scale have
been introduced by \citet{yhw14} to represent the angle in a
rectangular coordinate for the numbers of null and emission,
$\alpha=ArcTan(\frac{N_{\rm null}}{N_{\rm emission}})$, and their
square length, $s=\sqrt{N_{\rm null}^2+N_{\rm emission}^2}$. A
null-emission pair with a nulling degree of $45^\circ$ means that the
null and emission are of the similar length within a nulling cycle on
average, a nulling degree of $\sim0^\circ$ or $90^\circ$ means that a
pulsar has no nulling or is dominated by nulling.

% nulling properties
Despite the global features of nulling, a zoomed-in view of
transitions between emission and null exhibits diverse
variations. Pulse emission generally sets up abruptly, but it can
either decay exponentially to the nulling state, e.g. PSR B0818$-$41
\citep{bgg10}, or cease abruptly, e.g. PSR B0031$-$07
\citep{gjw14}. While for pulsars like PSR J1727$-$2739, the
transitions from emission to null can be either rapid or gradual
\citep{wwy+16}. These features of nulling might be related to the
failure of generation of particles near pulsar polar cap region, or
the loss of coherence for emissions \citep{fr82}, or the change of
emission mechanism \citep{zqh97}, or due to the geometry of view
\citep{hr07}.

%%%%%%%%%%%%%%%%%%%%%%%%%%%%%
% Why needs our research
Sensitive single pulse observations with long observation time are
needed for the improvement of our understanding of pulsar
nulling. Most of the previous observations were carried out at
frequencies lower than 1400~MHz. Nulling behaviors at high frequencies
may be different, because emission of high frequencies originates from
the low altitude in the magnetosphere where relativistic particles
responsible for the emission may go out in different bunches
\citep[e.g.][]{whw13}. Nulling observations at high frequencies can
verify the broadband feature and distinguish if nulling is physical or
geometrical origin and if it is caused by the global change of a
pulsar magnetosphere.

%%%%%%%%%%%%%%%%%%%%
\begin{figure*}
  \centering
  \includegraphics[angle=0,height = 0.8\textheight] {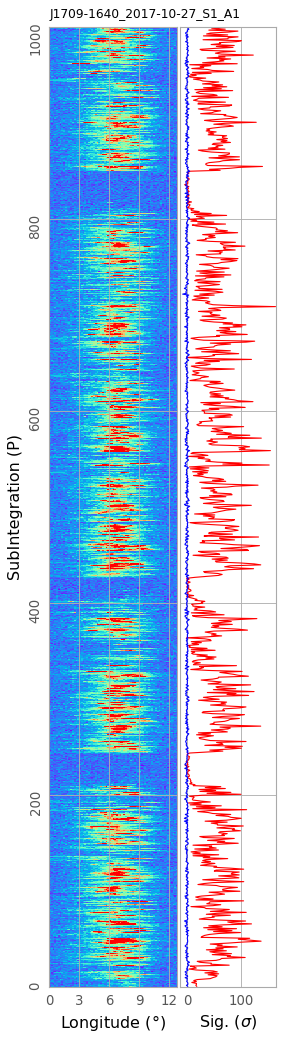}
  \includegraphics[angle=0,height = 0.8\textheight] {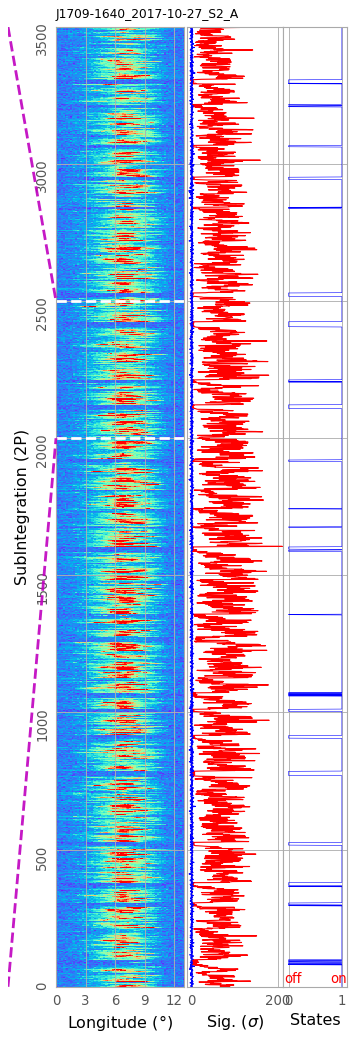}
  \includegraphics[angle=0,height = 0.8\textheight] {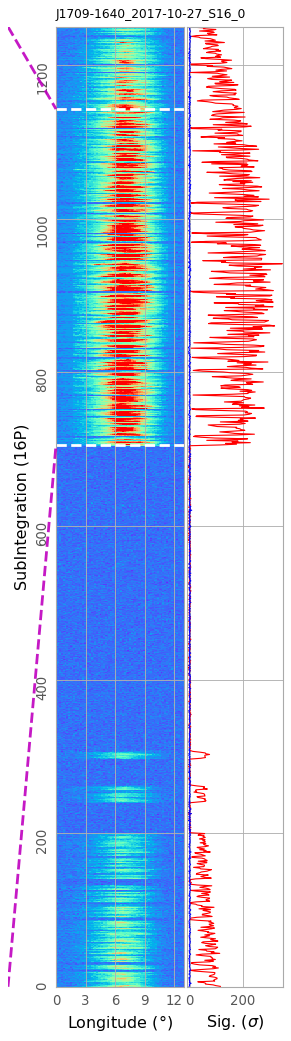}
  \caption{An example of pulse sequence taken on 2017-20-27 for PSR
    J1709-1640. {\it Left plots:} part of the observation displayed
    with single pulses. Left panel is for intensities of successive
    pulses. Right panel is for the significance of on pulse (red) and
    off pulse (blue) emissions for each pulse. {\it Middle plots:} a
    much longer part of the observation taken for nulling
    analysis. The subintegrations are formed every two pulses, with
    its emission (1) or null (0) state indicated. {\it Right plots:}
    the entire observation with subintegrations formed every 16
    pulses.}
  \label{fig:J1709phase-t}
\end{figure*}

In this paper, we investigate nulling behaviors of 20 pulsars from
very long time pulsar observations carried out at 2250~MHz by using
the Jiamusi 66-m telescope. This paper is organized as follows. In
Section 2, we briefly describe the observation details and data
reduction procedures. Nulling analysis methods and nulling behaviors
of individual pulsar are presented in Section 3. Discussion and
conclusions are given in Sections 4.

%%%%%%%%%%%%%%%%%%%%%%%%%%%%%%%%%%%%%%%%%%%%%%%
\section{Observations and Data reduction}

Observations of 20 pulsars were carried out by using the Jiamusi 66-m
telescope at Jiamusi Deep Space Station, China Xi'an Satellite Control
Center from 2015 June to 2018 September. The receiver used for
observations has a central frequency of 2250~MHz and bandwidth of
about 140~MHz. Intermediate frequency signals from the left and right
hand polarization were sampled, channelized and added together with a
digital backend. Its data products generally have 256 spectral
channels with a frequency resolution of 0.58MHz and a time resolution
of 0.2~ms, or turn to 128 spectral channels with a frequency
resolution of 1.16MHz and a time resolution of 0.1~ms. More details
can be found in \citet[]{hhp+16}.

Offline data processing are as follows. The total power is first
scaled with respect to flux-calibrator observations (e.g. 3C286, 3C48)
for each frequency channel. The data are then dedispersed and binned
to form single pulses for each frequency channel according to the
ephemeride of a pulsar with DSPSR \citep{vb11}. Radio frequency
interference is then identified and exercised from a 2-D frequency and
time domain through statistical analysis of off-pulse and whole-pulse
intensities with PSRCHIVE \citep{hvm04}. Further, the dynamic spectra
are formed from the data to estimate the influence of interstellar
scintillation by following \citet{whh+18}. In fact, most of the
observations are affected by interstellar scintillation due to small
dispersion measures (DMs), with decorrelation bandwidths of
scintillation of these pulsars larger than or comparable with the
observation bandwidth (140~MHz). To reduce its influence on our
results, only the scintillation enhanced blocks of data are chosen for
this nulling analysis.

Observational parameters of the 20 pulsars are listed in
Table~\ref{table:obs}. The number of frequency channels $N_{\rm ch}$
and time duration for each observation $T_{\rm obs}$ in minutes are
listed in column 5 and 6. For weak pulsars, consecutive $2^n$ pulses
are integrated. The number of subintegrations $N_{\rm sub}$ are listed
in column 7 with the number of $N_{\rm int}$ ($2^n$) pulses integrated
in each subintegration indicated in column 8. If one data is affected
by broad-band radio frequency interferences or faded by interstellar
scintillation, these corrupted data blocks are omitted. The number of
total data blocks $N_{\rm blk}$ is listed in column 9. Plots of pulse
sequences are presented in appendix~\ref{appendixA} and indicated in
column 10.

An example of pulse sequences is shown in
Figure~\ref{fig:J1709phase-t} for PSR J1709-1640, with some apparently
long nulls. A long block of scintillation enhanced data is chosen for
nulling analysis, as demonstrated in the middle and left plots.

%%%%%%%%%%%%%%%%%%%%%%%%%%%%%%%%%%%%%%%%%%%%%%%
\section{Nulling Analysis}

%***************************************
\subsection{Emission-null sequences and nulling fraction}

\begin{figure}
  \centering
  \includegraphics[angle=0, width=0.45\textwidth] {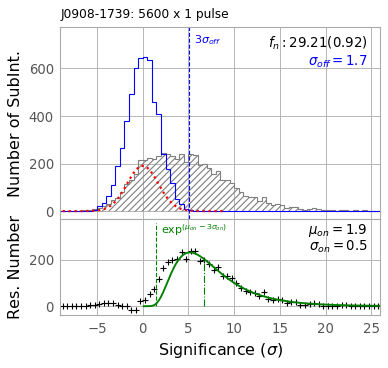}\\
  \includegraphics[angle=0, width=0.45\textwidth] {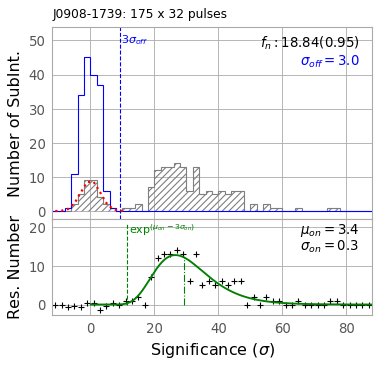}\\
  \caption{Histograms of the significance of on-pulses (grey hatched
    step) and off-pulses (blue no hatched step) for PSR J0908$-$1739
    taken on 2016-05-23. The upper plots are for statistics of single
    pulses, and the lower plots are for subintegrations of every 32
    pulses. Distribution of the off-pulse significance can be modeled
    by a normal function with a mean of zero and the standard
    derivation of $\sigma_{\rm off}$. The normal function is scaled
    with $f_{\rm n}$ and represented by red dotted line. It is subtracted
    from the distribution of on-pulse significance with the residuals
    indicated by black "+" dots in the panel. The residual
    distribution is modeled by a lognormal function with a mean of
    $\mu_{\rm on}$ and the standard derivation of $\sigma_{\rm
      on}$. Green solid, dash-doted and dashed lines are for the
    modeled distribution, the mean and the lower 3-sigma limit.}
  \label{fig:J0908NF}
\end{figure}

%%%%%%%%%%%%%%%%%%%%%%%%%%%%%%%%%%%%%%
\begin{table*}
  \centering
  \footnotesize
  %\scriptsize      
  \caption{Nulling Statistics of 20 pulsars.}
  \label{table:nf}
  \tabcolsep 1.5mm
  %\tiny
  \begin{tabular}{crclrrrcccc}
    \hline  
    \hline
JName &  $N_{\rm sub}$ & $f_{\rm n}$  & $P_{\rm cri}$ & $N_{\rm c}$ & $N_{\rm e}$ & $N_{\rm n}$ & $\alpha_1$&$\alpha_2$&$s_1$&$s_2$  \\
%\cline{8-9}  
      &          &($\%$) &  ($\%$)   &           &    (P)    &   (P)   & ($^\circ$)&($^\circ$)& (P) & (P)     \\
 (1)  & (2)      & (3)   &   (4)     &    (5)    &   (6)     &   (7)   &   (8)  &  (9)  & (10) & (11) \\
\hline
J0034$-$0721&   380& $12.4^\dagger<f_{\rm n}<37.4$ & 0.834  & -  &    -        &    -     &     - &  -  &  -  & -      \\
J0248$+$6021&  1312& $0.0<f_{\rm n}<34.7$       & 0.125  & -  &    -        &    -     &     - &  -  &  -  & -       \\
J0304$+$1932&   300& $8.9<f_{\rm n}<17.5$       & 0.012  &  16&  132.8(76.0)& 17.6(7.2)& 9.4(4.7)&  11.6(9.3)& 134(76)& 128(72) \\
J0332$+$5434& 22000& $0.0\pm0.0$              & 5.2e-04& -  &    -        &   -      &     - &  -  &  -  & -      \\
J0528$+$2200&   243& $8.1^\dagger<f_{\rm n}<22.3$ & 1.158  & -  &    -       &    -     &     - &  -  &  -  & -      \\
J0543$+$2329&  4582& $0.0<f_{\rm n}<16.6$       & 0.003  & -  &     -       &    -     &     - &  -  &  -  & -      \\
J0826$+$2637&  4745& $0.0<f_{\rm n}<3.4$        & 0.012  & -  &      -      &     -    &     - &  -  &  -  & -       \\
J0908$-$1739&   175& $18.8<f_{\rm n}<29.2$      & 0.008  &   2&1526.4(992.0)&512.0(32.0)&     - &  -  &  -  & -      \\
J0922$+$0638&  9600& $0.0<f_{\rm n}<0.06$       & 0.095  & -  &       -     &      -   &     - &  -  &  -  & -       \\
J0953$+$0755&  5850& $0.1<f_{\rm n}<11.8$       & 0.193  & -  &        -    &       -  &     - &  -  &  -  & -       \\
J1136$+$1551&  2078& $0.3<f_{\rm n}<11.5$       & 0.038  &  16& 374.8(374.0)&      4(0)&  1.4(1.4)&   2.0(1.7)&353(361)&218(159) \\
J1239$+$2453&  9000& $1.1<f_{\rm n}<2.8$        & 0.043  & 161& 106.8(107.2)&  2.4(1.0)&  3.7(6.1)&   3.7(6.2)&106(108)& 104(97) \\
J1509$+$5531&  4600& $2.0\pm0.1$              & 0.067  & 120&   36.0(35.3)&  1.7(1.2)& 8.2(11.3)&  8.2(12.2)&  36(35)&  36(35)\\
J1709$-$1640$^\ddag$&  3500& $2.1<f_{\rm n}<3.4$ & 0.007  &  36& 178.2(178.8)& 11.2(9.6)&26.3(31.3)& 22.4(26.0)&176(175)&180(180) \\
J1844$+$00  &   734& $0.3<f_{\rm n}<34.0$       & 4.7e-07&   1&  11616(4512)&    256(0)&     - &  -  &  -  & -      \\
J1932$+$1059& 73000& $0.03\pm0.01$            & 0.024  & 243& 294.5(430.1)&  1.0(0.2)&  1.6(3.1)&   1.7(3.1)&297(432)&291(431)  \\
J2022$+$5154& 24500& $0.12\pm0.01$            & 1.2e-03&  75& 313.0(346.7)&  1.2(0.4)&  2.7(7.4)&  3.3(10.5)&321(351)&310(345) \\
J2048$-$1616&   506& $2.1<f_{\rm n}<6.4$        & 0.112  &  30&   29.8(34.2)&  3.0(1.6)&14.7(13.5)& 14.4(14.8)&  27(30)&  31(35) \\
J2313$+$4253&  8000& $5.2\pm0.2$               & 6.2e-05& 219&   33.2(34.9)&  3.0(2.5)&21.9(27.1)& 20.4(24.0)&  34(35)&  34(35)  \\
J2321$+$6024&  4656& $8.5<f_{\rm n}<18.5$       & 0.053  & 393&   36.4(39.6)& 10.8(8.4)&25.8(20.2)& 25.5(19.0)&  40(39)&  40(39)  \\
\hline

\multicolumn{11}{l}{Note. $^\dagger$ distributions of the
  significances for emission and null pulses can not be
  resolved. $^\ddag$ analysis is for the block of data of
  2017-10-27$\_$A. }  \\

\multicolumn{11}{l}{In the table, $N_{\rm sub}$ is from
  Table~\ref{table:obs}, $f_{\rm n}$ is the nulling fraction or its upper or
  lower limits, $P_{\rm cri}$ is the probability for misidentification
  of nulls, } \\

\multicolumn{11}{l}{$N_c$ is the number of nulling cycles, $N_e$ and $N_n$ are the
  mean lengths of emission and null with their standard derivations
  given in the parentheses,}  \\

\multicolumn{11}{l}{($\alpha_1$,
  $\alpha_2$) and ($s_1$, $s_2$) are the nulling degrees and scales for
  $EN$ and $NE$ paris from observations.}\\
  \end{tabular}
\end{table*}

The emission or null state of pulsars are analyzed in a few steps.
Although most of the intense RFIs have already been exercised in a
previous data processing step, some pulsars are still affected by low
levels of RFI. To get a flat baseline, we perform a least square
fitting of a fifth order polynomial to the off-pulse regions for each
sub-integration, and subtract them from the data. An off-pulse window
with the same number of bins ($n_{\rm bins}$) as the on-pulse one is
selected. The significance of emission from both the on and off pulse
windows are calculated for each sub-integration,
\begin{eqnarray}
  \rm S_{on} &=&\frac{1}{\sigma \sqrt{W_{\rm eq}}} \sum_{i=1}^{n_{\rm bins}} I_{\rm on,i} \nonumber \\
  \rm S_{off} &=&\frac{1}{\sigma \sqrt{W_{\rm eq}}} \sum_{i=1}^{n_{\rm bins}} I_{\rm off,i}.
\end{eqnarray}
Here, $\sigma$ is the standard derivation of off-pulse emission,
$W_{\rm eq}$ is the equivalent width of a top-hat pulse with the same
area and peak height as the pulse profile integrated from all the
subintegrations. The significance sequences are obtained for both the
on and off pulses, which are then binned to form the histogram
distributions. The distribution of off-pulse significance can be well
modeled by a normal function with a mean of zero, an amplitude of $A_0$
and the standard deviation of $\sigma_{\rm off}$,
\begin{equation}
D_{\rm off}(x) = A_0 \exp \left(-\frac{x^2}{2 \sigma_{\rm off}^2} \right).
\end{equation}
Any excess out of this distribution for the on-pulse significance
indicates that some pulses are in nulling state. We fit the normal
function with a mean of zero and the standard deviation $\sigma_{\rm
  off}$ to the on-pulse distribution with significance less than zero
to obtain the amplitude $A_1$. The ratio is termed as the nulling
fraction $f_{\rm n}=A_1/A_0$, which has an uncertainty of
\begin{equation}
{\sigma}_{f_{\rm n}} = \sqrt{\left(\frac{\delta A_1}{A_0}\right)^2+\left(\frac{A_1 \delta A_0}{A_0^2}\right)^2}.
\end{equation}
Here, $\delta A_0$ and $\delta A_1$ represent the uncertainties
obtained from fitting. The scaled off-pulse distribution with $f_{\rm n}$ is
then subtracted from the distribution of on-pulse
significance. Residual distribution generally follows a lognormal
distribution,
\begin{equation}
D_{\rm on} (x)= \frac{B_0}{x}  \exp \left[-\frac{(\ln x-\mu_{\rm on})^2}{2 \sigma_{\rm on}^2} \right].
\end{equation}
%%%%%%%%%%%%%%%%%
Here, $\mu_{\rm on}$ and $\sigma_{\rm on}$ are the mean and standard
deviation of the normally distributed logarithm of the on-pulse
significance.

The value $3 \sigma_{\rm off}$ represents the upper limit for the
distribution of off pulse significance. If $\exp(\mu_{\rm on}-3
\sigma_{\rm on})>3 \sigma_{\rm off}$, it means that the pulses in the
emission and null states are separated. The estimated $f_{\rm n}$ represents
a real fraction of pulses in nulling states. If $\exp(\mu_{\rm on}-3
\sigma_{\rm on})<3 \sigma_{\rm off}$, the distributions for the
emission and null pulses can not be separated. The obtained $f_{\rm n}$ is
then overestimated, which represents an upper limit. When the possible
short nulls are ignored, every $2^n$ pulses are integrated for the
pulse sequence until $\exp(\mu_{\rm on}-3 \sigma_{\rm on})>3
\sigma_{\rm off}$. The so-obtained $f_{\rm n}$ represents a lower limit.

An example analysis of the nulling fraction is shown in
Figure~\ref{fig:J0908NF} for PSR J0908-1739. The distribution of
on-pulse significance is much broader, whose lower 3-sigma limit
$\exp(\mu_{\rm on}-3 \sigma_{\rm on})$ is smaller than $3 \sigma_{\rm
  off}$. The nulling pulses around zero are not well separated from
the emission pulses, and the so estimated $f_{\rm n}$ of 29.2\% represents
an upper limit. In the bottom panels, every 32 pulses are integrated
to improve the significance of pulses. The histograms for
subintegrations with significance around zero are separated from those
for the emission ones. The so estimated $f_{\rm n}=18.8$\% represents a
lower limit, because nulls shorter than 32 pulses can not be resolved.

It should be noted that nulling fractions can be estimated either from
the distributions of on and off pulse energy
\citep[e.g.][]{rit76,gjk12} or from the distributions of their
significance \citep{lbr+13}. In previous researches, the on and off
pulse energies were generally scaled by averages of energies for every
200 pulses to compensate for energy variation caused by interstellar
scintillation following \citet{rit76}. For long nulls (e.g. PSR
J1709$-$1640), such robust correction bias the off-pulse energy
distribution from the Gaussian, as noticed by
\citet{viv95}. Identification of the emission or null state of a pulse
or subintegration is often based on an artificially setting threshold
significance, e.g. S/N=3. If the energy distribution has a
$\sigma_{\rm off}>1$, the threshold S/N=3 is not enough to
discriminate the emission and null states. Our identification is based
on the distribution of pulse significance with the following
criteria,

(1) any pulse with $\rm S_{on} \ge 4 \sigma_{\rm off}$ is classified
as an emission state.

(2) any pulse with $\rm S_{on} \le 2 \sigma_{\rm off}$ is classified
as a null for non-detection.

(3) if a pulse is of 2 $\sigma_{\rm off} < \rm S_{on} \le 3
\sigma_{\rm off}$ and the adjacent pulses ahead and behind have $\rm
S_{\rm on}$ > 3 $\sigma_{\rm off}$, it is classified as an emission
state, otherwise a null.

(4) if a pulse has 3 $\sigma_{\rm off}$ < $\rm S_{on}$ < 4
$\sigma_{\rm off}$ and the adjacent in the two sides have $\rm S_{\rm
  on}$ < 3 $\sigma_{\rm off}$, it is termed as a null, otherwise an
emission state.

The identified emission or null state for the example pulse sequence
of PSR J1709-1640 is also shown in Figure~\ref{fig:J1709phase-t} with
diverse nulling behaviors. The left and middle panels of the
observation represent the normal states of the pulsar, which usually
stays in an on state with short nulls of several to several tens of
pulses. But occasionally, it shows a long null (as of 6368 pulses
for $\sim$1.2h) as shown by the right panels for the entire pulse
sequence. Moreover, the sequence shows moderate nulls of 640 and 560
pulses before it goes into the long null, similar to those in the
observation by \citet{njmk18}.

By this approach, we analyzed nulling fractions and states of 20
pulsars, as shown in Figure~\ref{fig:nfs} and listed in column 3 of
Table~\ref{table:nf}, with the number of pulses or subintegrations
employed for the analysis listed in column 2. For 5 PSRs J0332+5434,
J1509+5531, J1932+1059, J2022+5154, and J2313+4253, the $f_{\rm n}$ are
obtained since their distributions of on and off-pulse significance
are separated with single pulses. For the other pulsars, only upper
and lower limits of $f_{\rm n}$ are estimated. The upper limits are obtained
with single pulses whose distributions of on and off-pulse
significance are not resolved, and the lower limits are obtained with
subintegrations with the separated distributions of on and off-pulse
significance. The cumulative probability, $P_{\rm cri}$, for the
modeled on-pulse distribution at $3 \sigma_{\rm off}$ is calculated,
as listed in column 4 in Table~\ref{table:nf}, which is the
probability of misidentification of emission to null for the resolved
on and off pulses or subintegrations.

%%%%%%%%%%%%

%%***************************************
\subsection{Emission and null lengths}

%%%%%%%%%%%%%%%%%%%%%%%%%%%%%%%%%%%%%
\begin{figure}
  \centering
  \includegraphics[angle=0, width=0.48\textwidth] {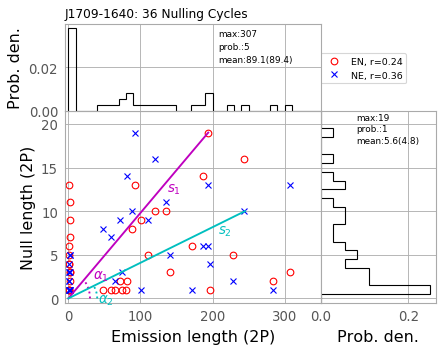}\\
  \caption{Distributions of emission and null lengths of PSR
    J1709$-$1640. Pairs for the emission and the next null are denoted
    as $EN$ and represented by red "o" sign, while pairs for the
    emission and the pre null are denoted as $NE$ and represented by
    blue "$\times$". Nulling degrees, $\alpha_1$ and $\alpha_2$, and
    nulling scale, $s_1$ and $s_2$, are indicated for $EN$ and
    $NE$. The correlation coefficients between emission and null
    lengths are shown in the top right corner. Histograms for the
    emission and null lengths are drawn in the top and right
    panels. Lengths of the longest emission or null state, their most
    probable and mean values are indicated in each panel. }
  \label{fig:J1709EN_dis}
\end{figure}

\begin{figure}
  \centering
  \includegraphics[angle=0, width=0.4\textwidth] {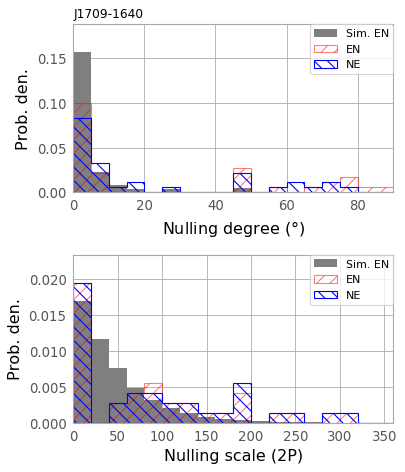}\\
  \caption{Histograms of nulling degrees and nulling scales of
    emission-null length pairs of PSR J1709$-$1640. The left hatched
    red step and right hatched blue step represent distributions for
    the $EN$ and $NE$ length pairs, respectively. The grey bars
    are for the $EN$ length pairs of randomly distributed
    emission-null sequence from simulation.  }
  \label{fig:J1709EN_cor}
\end{figure}

The emission and null lengths represent the durations for a pulsar
staying contiguously at a given state. A emission-null (1-0) sequence
is composed of inter-changing emission and null states with lengths of
($N_{e_1}$, $N_{n_1}$, $N_{e_2}$, $N_{n_2}$, $N_{e_3}$, ...). The
number of state changing cycles, $N_c$, is listed in column 5 of
Table~\ref{table:nf} for each pulsar. Its ratio, $N_c/N_{\rm sub}$,
represents the nulling cadence during one observation. Distributions
of emission and null lengths vary a lot, their means, <$N_e$> and
<$N_n$>, together with standard derivations given in parentheses are
listed in columns 6 and 7 of Table~\ref{table:nf}. Previously, it was
suggested that more emission and null states tend to have small
lengths and exponential functions were usually employed to model both
distributions \citep[e.g.][]{gjk12}.

An example of distributions of emission and null lengths is shown in
Figure~\ref{fig:J1709EN_dis} for PSR J1709-1640, with length pairs of
$EN$ and $NE$ represented by red ``o''s and blue ``$\times$''s in the
2-D length plan (bottom-left). Histograms for the emission and null
lengths are shown in the top and right panels, which do not follow
exponential functions. Figure~\ref{fig:J1709EN_cor} shows the
distributions of nulling degrees and scales for both the $EN$ and $NE$
length pairs of PSR J1709$-$1640. It is apparent from the distribution
of nulling degrees that nulls can have variable portions within a
nulling cycle, but more cycles tend to have small nulling
degrees. There are more cycles tending to have small nulling scales,
but the distribution of which does not decay exponentially with the
scale length.

%angle - square treatment:
The duration of an emission state might be related to the length of
its prior or post nulls. From one emission-null sequence, the emission
and the next null length pairs [($N_{e_1}$, $N_{n_1}$), ($N_{e_2}$,
  $N_{n_2}$), ...], named as $EN$, can be formed together with the
emission and the pre null length pairs [($N_{e_2}$, $N_{n_1}$),
  ($N_{e_3}$, $N_{n_2}$), ...], named as $NE$.  As introduced by
\citet{yhw14}, interaction between the states of emission and null can
be demonstrated in terms of nulling degrees, $\alpha_1$ and
$\alpha_2$, and nulling scales, $s_1$ and $s_2$, that correspond to
emission and null length pairs from $EN$ and $NE$. Their means and the
standard derivations given in parentheses are listed in columns 8 to
11 of Table~\ref{table:nf}.

By this approach, the emission and null lengths are analyzed for 10 of
the 20 pulsars that have well distinguished emission and null states,
as listed in Table~\ref{table:st} and shown in
Figure~\ref{fig:dis-cor}.

%%%%%%%%%%%%%%%%%%%%%%%%%%%%%%%%%%%%%%%%%
\begin{table*}
  \centering
  %\small
  \footnotesize
  %\scriptsize      
  \caption{Results for correlations and randomness tests.}
  \label{table:st}
  \tabcolsep 1.5mm
  %\tiny
  \begin{tabular}{crrcccccc}
    \hline  
    \hline

 JName  &\multicolumn{2}{c}{Corr. Coeff.} & \multicolumn{6}{c}{K-S Statistic Deviative / p-value}  \\
  \cline{4-9}
        &  $EN$  & $NE$  &$\alpha_1$vs $\alpha_2$&$\alpha_1$vs $\alpha_1^s$&$\alpha_2$vs $\alpha_2^s$&$s_1$vs $s_2$&$s_1$vs $s_1^s$&$s_2$vs $s_2^s$ \\
 (1) &   (2)     &  (3)   &  (4) & (5) & (6) & (7) & (8) & (9) \\
\hline
J0304$+$1932& 0.62&-0.14&  0.15/0.98& 0.22/0.44   & 0.13/0.96   & 0.11/1.00& 0.36/3.0e-02& 0.38/2.7e-02\\
J1136$+$1551&  -  &  -  &  0.19/0.95&0.31/6.8e-02& 0.43/3.3e-03& 0.19/0.95& 0.31/6.6e-02& 0.43/3.2e-03\\
J1239$+$2453&-0.06&-0.06&  0.03/1.00& 0.24/9.6e-09& 0.23/5.5e-08& 0.02/1.00& 0.21/1.5e-06& 0.20/3.4e-06\\
J1509$+$5531& 0.00& 0.05&  0.05/1.00& 0.24/1.7e-06& 0.23/3.2e-06& 0.02/1.00& 0.14/2.2e-02& 0.14/1.9e-02\\
J1709$-$1640& 0.24& 0.36&  0.19/0.51&0.37/1.3e-04& 0.42/6.4e-06& 0.17/0.71& 0.32/1.5e-03& 0.32/1.5e-03\\
J1932$+$1059&-0.04& 0.00&  0.02/1.00&0.67/9.1e-91& 0.68/8.2e-93& 0.01/1.00& 0.67/9.1e-91& 0.68/1.1e-91\\
J2022$+$5154& 0.17&-0.12&  0.08/0.97&0.34/3.7e-08& 0.33/7.7e-08& 0.04/1.00& 0.30/2.0e-06& 0.31/5.6e-07\\
J2048$-$1616& 0.30&-0.03&  0.10/0.99& 0.35/1.2e-03& 0.39/2.4e-04& 0.09/1.00& 0.29/1.2e-02& 0.28/2.0e-02\\
J2313$+$4253& 0.17& 0.22&  0.07/0.68& 0.23/3.8e-10& 0.24/2.7e-11& 0.05/0.95& 0.24/5.8e-11& 0.24/5.8e-11\\
J2321$+$6024&-0.04& 0.12&  0.05/0.74& 0.33/2.7e-38& 0.37/6.4e-48& 0.03/0.97& 0.14/2.1e-07& 0.14/1.2e-06\\
\hline
  \end{tabular}
\end{table*}

1) A few pulsars have dominant emission states and nulls generally
last for short durations, for example, PSRs J1932+1059 and J2022+5154.

2) The nulling degrees are in the range from about 1.4 to 26.3
degrees.

3) The nulling scales can have lengths from 31 to 353 pulses with
comparable deviations.

4) For most pulsars, the distributions of emission and null lengths
cannot be well-modeled by the exponential functions for the stochastic
Poisson processes except for the emission lengths of PSRs J1509+5531
and J1932+1059.

%%***************************************
\subsection{Interactions between emission and null}

For nulling pulsars, it was intuitively thought that the duration of
an emission state might be dependent on the duration of the preceding
nulling state, or vice versa. Correlations between the emission and
null lengths are examined, and correlation coefficients are listed in
columns 2 and 3 of Table~\ref{table:st} for the $EN$ and $NE$ length
pairs. The correlation coefficients for $EN$ vary from -0.04 to 0.62
with a pair number weighted average of 0.03. While for $NE$ length
pairs, the coefficients vary from -0.14 to 0.36 with a weighted
average of 0.07. It means that the duration of an emission or null
state can be correlated with the duration of its preceding null or
emission for some pulsar, but only with a marginal significance.

To further quantify the difference between the $EN$ and $NE$, two
sample K-S tests are carried out for both the nulling degrees
($\alpha_1$ vs $\alpha_2$) and nulling scales ($s_1$ vs $s_2$). Here,
the tests are performed on the unbinned data, which return p-values
and the statistic deviatives that represent the maximum differences
between two cumulative distribution functions, as listed in columns 4
and 7 of Table~\ref{table:st}. If a p-value is larger than the
significance level 0.1, the hypothesis that length pairs of $EN$ and
$NE$ come from the same distribution cannot be rejected. For 10 of the
20 pulsars with $N_c>10$, the afore-mentioned statistics and K-S tests
are performed. We found:

1) For $\alpha_1$ and $\alpha_2$, the K-S tests have maximum statistic
deviative of 0.19 and p-values larger than 0.51, much larger
than the significance level of 0.1.

2) For $s_1$ and $s_2$, the K-S tests have maximum statistics
deviative of 0.19 and p-values larger than 0.71, also larger than the
significance level.

In summary, no significant difference is found between $EN$ and $NE$
length pairs from the distributions of $\alpha$ and $s$. Both the $EN$
and $NE$ may come from the same processes.

%%***************************************
\subsection{Randomness tests}

% simulation
To examine if the emission and null interact randomly, the statistical
analysis is done as following. A sequence with the same size as
observations of a pulsar is first simulated from a uniform
distribution $[0, 1)$. A threshold of nulling fraction is then set to
  the sequence, any value below that is denoted as null and vice visa
  for those above the threshold. The $EN$ and $NE$ length pairs are
  extracted from the simulated 1-0 sequence. If observations of a
  pulsar have more than 10000 subintegrations or pulses, 1000 sets of
  randomly distributed 1-0 sequences are simulated to give average
  estimates for $\alpha^s$ and $s^s$. Otherwise, 10000 sets of
  sequences are simulated for the estimation. From the $EN$ and $NE$
  length pairs of these simulated sequences, nulling degrees and
  scales are calculated, which are termed as $\alpha_1^s$,
  $\alpha_2^s$, $s_1^s$ and $s_2^s$.

Two sample K-S tests are taken on the observed and simulated sequences
for $\alpha_1$ vs $\alpha_1^s$ and $\alpha_2$ vs $\alpha_2^s$ for the
occurrence of nulls. The statistic derivatives and p-values are listed
in columns 5 and 6 of Table~\ref{table:st} for the $EN$ and $NE$
length pairs. Same is done for $s_1$ vs $s_1^s$ and $s_2$ vs $s_2^s$
to test the duration of nulling cycles, as listed in columns 8 and 9
of Table~\ref{table:st}. The simulated distributions are shown in
Figure~\ref{fig:J1709EN_cor} as an example and in
Figure~\ref{fig:dis-cor} for the 10 pulsars.

1) For the occurrence of nulls, the randomness hypothesis is rejected
at high significance levels (p-values smaller than 0.1) for 9 pulsars
in Table~\ref{table:st} from both $EN$ and $NE$ except for PSR
J0304+1932.

2) For the duration of nulling cycles, the randomness hypothesis is
rejected at high significance levels (p-values smaller than 0.1) for
all the 10 pulsars from both $EN$ and $NE$.

In summary, emission and null do not interact randomly in the aspects
of both the occurrence and duration for 9 pulsars. The occurrence of
null can be random but its duration is not random for PSR J0304+1932.

%***************************************
\subsection{Nulling periodicity}

%%%%%%%%%%%%%%%%%%%%%%%%%
\begin{table}
  \centering
  \footnotesize
  %\scriptsize      
  \caption{Nulling periocidity and intensitiy variations of 10 pulsars.}
  \label{table:pI}
  \tabcolsep 3.0mm
  %\tiny
  \begin{tabular}{cccc}
    \hline  
    \hline
JName &  null freq. &$\frac{<I_2>}{<I_1>}$&$\frac{<I_{-1}>}{<I_{-2}>}$  \\
%\cline{8-9}  
      &    (1/P) &       &    \\
\hline
J0304$+$1932& 0.011& 1.16& 0.74\\
J1136$+$1551&  -   & 1.52& 1.17\\
J1239$+$2453&  -   & 0.96& 0.90\\
J1509$+$5531& 0.043& 1.64& 0.67\\
J1709$-$1640$^\ddag$& 0.005& 1.73& 0.71\\
J1932$+$1059&  -   & 1.36& 0.78\\
J2022$+$5154&  -   & 1.18& 0.78\\
J2048$-$1616&  -   & 1.18& 0.96\\
J2313$+$4253& 0.022& 1.03& 0.94\\
J2321$+$6024&  -   & 1.28& 0.86\\
\hline
\multicolumn{4}{l}{Note. $^\ddag$ analysis is for the block of data, 2017-10-27$\_$A. }    \\
\multicolumn{4}{l}{<$I_1$>,  <$I_2$>, <$I_{-1}$> and <$I_{-2}$>, mean intensities of the first, the }     \\
\multicolumn{4}{l}{second, the last and the one but last subintegrations of emission. } \\
  \end{tabular}
\end{table}

Blocks of emission and null pulses are generally arranged orderly
within a sequence, some of which show quasi-periodic variations, as
demonstrated in Figure~\ref{fig:J1709phase-t}. To quantify this
quasi-periodic pattern, a Fourier transform is performed on the
emission-null (1-0) time sequence instead of the sequence for the
total intensity. This is because the method concentrates only on the
states of emission and eliminates the periodicities caused by
sub-pulse drifting \citep[e.g.][]{hr07,bmm17}, mode-changing
\citep[e.g.][]{ymw+20} and amplitude modulations
\citep[e.g.][]{bmm20}. The discrete Fourier transformation (DFT)
length depends on the length of an observation, which is chosen to
ensure that the sequence has more than or equal to two but less than
four DFT lengths. For a given sequence, the DFT process repeats by
sliding 8, 16, or 32 points until the end. The power of all the DFTs
from all the observations of a pulsar are finally averaged to form the
entire spectra.

\begin{figure}
  \centering
  \includegraphics[height=9.2cm,width = 0.3\textwidth] {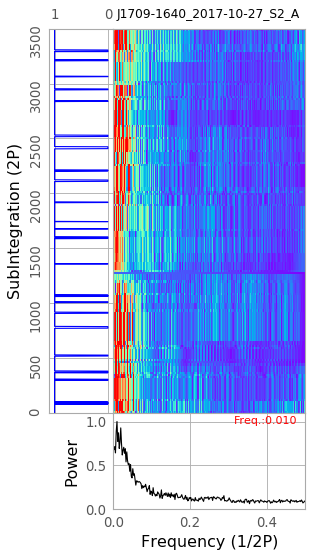}
  \caption{Sliding Fourier transform of the emission-null (1-0)
    sequence for PSR J1709$-$1640. The emission or null state of each
    subintegration is indicated in the left panel. Discrete Fourier
    transformation power from a box car of 1024 points is plotted in
    the main panel, which slides across the state sequence every 32
    points. The average power is shown in the bottom panel.}
  \label{fig:J1709EN-period}
\end{figure}

The DFT of the state sequence of one observation of PSR J1709$-$1640
is shown in Figure~\ref{fig:J1709EN-period} as an example. It is
apparent that the 1-0 sequence is modulated across a broad range of
periods. There might be a significant periodicity for part of the
sequence. But it could be averaged out for the whole sequence and lead
to the most probable modulations of about 200 periods. The 1-0
sequences of the 10 pulsars show three kinds of periodic
modulations, as shown in Figure~\ref{fig:period}.

1) Four PSRs J0304+1932, J1509+5531, J1709-1640 and J2313+4253 have
quasi-periodic nulling, with nulling frequencies listed in column 2 of
Table~\ref{table:pI}. These behaviors might be related to the
carousel beam patterns, and the nulls represent empty passes of sight
lines through the patterns \citep[e.g.][]{hr07}.

2) Emissions of PSRs J1239+2453, J2022+5154 and J2048$-$1616 are
modulated by nulls with long periodicities, and the modulation is very
strong for PSR J2321+6024. These pulsars occasionally exhibit frequent
nulls, which might result from the change of pulsar emission
mechanisms. It can be caused either by the failure of particle
production in the polar cap region of a pulsar magnetosphere or by the
lose of coherence for the relativistic particles
\citep[e.g.][]{fr82,zqh97}.

3) PSRs J1136+1551 and J1932+1059 tend to have featureless
modulations, i.e., with a `white spectrum'. These pulsars null at a
wide range of pulsar periods with no preferred periodicities.

%***************************************
\subsection{Variations of pulse intensity during emission}

\begin{figure}
  \centering
  \includegraphics[angle=0,width = 0.357\textwidth] {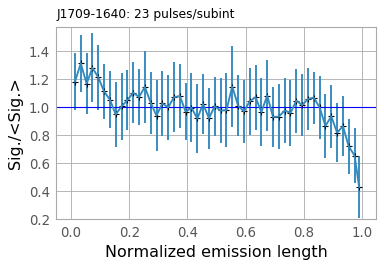}
  \includegraphics[angle=0,width = 0.357\textwidth] {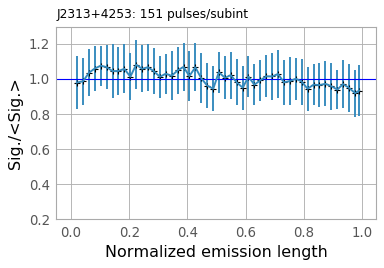}\\
  \caption{Tendency of pulse intensity variation during the emission
    states for PSRs J1709$-$1640 and J2313+4253, normalized by its
    average intensity and emission length. Black dots with error bars
    represent the average pulse intensity. }
  \label{fig:BEV}
\end{figure}

Pulse intensities vary during an emission state. For emission state
lasting for more than or equal to 5 pulses or subintegrations, the
pulse intensity sequences are first normalized by the averages and
sequence lengths, and then averaged across the normalized
length. Systematic variation of pulse intensities, if exists, should
be revealed. Integrated pulse profiles are also obtained for the
first, the second, the last two subintegrations of emission on states,
as labeled by $E_1$, $E_2$, $E_{-2}$ and $E_{-1}$,
respectively. Intensity ratios, $<I_2>/<I_1>$, for $E_2$ to $E_1$, and
$<I_{-1}>/<I_{-2}>$ for $E_{-1}$ with respect to $E_{-2}$, are listed
in columns 2 and 3 of Table~\ref{table:pI}.

\begin{figure}
  \centering
  \includegraphics[angle=0,width = 0.35\textwidth] {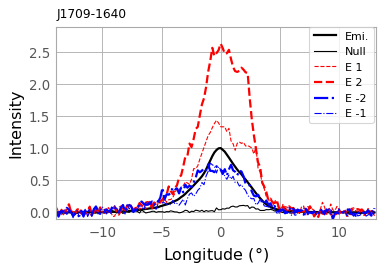}
  \includegraphics[angle=0,width = 0.35\textwidth] {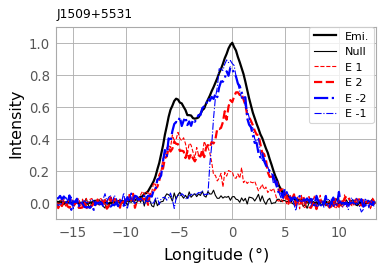}\\
  \caption{Pulse profiles of PSRs J1709$-$1640 and J1509+5531 for the
    emission and nulls are represented by the thick solid and thin
    solid lines in black. Pulse profiles for the first and second
    subintegration of the emission on states, $E_1$ and $E_2$, are
    represented by dashed lines in red, those for the last and the
    last but one subintegrations, $E_{-1}$ and $E_{-2}$, are
    represented by dash-doted lines in blue. }
  \label{fig:BEVprof}
\end{figure}

The tendencies of intensity variations for the emission states are
shown in Figure~\ref{fig:BEV} for PSRs J1709$-$1640 and J2313+4253 as
examples. PSR J2313+4253 is a representative of most pulsars, whose
pulse intensity increases abruptly from a null at the beginning and
diminish quickly to null at the end, and the intensities of these
pulses vary little during emission states. PSR J1709$-$1640
exceptionally sets up to a large intensity than the average at the
start of the emission state, and gradually diminish to null at the
end.

\begin{figure}
  \centering
  \includegraphics[angle=0,width = 0.45\textwidth] {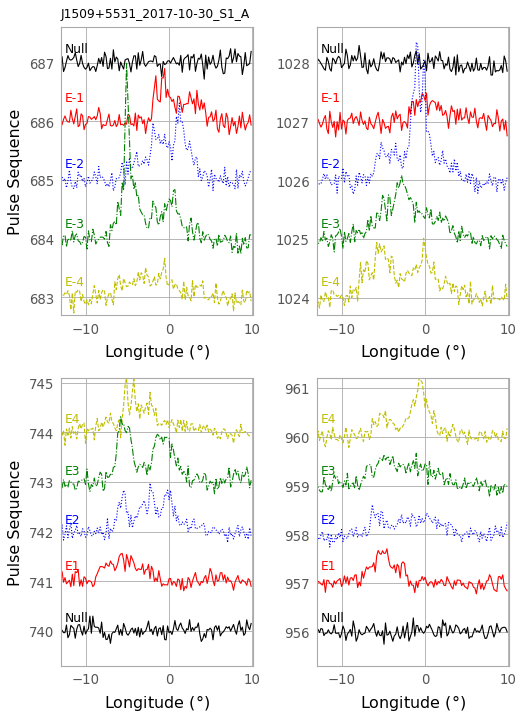}
   \caption{Pulse sequences for the emission state set up ($E_1$ -
     $E_4$) and the transition to null ($E_{-4}$ to $E_{-1}$) for PSR
     J1509$+$5531. Null pulses are represented by the solid black
     lines. }
  \label{fig:J1509evol}
\end{figure}

Variations of pulse shapes and intensities during the transitions from
emission to null states and vice versa are shown in
Figure~\ref{fig:profs}. For PSRs J0304+1932, J1239+2453, J2048$-$1616
and J2321+6024, the first two subintegrations, $E_1$ and $E_2$, and
the last two subintegrations, $E_{-2}$ and $E_{-1}$, are weaker than
the average. For PSRs J1136+1551, J1932+1059 and J2022+5154, these
pulses near the transitions are stronger than the average. PSRs
J1709$-$1640 and J1509+5531 show exceptional transitions, as
demonstrated in Figure~\ref{fig:BEVprof}. The intensities of $E_1$ and
$E_2$ of PSR J1709$-$1640 are stronger than the average, and $E_2$ is
even much stronger than $E_1$, but intensities of $E_{-2}$ and
$E_{-1}$ are weaker than the average with $E_{-1}$ weaker than
$E_{-2}$. Null-to-emission transition of PSR J1509$+$5531 starts with
the leading component of the pulse profile emerging first, and
emission-to-null transition also takes place first with the leading
component, as shown in Figure~\ref{fig:J1509evol} for the four
transitions. The phase dependency of transitions between emission and
null implies that nulling of the pulsar is not caused by global change
of pulsar magnetosphere, but just by density patch changes of
relativistic particles in parts of a magnetosphere.
%%%%
%***************************************
%%%%

%%%%%%%%%%%%%%%%%%%%%%%%%%%%%%%%%%%%%%%%%%%%%%%
\subsection{Notes for individual pulsar}

\subsubsection{PSR J0034-0721}

It is a famous pulsar with remarkable intensity and phase modulations,
its nulling state was first noticed by \citet{ht70}. The nulls
typically lasted for several tens to hundreds of periods comparable
with the duration of the emission states. There was at least one null
separating two modes with different drifting rates
\citep{smk05,iwjc20}. Its nulling was found to be simultaneous across
a broad range of frequencies \citep{gjk+14}, with a nulling fractions
estimated to be 43\% at 303~MHz \citep{gjk+14}, 44.6\% at 326.5~MHz
\citep{viv95}, 31.3\% at 333~MHz \citep{bmm17}, 44\% at 607~MHz
\citep{gjk+14}, 22.8\% at 618~MHz \citep{bmm17}, 37.7\% at 645~MHz
\citet{big92} and 43\% at 1380~MHz \citep{gjk+14}. A periodicity of 75
rotations for the nulling was identified \citep{bmm17}.

We made a long observation of 200 minutes at 2250~MHz, and every 32
pulses are integrated to form subintegrations, as shown by two blocks
of data in Figures~\ref{fig:J0034_20151212}. The nulling fraction is
estimated to be in the range of 12.4\% to 37.4\%, which is consistent
with the low-frequency estimates at 333 and 618 MHz, but smaller than
those from \citet{viv95} and \citet{gjk+14}. The discrepancy is caused
by the fact that their nulling fractions were estimated by using
extreme values instead of the total intensities of the pulses.

\subsubsection{PSR J0248+6021}

This is the first time to report nulling for this pulsar. Every 32
pulses are integrated to form subintegrations for two observations of
33 and 120 minutes, respectively, as shown in
Figures~\ref{fig:J0248_20150616-20160811}. The nulling fraction is
estimated to be in the range of 0.0\% to 34.7\%, as estimated from
subintegrations of every 32 pulses and single pulses.

\subsubsection{PSR J0304+1932}

The pulse emission is modulated by drifting subpulses and nulls. Its
nulling fraction was found to be 13\% and 14\% from observations taken
at 327~MHz \citep{rr09, hr09}, 8.7$\pm$1.2\% at 333~MHz \citep{bmm17},
10\% at 430~MHz \citep{ran86} and 6.1\% at 618~MHz \citep{bmm17}. The
nulling is not random \citep{rr09}, and has a periodicity of
128$\pm$32 or 103$\pm$34 rotations \citep{hr09,bmm17}.

We made one observation at 2250~MHz for 56 minutes, every 8 pulses are
integrated to separate the emission and null pulses, as shown in
Figure~\ref{fig:J0304_20151214}. It typically nulls for about 16
pulses, and an estimated nulling fraction is in the range between
8.9\% and 17.5\%, consistent with those at low frequencies. Occurrence
of its nulling is random, but the duration is not random as shown by
the distributions of nulling scales, confirming the Wald-Wolfowitz
runs test results by \citet{rr09}. Quasi-periodic modulation of about
90 rotations is marginally found, which agrees with those of
\citet{hr09} and \citet{bmm17}.

\subsubsection{PSR J0332+5434}

Its nulling was first reported by \citet{rit76} with a nulling
fraction below 0.25\% at 408~MHz. We made two observations for 180 and
84 minutes respectively at 2250~MHz. Only two pulses may null among
the 22000 periods, as shown in Figures~\ref{fig:J0332_20160221} and
\ref{fig:J0332_20160224}. Hence, its nulling fraction is about 0\%,
consistent with the low frequency estimation.

\subsubsection{PSR J0528+2200}

It has drifting and nulling behaviors, with a nulling fraction of
about 25\% and 28\% at 327~MHz \citep{hr09,rr09}, 14.4\% at 333~MHz
\citep{bmm17} and 25\% at 610~MHz \citep{rit76}. The nulls are not
random \citep{rr09}, with two bright, probably harmonically related
low-frequency structures as shown by \citet{hr09}.

We made two observations for 29 and 32 minutes at 2250~MHz, every 4
pulses are integrated to form subintegrations. The significance
distributions for the on and off pulses remain non-separable, as
shown in Figures~\ref{fig:J0528_20150618-20160127}. The nulling
fraction is estimated to be in the range of 8.1\% to 22.3\%, roughly
consistent with those at low frequencies.

\subsubsection{PSR J0543+2329}

No reports on nulling before and here is the first. We made two
observations for 50 and 255 minutes at 2250~MHz. Every 16 pulses are
folded to form subintegrations, as shown in
Figures~\ref{fig:J0543_20150625} and \ref{fig:J0543_20171101}. The
nulling fraction is estimated to be in the range of 0.0\% to 16.6\%.

\subsubsection{PSR J0826+2637}

Its nulling was first found by \citet{rit76} and was not random
\citep{rr09}. The nulling fraction was estimated to be 7\% at 327~MHz
\citep{hr09}, but $\le5$ at 430~MHz \citep{rit76}. Recently, the
prominent `Quiet' and `Bright' modes were identified, which might be
related to the change of the plasma formation front
\citep[e.g.][]{syh+15, row20}. In the bright mode, the nulls had a
fraction of a few percent, but more than 90\% in the quiet mode
\citep{bm19}.

We made three long observations at 2250~MHz, from which continuous
blocks of pulses that are RFI free and scintillation enhanced (they
might also be bright modes) are chosen for the nulling analysis. Every
4 pulses are integrated to form subintegrations, as shown in
Figures~\ref{fig:J0826_20151112}, \ref{fig:J0826_20151114} and
\ref{fig:J0826_20151115}. The lower and upper limits for the nulling
fraction are estimated to be 0.0\% and 3.4\% from the entire 4745
subintegrations, which agrees with those at low frequencies for the
bright mode.

\subsubsection{PSR J0908-1739}

Its nulling fraction was found to be 26.8\% at 333~MHz and 25.7\% at
618~MHz \citep{bmm17}. We made one observation for 37.4 minutes at
2250~MHz, and every 32 pulses are integrated to form 175
subintegrations. The two significant nulls are shown in
Figure~\ref{fig:J0908_20160523}. The nulling fraction is estimated to
be in the range of 18.8\% to 29.2\%, consistent with those at low
frequencies.

\subsubsection{PSR J0922+0638}

A nulling fraction below 0.05\% was roughly estimated at 430~MHz
\citep{wab+86}. We made three observations at 2250~MHz, every 2 pulses
are folded, as shown in Figures~\ref{fig:J0922_20150714} and
\ref{fig:J0922_20150817-20150818}. The lower and upper limits of the
nulling fraction are estimated to be 0.0\% and 0.06\% from a total of
9600 subintegrations, consistent with those at low frequencies.

\subsubsection{PSR J0953+0755}

This pulsar has interpulse emission with a nulling fraction below 5\%
at 326.5 and 430~MHz \citep{rit76,viv95}. We made one observation for
198 minutes at 2250~MHz, every 8 pulses are integrated to form 5850
subintegrations, as shown in Figure~\ref{fig:J0953_20151212}. The
lower and upper limits of the nulling fraction are estimated to be
0.09\% and 11.8\%, consistent with those at low frequencies.

\subsubsection{PSR J1136+1551}

Quasi-periodic intensity modulations was found for this bright pulsar
\citep{th71}, and the nulling was identified by \citet{bac70}. The
nulling fraction is about 20\% at 327~MHz \citep{bgk+07,rr09,hr09},
13.7\% at 333~MHz \citep{bmm17}, 14\% at 408~MHz \citep{rit76} and
11.9\% at 618~MHz \citep{bmm17}. One of its double components missed
at certain times, remarked as being partial-nulls \citep{hr09}. The
nulling is not random \citep{rr09}, and has a periodicity of about 30
rotations \citep{hr07}. Multi-frequency observations showed that the
nulls do not always occur simultaneously, some pulses just null at
low frequencies \citep{bgk+07}.

We made three observations at 2250~MHz. The second observation was
affected by RFI and was split into 4 continuous blocks. Every 4
pulses of the observations are folded to form subintegrations, as
shown in Figures~\ref{fig:J1136_20150919-20180916},
\ref{fig:J1136_20171025AB} and \ref{fig:J1136_20171025CD}. The nulling
fraction is estimated to have a lower limit of 0.3\% and an upper
limit of 11.5\%, roughly consistent with those at low frequencies. The
occurrence and duration of nulls are not random, confirming the tests
at low frequencies. But no significant periodicity of nulling is
found, which means that the physical origins for its nulling should be
frequency dependent.

\subsubsection{PSR J1239+2453}

It is also among the first set of pulsars reported to null
\citep{bac70}. The amplitudes of its individual components were found
to have different modulations \citep{th71}, together with mode
changing and sub-pulse drifting at multiple drift rates
\citep[e.g.][]{wes06, njmk17}. The nulling fraction was estimated to
be 7\% at 325~MHz \citep{njmk17}, 6\% at 327~MHz \citep{hr09}, 2\% at
333~MHz \citep{bmm17}, 6\% at 408~MHz \citep{rit76}, 4\% at 610~MHz
\citep{njmk17} and 3.1\% at 618~MHz \citep{bmm17}. Observations at 325
and 610~MHz show that pulses null simultaneously at these two
frequencies \citep{njmk17}, but no periodicity was found for the
nulling \citep{bmm17}.

We made four observations at 2250~MHz, every 2 pulses are integrated
to form subintegrations, as shown in
Figures~\ref{fig:J1239_20151212-20151216}, and
\ref{fig:J1239_20151219-20160218}. Each observation demonstrates
significant normal and abnormal states. The abnormal emissions are
relatively strong, and are less affected by nulling. The most
prevalent nulls are one or two pulses with a nulling fraction
estimated to be in the range of 1.08\% to 2.8\%. The nulling fraction
is consistent with those at 333 and 618~MHz from
\citet{bmm17}. Fourier spectra of the emission-null sequences do not
show any significant periodicity, which agrees with \citet{bmm17}.

\subsubsection{PSR J1509+5531}

The nulling was first reported by \citet{njmk17} from simultaneous
observations at 327 and 610~MHz, with a nulling fraction of 7\% at
610~MHz.

We made one long observation at 2250~MHz. To avoid RFI and
interstellar scintillation, two blocks of single pulse data are chosen
for the nulling analysis, as shown in
Figures~\ref{fig:J1509_20171030AB}. Nulls occur for one or two pulses,
with a nulling fraction estimated to be 2.0\%, smaller than those at
low frequencies. The profile components are not synchronous for
null-to-emission and emission-to-null transitions, with the leading
component emerging and diminishing first. The nulls are not random,
and have a periodicity of about 23 rotations.

\subsubsection{PSR J1709-1640}

The nulling of this pulsar was first reported by \citet{wse07}. It
usually stays in an active state with short nulls of less than 150
pulses. The nulling fraction is 3.7\% at 333~MHz and 4.9\% at 618~MHz
\citep{bmm17}. Occasionally, it switches to an inactive state for long
nulls of 1 to 4.7 hours \citep{njmk18}. The nulls are found to be
concurrent between 327 and 610~MHz.

We carried out a long observation of 220 minutes at 2250~MHz, which
shows nulls for short ($\sim$0.01 hour), moderate ($\sim$0.1 hour) and
long ($\sim$1.2 hour) durations, see
Figure~\ref{fig:J1709_20171027-20171103}. Details of its nulling
behaviors have been discussed in the Section 3 already. The nulling
fraction is 35.6\% for the whole data, but in the given scintillation
enhanced and RFI free data it is only between 2.1\% and 3.4\%,
consistent with those at low frequencies in the active state. The
nulls are not random, and have a periodicity of about 200 rotations.

\subsubsection{PSR J1844+00}

This is the first time to report nulling of this pulsar. We made one
observation for 180 minutes at 2250~MHz, every 32 pulses are
integrated to form subintegrations, as shown in
Figures~\ref{fig:J1844_20180809}. Significant nulls of 256 pulses are
detected. Its nulling fraction is estimated to be in the range between
0.3\% and 34.0\%.

\subsubsection{PSR J1932+1059}

It has an interpulse, and was first reported to null by \citet{bac70},
with a nulling fraction below 1\% at 408~MHz \citep{rit76}.

We made one long observation for 275 minutes at 2250~MHz, and data are
separated into four blocks to avoid RFI, as shown in
Figures~\ref{fig:J1932_20160221AB} and \ref{fig:J1932_20160221CD}. It
typically nulls for one pulse and the duration of emission roughly
follows exponential distributions. Its nulling fraction is estimated
to be 0.03\%, consistent with the low-frequency one. The occurrence and
duration of nulls are not random. Its average emission remains stable
and no nulling periodicity is found.

\subsubsection{PSR J2022+5154}

Its nulling was first reported by \citet{rit76}. The nulls generally
lasted for one or two periods with a nulling fraction below 5\% at
408~MHz \citep{rit76}, or 1.4\% at 610~MHz \citep{gjk12}.

We made three observations at 2250~MHz, as shown in
Figures~\ref{fig:J2022_20150615-20171101} and
\ref{fig:J2022_20171106}. It typically nulls for one pulse with a
nulling fraction of 0.12\%, consistent with that at 408~MHz but
smaller than that at 610~MHz. The occurrence and duration of nulls are
not random, but no periodicity of nulling can be identified.

\subsubsection{PSR J2048-1616}

Fluctuation of its three components was investigated by \citet{th71},
which showed sub-pulse drifting and nulling. The nulls were found to
be broadband \citep{njmk17} with a nulling fraction of 14\% at 325~MHz
\citep{njmk17}, 10\% at 326.5~MHz \citep{viv95}, 8.3\% at 333~MHz
\citep{bmm17}, 10\% at 408~MHz \citep{rit76}, 17\% at 610~MHz
\citep{njmk17}, 9.0\% at 618~MHz \citep{bmm17} but 22\% at 1308~MHz
\citep{njmk17}. These nulls showed periodic variations of about every 51
rotations \citep{bmm17}.

We made two observations for 30 and 300 minutes, respectively, at
2250~MHz. The longer observation was seriously affected by RFI, and
only a block of 116 pulses is chosen for the analysis. Every 2 pulses
of the observations are folded, as shown in
Figures~\ref{fig:J2048_20150618-20160808}. The nulls usually last for
less than 5 pulses with a nulling fraction in the range between 2.1\%
and 6.4\%, smaller than those at low frequencies. Its nulls are not
random, but no significant periodicity is identified for the nulling.

\subsubsection{PSR J2313+4253}

It was reported to null with a nulling fraction of 3.7\% at 333~MHz by
\citet{bmm17}, and the nulls have at a periodicity of about every
32 rotations.

We made one observation at 2250~MHz, two blocks of data are chosen for
the nulling analysis, as shown in
Figure~\ref{fig:J2313_20171025AB}. It typically nulls for one pulse
with a nulling fraction of 5.2\%, a bit larger than that at a low
frequency. Fourier spectra analysis demonstrates that its emission is
modulated by nulls with a periodicity of about 45 rotations, which
agrees with the low-frequency measurement.

\subsubsection{PSR J2321+6024}

The frequent nulls were first reported by \citet{rit76}, which
correlated with the drifting and the mode-changing \citep{wf81}. The
nulling fraction is about 35\% at 303~MHz \citep{gjk+14}, 25\% at
408~MHz \citep{rit76}, 33\% at 607~MHz \citep{gjk+14}, 29\% at 610~MHz
\citep{gjk12}, 31\% at 1380~MHz and $>30$\% at 4850~MHz
\citep{gjk+14}. Simultaneous observations by \citet{gjk+14} showed
that the nulls were highly concurrent across a broad frequency range,
and 1-3\% of the pulses deviate from this behavior.

We made two observations for 5 and 7 hours. The first observation and
4 blocks of the second observation are chosen for the nulling
analysis, as shown in Figures~\ref{fig:J2321_20151217},
\ref{fig:J2321_20171030AB} and \ref{fig:J2321_20171030CD}. It
frequently nulls with a nulling fraction in the range between 8.5\%
and 18.5\%, smaller than those previous estimates by about 30\%. This
is because the less significant emitting pulses can not be modeled by
the significance distributions of off pulses, as shown by the excess
of the residual distribution in Figure~\ref{fig:nfs}, but the previous
researches termed these as nulls. These nulls are not random, and the
emission are modulated by nulls at long periodicities.

%%%%%%%%%%%%%%%%%%%%%%%%%%%%%%%%%%%%%%%%%%%%%%%
\section{Discussion and conclusions}

In this paper, we investigate nulling behaviors of 20 pulsars by
employing single pulse observations at 2250~MHz from JMS 66~m
telescope. This is the first time to report nulling of three pulsars,
PSRs J0248+6021, J0543+2329 and J1844+00. A set of consistent methods
is proposed for nulling analysis. The nulling fractions are first
estimated for these pulsars at relative higher frequency. The
emission-null (1-0) sequences are then constructed from which
the emission and null lengths, their interactions, randomness and
the periodicity are analyzed.

In general, the distributions of emission and null lengths cannot be
well modeled by exponential functions for most of the pulsars. No
significant difference is found between the emission-null and
null-emission length pairs from the tests of both the nulling degrees
and nulling scales. The K-S tests of nulling degrees and scales from
the simulated and observed sequences reject the hypothesis that the
occurrence and duration of nulls are characterized by random variables
at high significance levels (p-values smaller than 0.1) for most of
the pulsars except PSR J0304+1932. Emission-null sequences of the
pulsars exhibit quasi-periodic, low frequency and featureless
modulations, which might relate to the carousel beam patterns, changes
of emission mechanisms and random nulls.

During the switching between emission and null states, pulse
intensities show diverse variations, gradual rise with gradual decay,
erratic rise with gradual decay, abrupt rise with abrupt ceasing,
etc. PSR J1709$-$1640 shows the most significant rise together with
the prominent decaying. PSR J1509+5531 demonstrates significant
transitions during the null-to emission and the emission-to-null
within the pulse window.

In the following, we discuss some fundamental issues of pulsar nulling.

\begin{figure}
  \centering
  \includegraphics[angle=0,width = 0.43\textwidth] {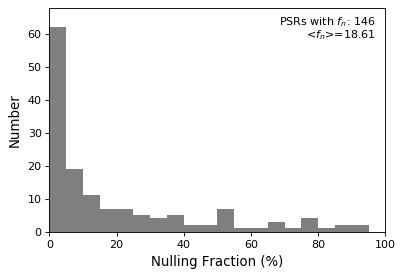} \\
  \caption{Distribution of the average nulling fraction $f_{\rm n}$ of 146
    of the 214 known nulling pulsars. The averages are listed in
    Table~\ref{table:null}. }
  \label{fig:NFdis}
\end{figure}
%%%%%%%%%%%%%%%%%%%%%%%%%%%%%%%%%%%%%%%%%%%%%%%%%%
\subsection{Known nulling pulsars and the average nulling fraction}

So far, 214 pulsars have been reported to null by observations from
303 to 4850~MHz, as collected from literatures in
appendix~\ref{appendixC}. 146 of them have nulling fraction
measurements as listed in Table~\ref{table:null}, and nulling
fractions of 79 pulsars were estimated at multi-frequencies with some
from simultaneous observations \citep[e.g.][]{gjk+14,njmk17}. These
nulling fractions are grouped into 5 frequency bands, 303-333,
408-430, 607-645, 1308-1518 and 2000-4850~MHz and then averaged. For
most pulsars, these nulling fractions are roughly consistent among
observations (within a factor of 2). But nulling fractions vary
significantly for PSRs J0630$-$2834, J0820$-$1350, J0828$-$3417,
J1559$-$4438, J1709$-$1640 and J1901$-$0906.

Averages of the available nulling fractions at these frequency bands
are calculated for each of the 146 nulling pulsars, as listed in
Table~\ref{table:null}. Its distribution is shown in
Figure~\ref{fig:NFdis}. It is apparent that more than half the pulsars
have nulling fractions less than 10 percent.

%%%%%%%%%%%%%%%%%%%%%%%%%%%%%%%%%%%%%%%%%%%%%%%%%%
\subsection{Frequency dependence of nulling}

\begin{figure}
  \centering
  \includegraphics[angle=0,width = 0.49\textwidth] {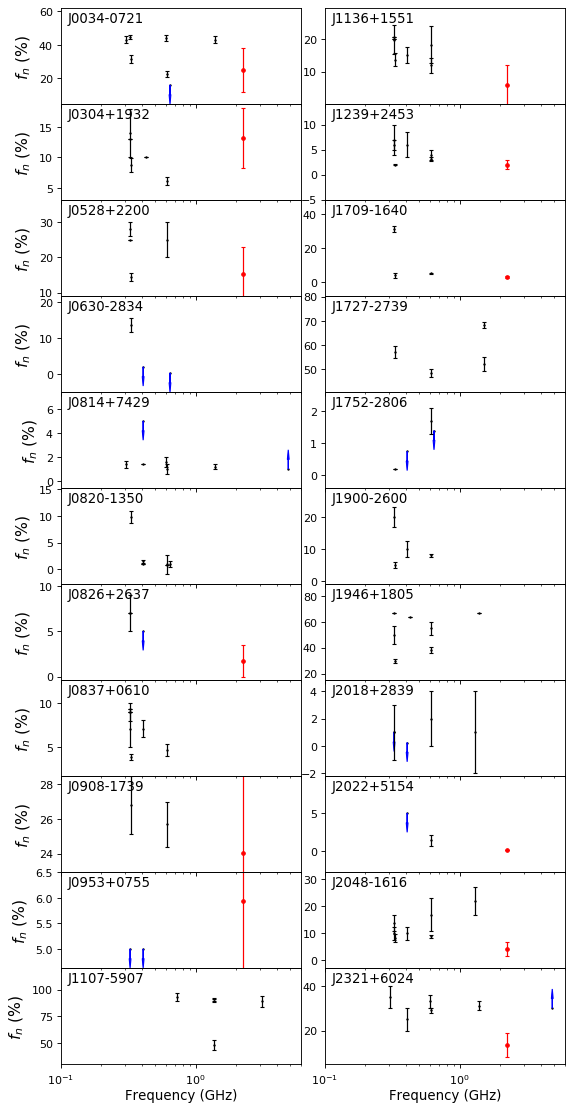} \\
  \caption{Frequency dependencies of nulling fractions of 22
    pulsars. Nulling fractions of these pulsars were estimated at at
    least 3 frequencies. Red dots represent nulling fractions
    estimated at 2250~MHz from the present work, blue arrows are for
    the upper or lower limits. Values of nulling fractions and their
    references are listed in Table~\ref{table:null}.}
  \label{fig:NFfreq}
\end{figure}

\begin{figure}
  \centering
  \includegraphics[angle=0,width = 0.45\textwidth] {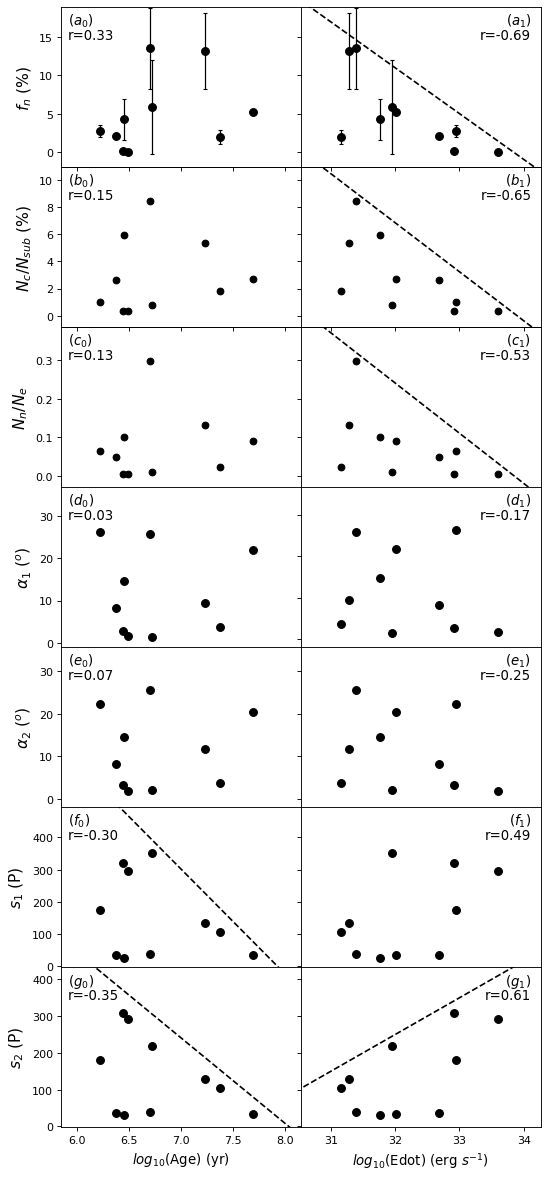}
  \caption{Dependencies of nulling fraction, nulling cadence
    ($N_c/N_{\rm sub}$), nulling to emission length ratios
    ($N_n$/$N_e$), nulling degrees ($\alpha_1$ and $\alpha_2$) and
    nulling scales ($s_1$ and $s_2$) on age and Edot for 10
    pulsars. Upper limits are plotted for $f_{\rm n}$, $N_c/N_{\rm sub}$,
    $N_n$/$N_e$ and $s_2$ with respect to Edot, and for $s_1$ and
    $s_2$ with respect to age. Correlation coefficients are indicated
    in the corners of the panels. Values of $f_{\rm n}$, $N_c$, $N_{\rm
      sub}$, $N_e$, $N_n$, $\alpha_1$, $\alpha_2$, $s_1$ and $s_2$ are
    listed in Tables~\ref{table:nf} and ~\ref{table:st}.}
  \label{fig:NF-Ratio_Age-Edot}
\end{figure}

For 22 pulsars with nulling fractions available at at least 3
frequency bands, their frequency dependencies are shown in
Figure~\ref{fig:NFfreq}. It is apparent that the nulling fractions are
generally consistent across a wide rage of frequencies for PSRs
J0814+7429, J0908$-$1739, J0953+0755 and J2018+2839. While for PSRs
J0630-2834, J0820-1350, J1107-5907, J1709-1640 and J1946+1805, nulling
fractions vary significantly across frequencies. For PSRs like
J1136+1551 and J2048-1616, the so-estimated nulling fractions at high
frequencies are smaller than the low-frequency ones. These diverse
properties mean that the broad-band nulls are produced by physical
processes with a global change of pulsar magnetosphere. Maybe the
distributions of relativistic particles are also changed. A combined
effects of the distributions of relativistic particles and pulsar
geometry can lead to various nulling features.

These effects lie in the sense that emissions of different frequencies
originate from different heights of pulsar magnetosphere
\citep[e.g.][]{whw13}. Relativistic particles emitting at different
heights are generated from different polar cap regions through
sparking process. If nulls of a pulsar from simultaneous observations
are frequency dependent, it means that parts of the polar cap region
are inactive so that nulling occurs at certain frequencies. In other
words, frequency dependent nulling can help us to distinguish the
sparking polar cap regions. Any global change of pulsar magnetosphere
will cause simultaneous nulling across a broad frequency band. PSR
J1709$-$1640 has nulls lasting for minutes to hours, and observations
in so limited time can cause different estimates of nulling
fractions. Moreover, different modes might also cause different
nulling fractions. For example, PSRs J0826+2637 and J1107$-$5907 have
two modes with quite different nulling fractions \citep{yws+14, bm19}.

%***************************************
\subsection{Dependence on pulsar parameters}

\begin{figure}
  \centering
  \includegraphics[angle=0,width = 0.45\textwidth] {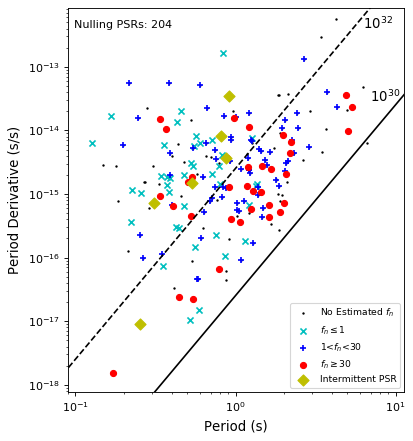}
  \caption{Period derivative against period for 204 Nulling pulsars.
    204 of the 214 known nulling pulsars have period and period
    derivative measurements. These pulsars are classified into five
    categories, i.e., those without nulling fraction estimation (small
    black dots), with nulling fractions less than or equal to 1\%
    (light blue `x'), with nulling fraction larger than 1\% but less
    than 30\% (blue `+'), with nulling fractions larger or equal to
    30\% (large red dots) and intermittent pulsars. The two sloping
    lines are with the constant rates of loss of rotational kinetic
    energy Edot=$10^{32}$, $10^{30}$ erg s$^{-1}$. }
  \label{fig:NPSRdis}
\end{figure}

Our observations of 10 pulsars have multiple nulling cycles
($N_c$>10), as listed in Table~\ref{table:nf}.
% figure
Figure~\ref{fig:NF-Ratio_Age-Edot} shows correlations of nulling
fraction, the cadence of occurrence of nulling $N_c/N_{\rm sub}$, null
to emission length ratios $N_n$/$N_e$, nulling degrees ($\alpha_1$ and
$\alpha_2$) and nulling scales ($s_1$ and $s_2$) with respect to
pulsar age and Edot for these pulsars. Here, nulling fractions and
their uncertainties are taken as the mean and half range of upper and
lower limits when both limits are available. Pulsars with large ages
tend to have large nulling fractions with correlation coefficient of
0.33, and to have small nulling scales with correlation coefficients
of -0.30 and -0.35. Little correlations are found for the other
nulling parameters with respect to pulsar age. Pulsars with large
energy loss rate tend to have small nulling fractions, small nulling
cadence and small null-to-emission length ratios with correlation
coefficients of -0.69, -0.65 and -0.53, but tend to have large nulling
scales with correlation coefficients of 0.49 and 0.61.

Distribution of the nulling pulsars on the period and period
derivative diagram is shown in Figure~\ref{fig:NPSRdis}. It is obvious
that periods of these nulling pulsars range from 0.1 to 7 seconds and
period derivatives from $10^{-18}$ to $10^{-12}$. The line with
constant rate of loss of rotational kinetic energy (Edot) of $10^{30}
\rm erg~s^{-1}$ roughly represents the lower boundary for these
nulling pulsars, which is in parallel with the lines with constant
accelerating potential above pulsar polar cap. It is generally
believed that pulsars are formed from the upper left part of the
diagram and move down to the right by crossing the constant Edot lines
when pulsars evolve. Particle acceleration process is reduced in the
magnetosphere of aged pulsars, so that pulsars start to null with a
nulling fraction increasing and eventually reach the off state. This
scenario is supported by the distribution of pulsars with different
nulling fractions in Figure~\ref{fig:NPSRdis}. It is obvious that
pulsars with nulling fractions $\le1\%$ are generally located in the
upper left part of the diagram, pulsars with nulling fractions >$30\%$
are located in the lower right, as divided by the Edot line of
$10^{32} \rm erg~s^{-1}$. The intermittent pulsars are exceptions
\citep{lsf+17}.

\begin{figure}
  \centering
  \includegraphics[angle=0,width = 0.49\textwidth] {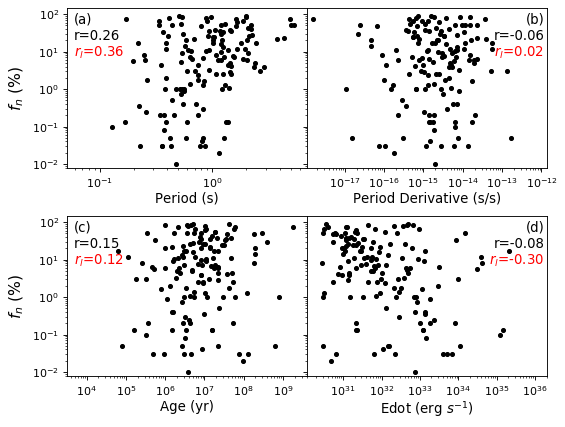}
  \caption{Correlations of the nulling fraction with respect to pulsar
    period, period derivative, age and Edot. Coefficients for linear
    and logarithmic correlations are indicated in each panel by r and
    $r_l$. The nulling fraction of each pulsar is taken as the average
    as listed in Table~\ref{table:null}. Values of pulsar period,
    period derivative, age and Edot are obtained from ATNF pulsar
    catalogue \citep{mhth05}. }
  \label{fig:NFcor}
\end{figure}

Figure~\ref{fig:NFcor} shows the correlations of nulling fraction with
respect to pulsar period, period derivative, age and Edot for this
most complete sample of nulling pulsars by now. The nulling fraction
is positively correlated with pulsar period with correlation
coefficients of r=0.26 and $r_l$=0.36 for both parameters in linear
and logarithmic scales. No correlation is found for period
derivatives. A marginal positive correlation is found between nulling
fraction and pulsar ages with r=0.15 and $r_l$=0.12, which agrees with
the understanding that old pulsars tend to have large nulling
fraction. Moreover, the conventional understanding that pulsars with
large Edot tend to have small nulling fraction is also confirmed with
correlation coefficients of r=-0.08 and $r_l$=-0.30. In summary, large
nulling fractions are more related to long period than to large age
and small Edot, in favor of the conclusion obtained by \citet{big92}.

%%%%%%%%%%%%%%%%%%%%%%%%%%%%%%%%%%%%%%%%%%%%%%%
%\section{Conclusions}

Data for emission-null sequences of all pulsars presented in this paper
are available at http://zmtt.bao.ac.cn/psr-jms/.

%%%%%%%%%%%%%%%%%%%%%%%%%%%%%%%%%%%%%%%%%%%%%%%
\begin{acknowledgements}
%
% We also thank the referee for helpful comments.
%
The authors are partially supported by the National Natural Science
Foundation of China through grants (No. 11988101, No. 11873058), the
Key Research Program of the Chinese Academy of Sciences (Grant
No. QYZDJ-SSW-SLH021), and the Open Project Program of the
Key Laboratory of FAST, NAOC, Chinese Academy of Sciences.
\end{acknowledgements}

\bibliographystyle{aa}
\bibliography{psr_Nulling}

\appendix

\section{Pulse sequences and null or emission states of 20 pulsars}
\label{appendixA}

We present plots here for observations of 20 pulsars listed in
Table~\ref{table:obs} as online appendix. It shows pulse intensities
of successive subintegrations formed every $2^n$ pulses, the
significances of on pulse (red) and off pulse (blue) emissions for
each subintegration, the emission (1) or null (0) state of the
subintegration, as in Figures~\ref{fig:J1709phase-t}. Here, $n$ ranges
from 0 to 5 depending on the significances of pulses.

\clearpage

%------------------------------J0034
\begin{figure}
  \centering
  \includegraphics[height=10.5cm,width=0.235\textwidth]{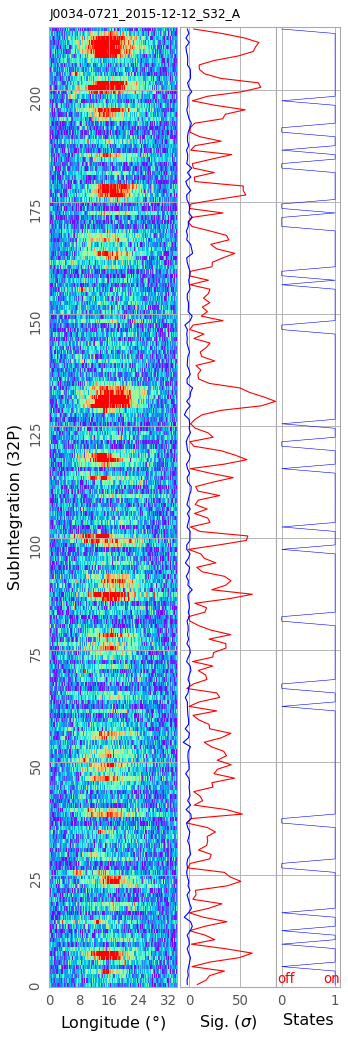}
  \includegraphics[height=10.5cm,width=0.235\textwidth]{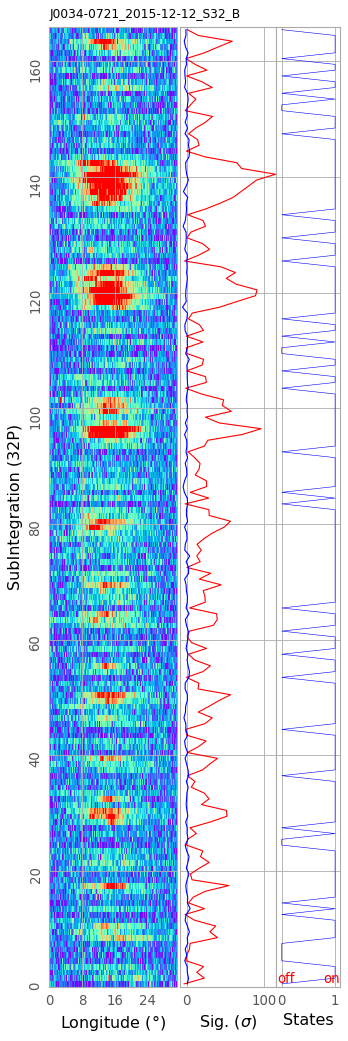} \\
  \caption{PSR J0034-0721 observed on 2015-12-12 (A and B) with every 32 pulses integrated.}
  \label{fig:J0034_20151212}
\end{figure}

%------------------------------J0248
\begin{figure}
  \centering
  \includegraphics[height=10.5cm,width=0.235\textwidth]{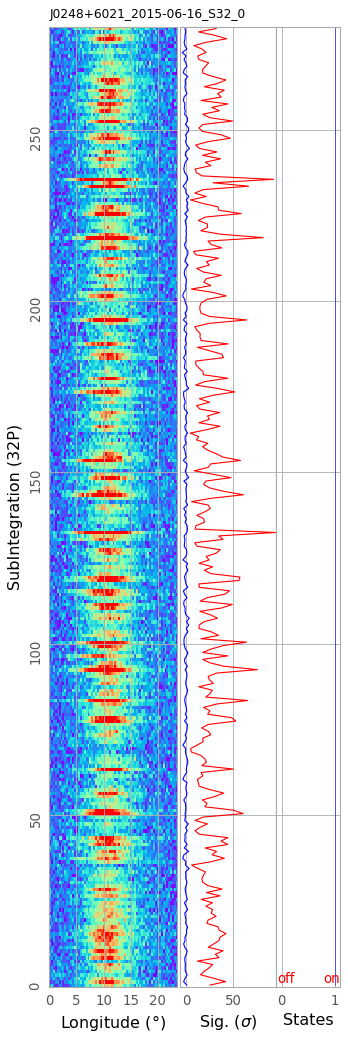}
  \includegraphics[height=10.5cm,width=0.235\textwidth]{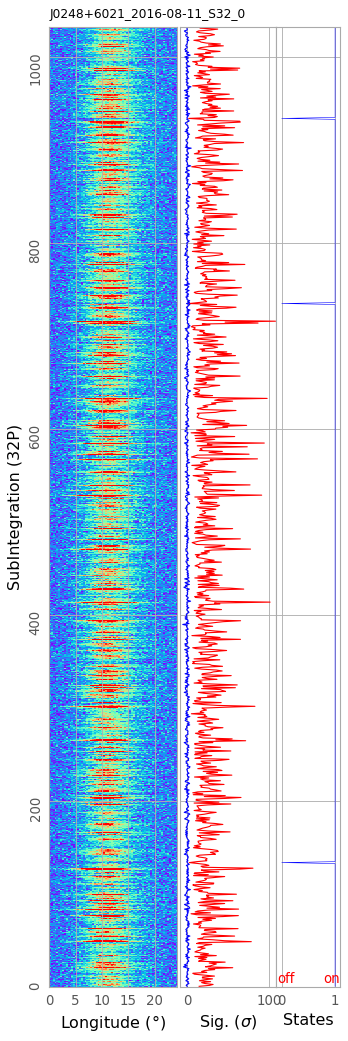} \\
  \caption{PSR J0248+6021 observed on 2015-06-16 and 2016-08-11 with every 32 pulses integrated. }
  \label{fig:J0248_20150616-20160811}
\end{figure}

%------------------------------J0304
\begin{figure}
  \centering
  \includegraphics[height=10.5cm,width=0.235\textwidth]{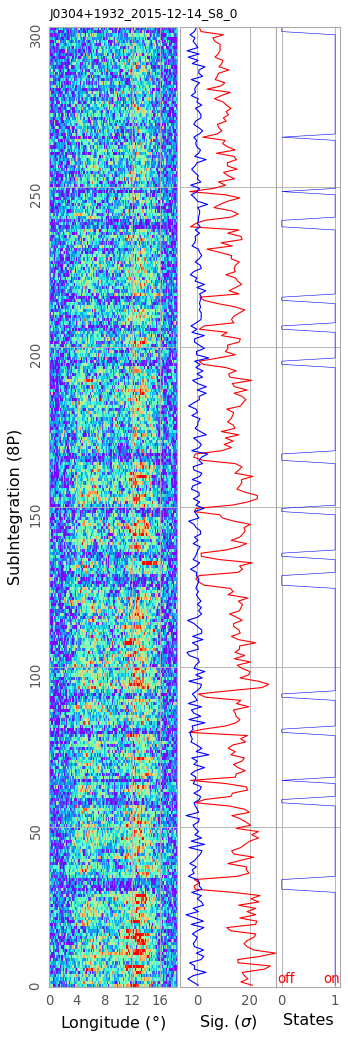}
  \caption{PSR J0304+1932 observed on 2015-12-14
  with 8 pulse integrated for each subintegration.}
  \label{fig:J0304_20151214}
\end{figure}

%------------------------------J0332
\begin{figure}
  \centering
  \includegraphics[height=10.5cm,width=0.235\textwidth]{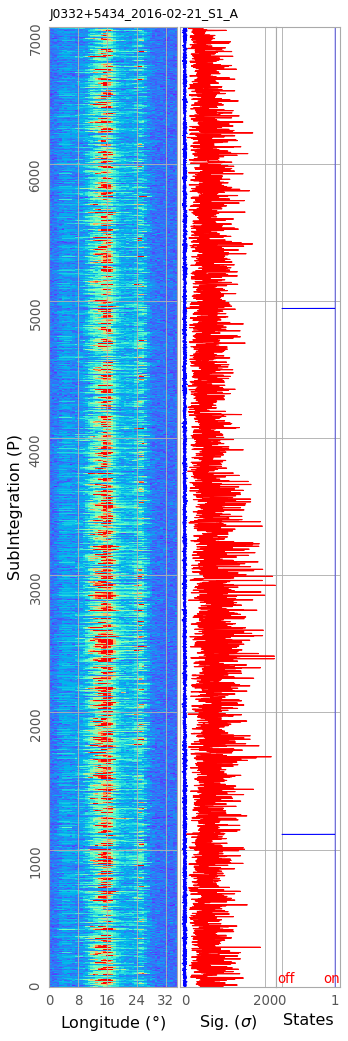}
  \includegraphics[height=10.5cm,width=0.235\textwidth]{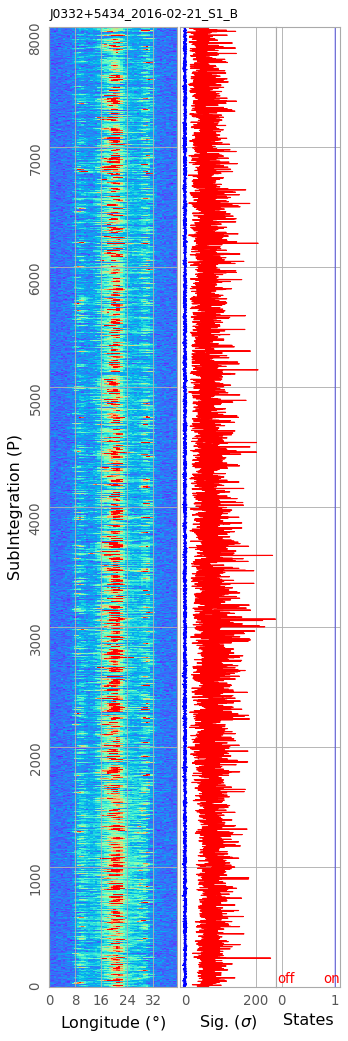} \\
  \caption{PSR J0332+5434 observed on 2016-02-21 (A and B) with single pulses. }
  \label{fig:J0332_20160221}
\end{figure}
%------------------------------
\begin{figure}
  \centering
  \includegraphics[height=10.5cm,width=0.235\textwidth]{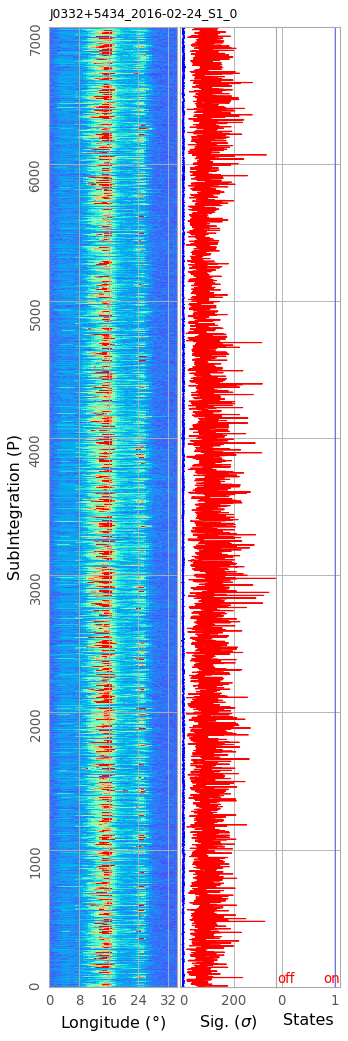} 
  \caption{PSR J0332+5434 observed on 2016-02-24 with single pulses. }
  \label{fig:J0332_20160224}
\end{figure}

%------------------------------J0528
\begin{figure}
  \centering
  \includegraphics[height=10.5cm,width=0.235\textwidth]{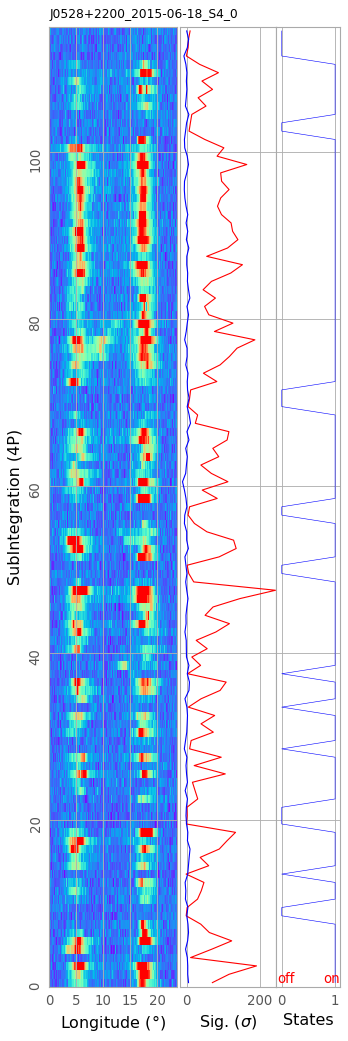}
  \includegraphics[height=10.5cm,width=0.235\textwidth]{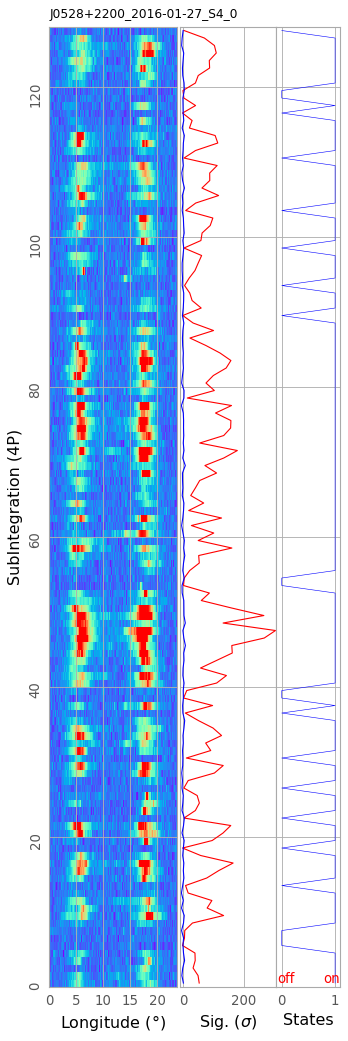} \\
  \caption{PSR J0528+2200 observed on 2015-06-18 and 2016-01-27 with subintegrations
  formed every 4 pulses.}
  \label{fig:J0528_20150618-20160127}
\end{figure}

%------------------------------J0543
\begin{figure}
  \centering
  \includegraphics[height=10.5cm,width=0.235\textwidth]{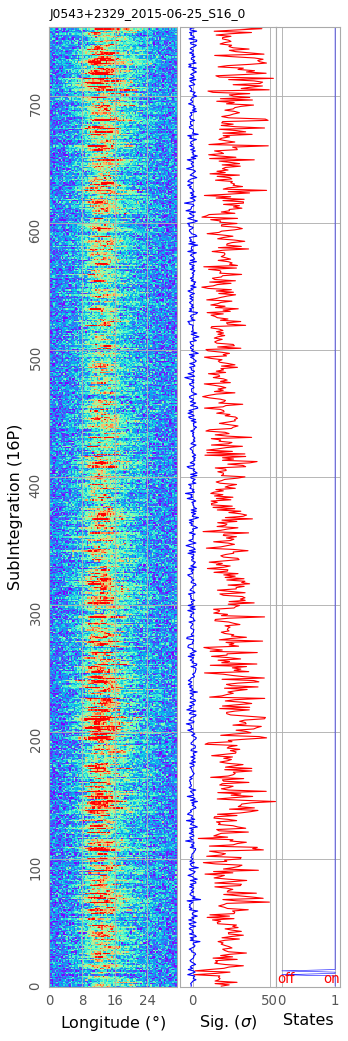}
  \caption{PSR J0543+2329 observed on 2015-06-25 with 16 pulses integrated for each subintegration.}
  \label{fig:J0543_20150625}
\end{figure}
%------------------------------
\begin{figure}
  \centering
  \includegraphics[height=10.5cm,width=0.235\textwidth]{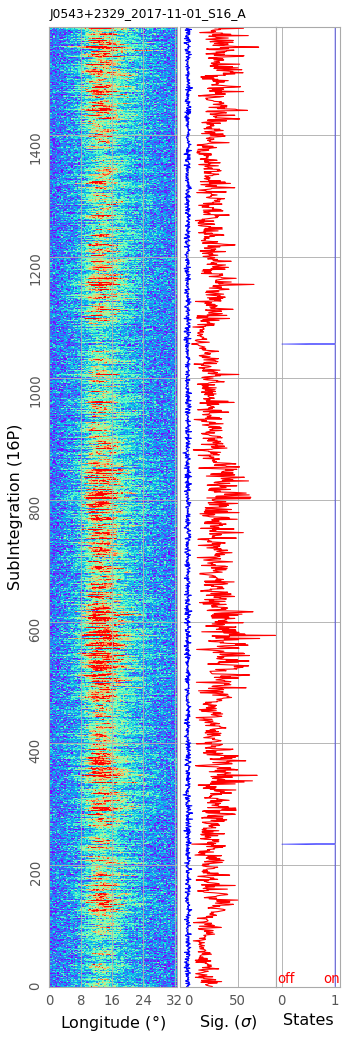}  
  \includegraphics[height=10.5cm,width=0.235\textwidth]{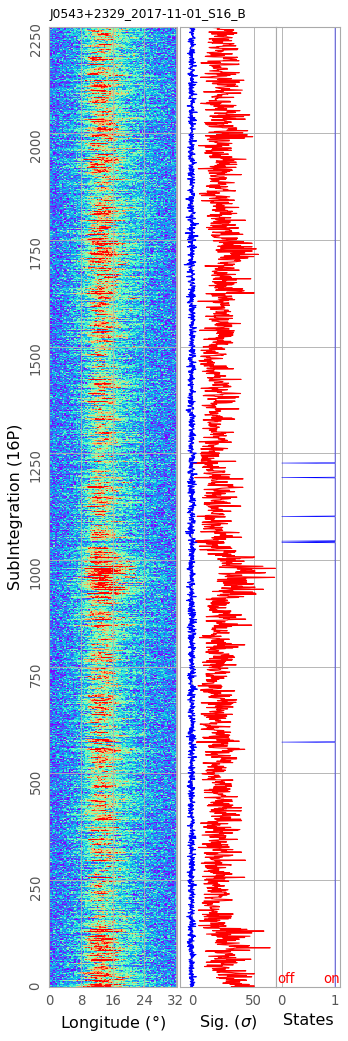} \\
  \caption{PSR J0543+2329 observed on 2017-11-01 (A and B) with 16 pulses integrated for each subintegration.}
  \label{fig:J0543_20171101}
\end{figure}

%------------------------------J0826
\begin{figure}
  \centering
  \includegraphics[height=10.5cm,width=0.235\textwidth]{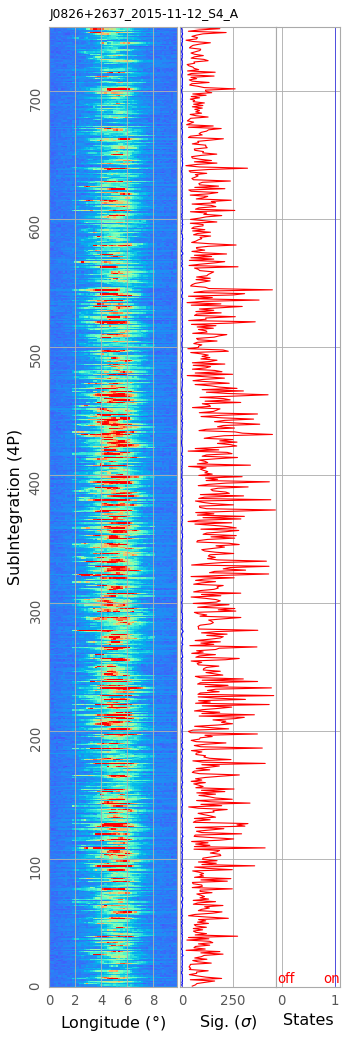}
  \includegraphics[height=10.5cm,width=0.235\textwidth]{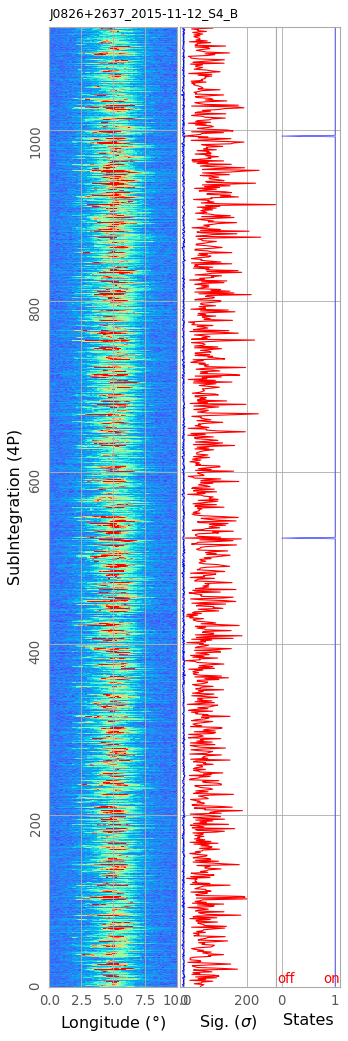}  
  \caption{PSR J0826+2637 observed on 2015-11-12 (A and B) with every 4 pulses folded. }
  \label{fig:J0826_20151112}
\end{figure}
%------------------------------
\begin{figure}
  \centering
  \includegraphics[height=10.5cm,width=0.235\textwidth]{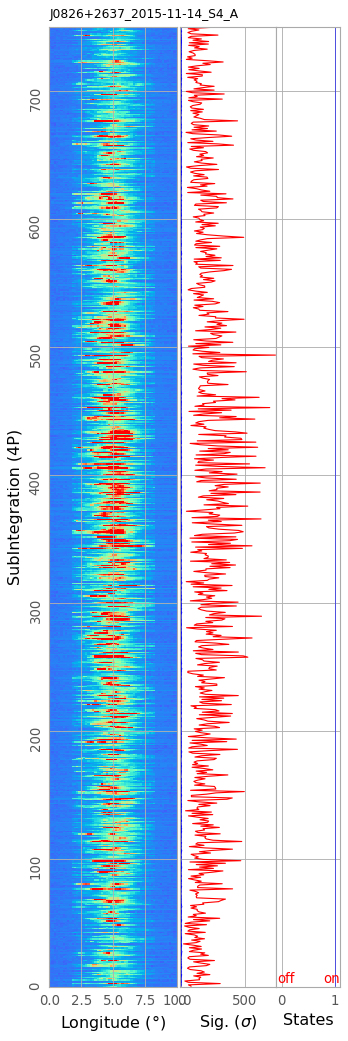}
  \includegraphics[height=10.5cm,width=0.235\textwidth]{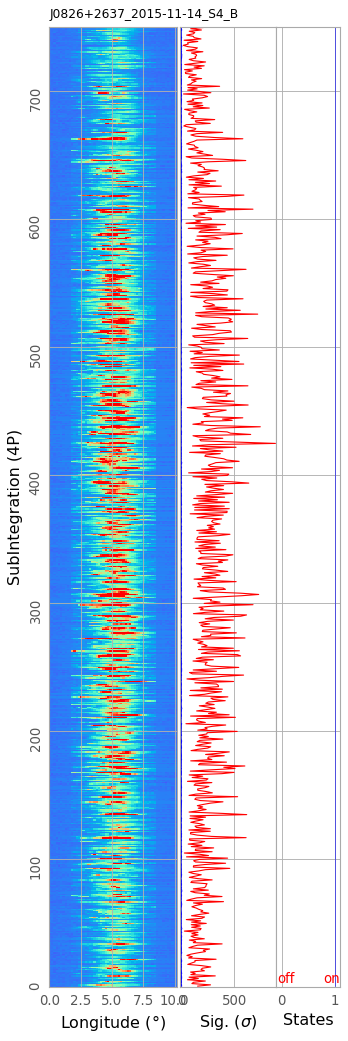}  
  \caption{PSR J0826+2637 observed on 2015-11-14 (A and B) with every 4 pulses folded. }
  \label{fig:J0826_20151114}
\end{figure}
%------------------------------
\begin{figure}
  \centering
  \includegraphics[height=10.5cm,width=0.235\textwidth]{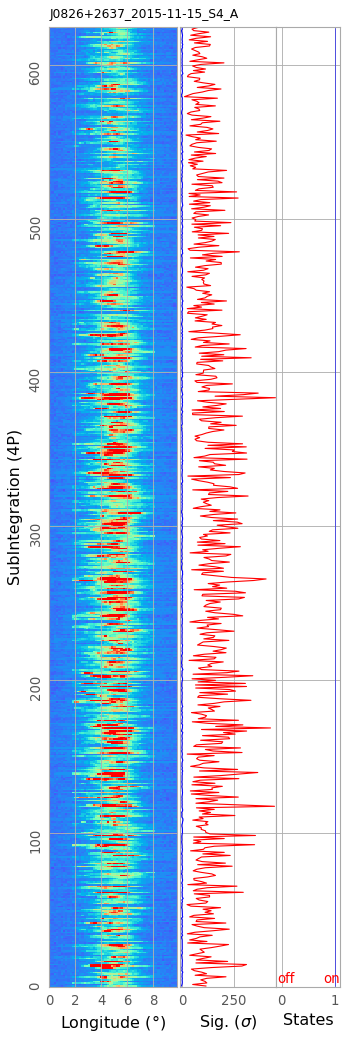}
  \includegraphics[height=10.5cm,width=0.235\textwidth]{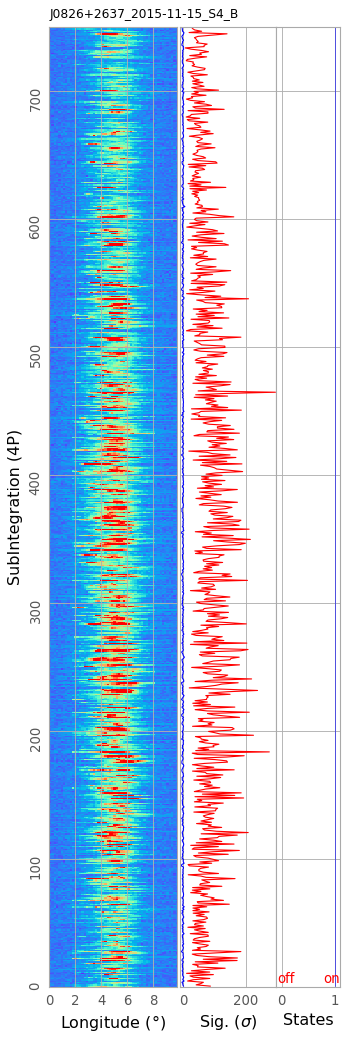}  
  \caption{PSR J0826+2637 observed on 2015-11-15 (A and B) with every 4 pulses folded. }
  \label{fig:J0826_20151115}
\end{figure}

%------------------------------J0908
\begin{figure}
  \centering
  \includegraphics[height=10.5cm,width=0.235\textwidth]{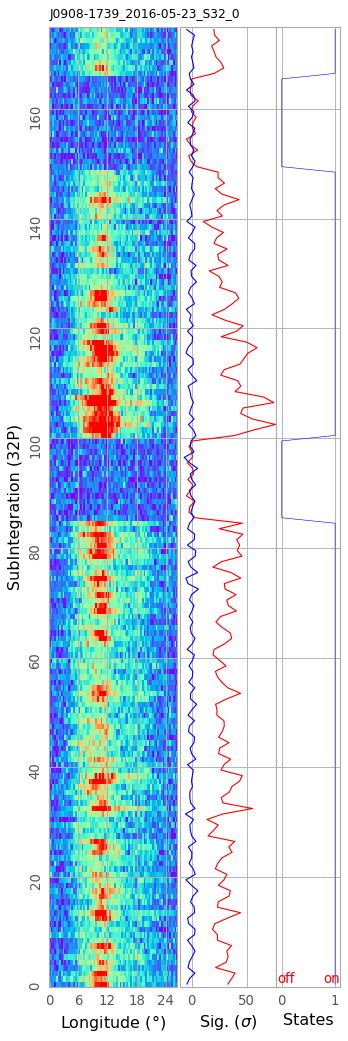} 
  \caption{PSR J0908-1739 observed on 2016-05-23
  and integrated every 32 pulses. }
  \label{fig:J0908_20160523}
\end{figure}

%------------------------------J0922
\begin{figure}
  \centering
  \includegraphics[height=10.5cm,width=0.235\textwidth]{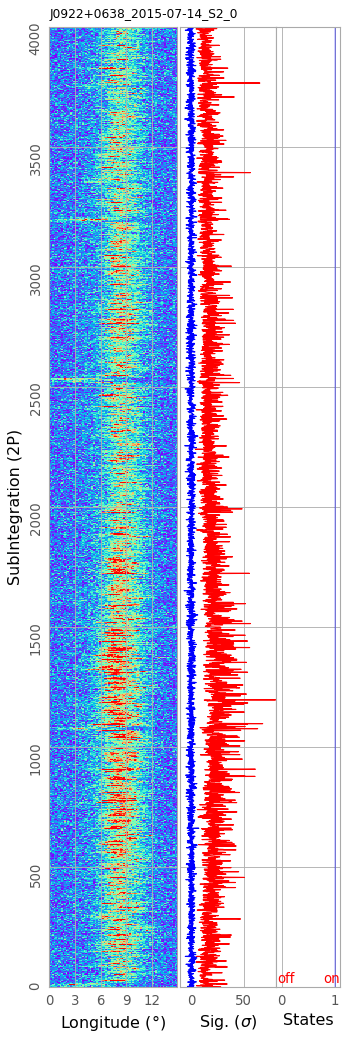}
  \includegraphics[height=10.5cm,width=0.235\textwidth]{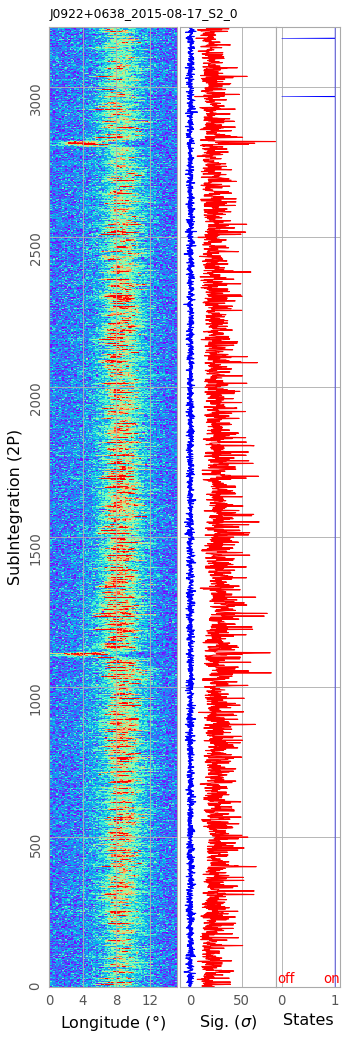}
  \caption{PSR J0922+0638 observed on 2015-07-14 (A and B)
  with subintegrations formed every 2 pulses. }
  \label{fig:J0922_20150714}
\end{figure}
%------------------------------
\begin{figure}
  \centering
  \includegraphics[height=10.5cm,width=0.235\textwidth]{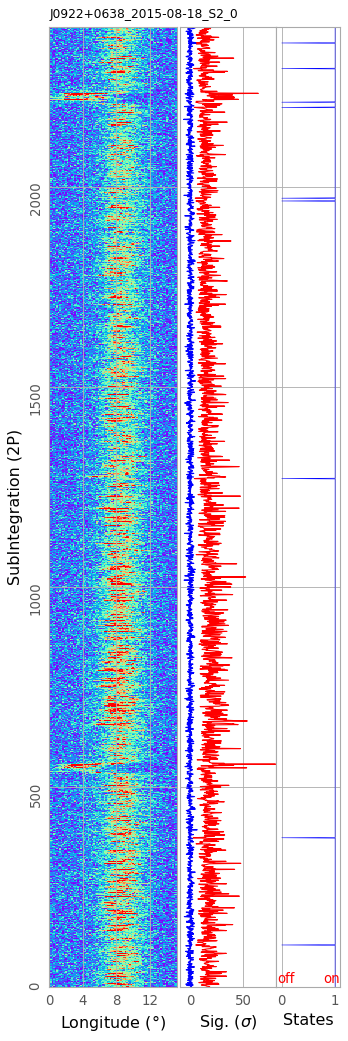}  
  \caption{PSR J0922+0638 observed on 2015-08-18 with every 2 and pulses folded.}
  \label{fig:J0922_20150817-20150818}
\end{figure}

%------------------------------J0953
\begin{figure}
  \centering
  \includegraphics[height=10.5cm,width=0.235\textwidth]{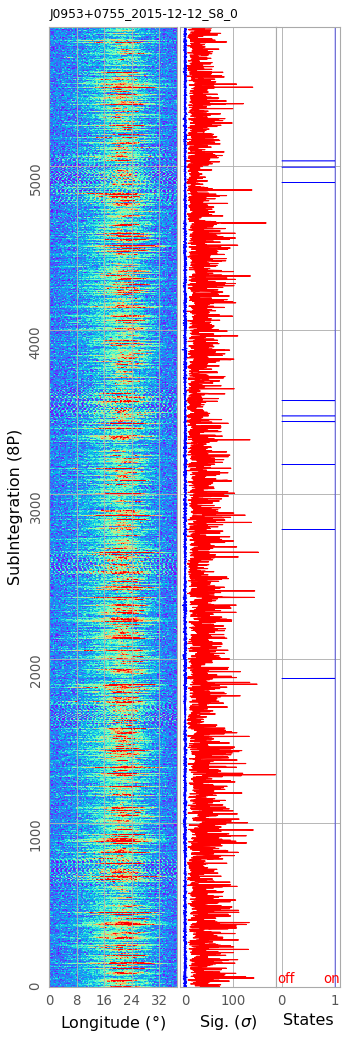}
  \caption{PSR J0953+0755 observed on 2015-12-12 with every 8 pulses folded. }
  \label{fig:J0953_20151212}
\end{figure}

%------------------------------J1136
\begin{figure}
  \centering
  \includegraphics[height=10.5cm,width=0.235\textwidth]{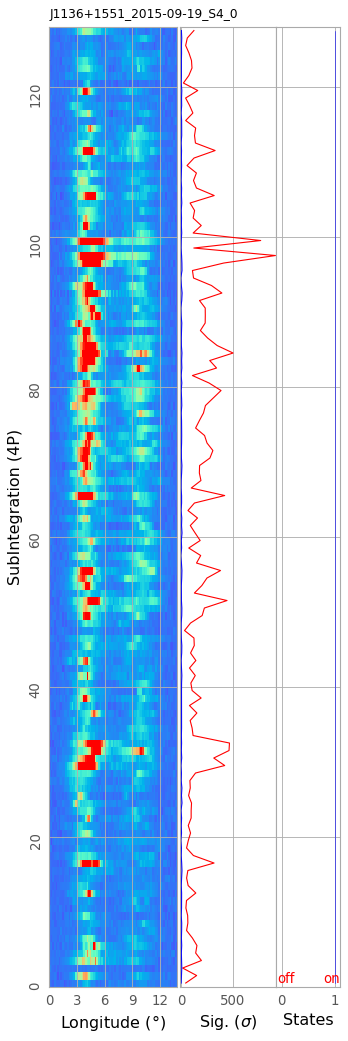}
  \includegraphics[height=10.5cm,width=0.235\textwidth]{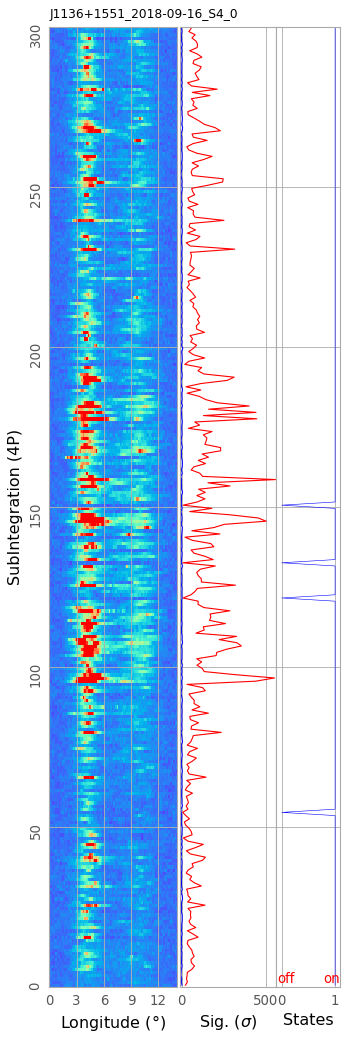}       
  \caption{PSR J1136+1551 observed on 2015-09-19 and 2018-09-16 with 4 pulses folded for each subintegration. }
  \label{fig:J1136_20150919-20180916}
\end{figure}

\clearpage
%------------------------------
\begin{figure}
  \centering
  \includegraphics[height=10.5cm,width=0.235\textwidth]{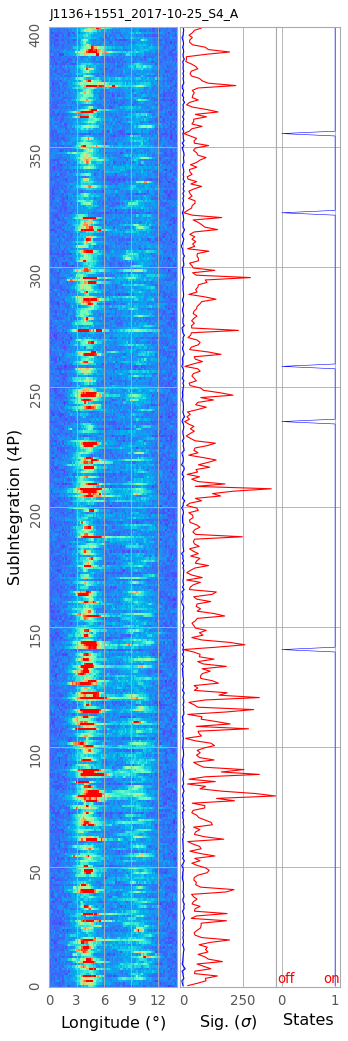}
  \includegraphics[height=10.5cm,width=0.235\textwidth]{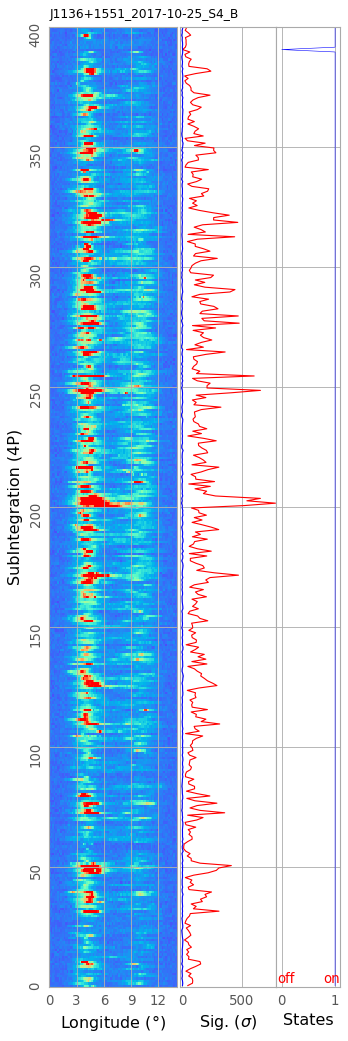}  
  \caption{PSR J1136+1551 observed on 2017-10-25 (A and B)
  with 4 pulses folded for each subintegration. }
  \label{fig:J1136_20171025AB}
\end{figure}
%------------------------------
\begin{figure}
  \centering
  \includegraphics[height=10.5cm,width=0.235\textwidth]{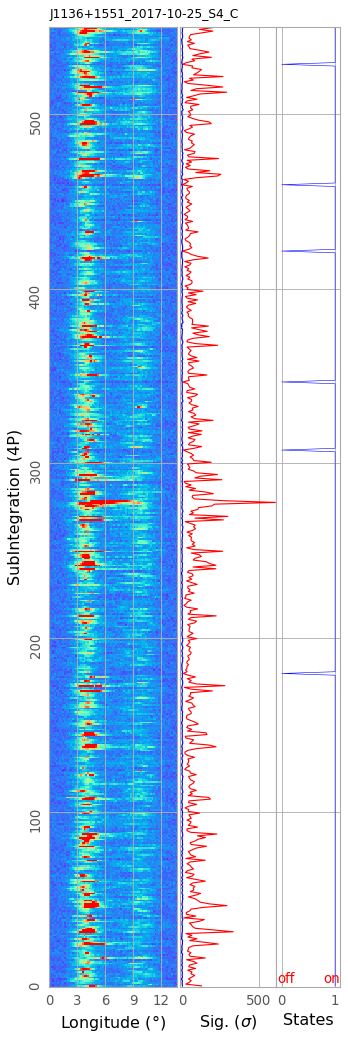}
  \includegraphics[height=10.5cm,width=0.235\textwidth]{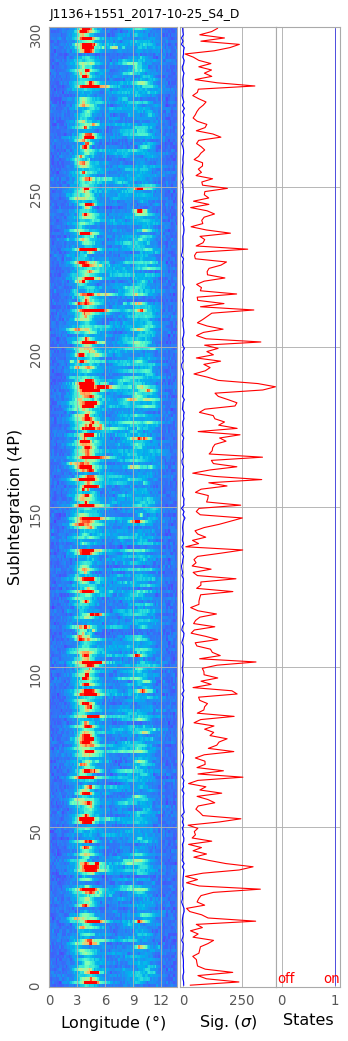}  
  \caption{PSR J1136+1551 observed on 2017-10-25 (C and D)
  with 4 pulses folded for each subintegration. }
  \label{fig:J1136_20171025CD}
\end{figure}

%------------------------------J1239
\begin{figure}
  \centering
  \includegraphics[height=10.5cm,width=0.235\textwidth]{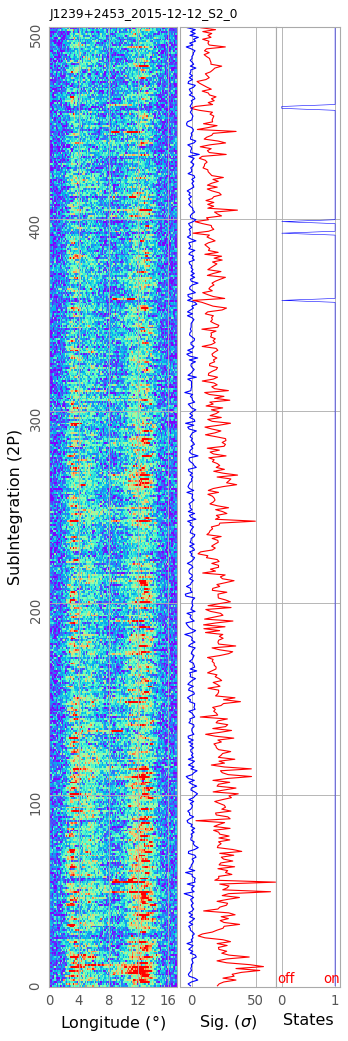}
  \includegraphics[height=10.5cm,width=0.235\textwidth]{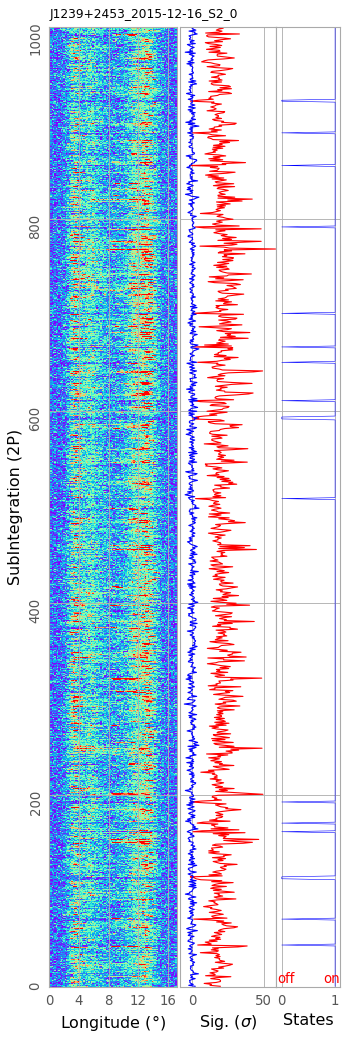}       
  \caption{J1239+2453 observed on 2015-12-12 and 2015-12-16 with every 2 pulses folded. }
  \label{fig:J1239_20151212-20151216}
\end{figure}
%------------------------------
\begin{figure}
  \centering
  \includegraphics[height=10.5cm,width=0.235\textwidth]{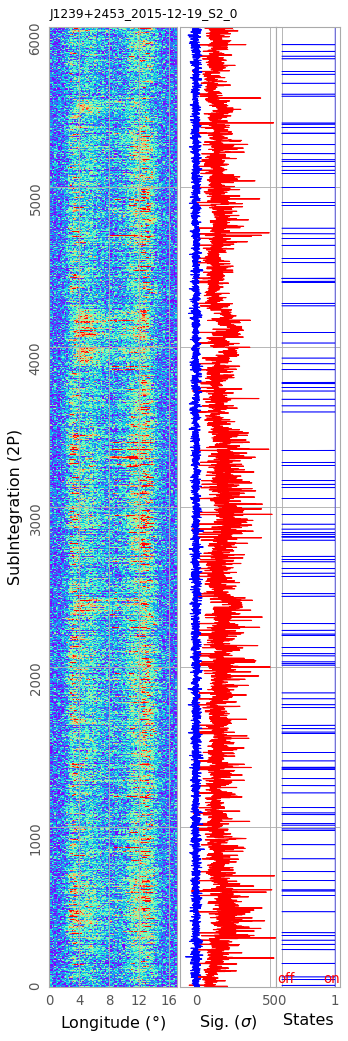}
  \includegraphics[height=10.5cm,width=0.235\textwidth]{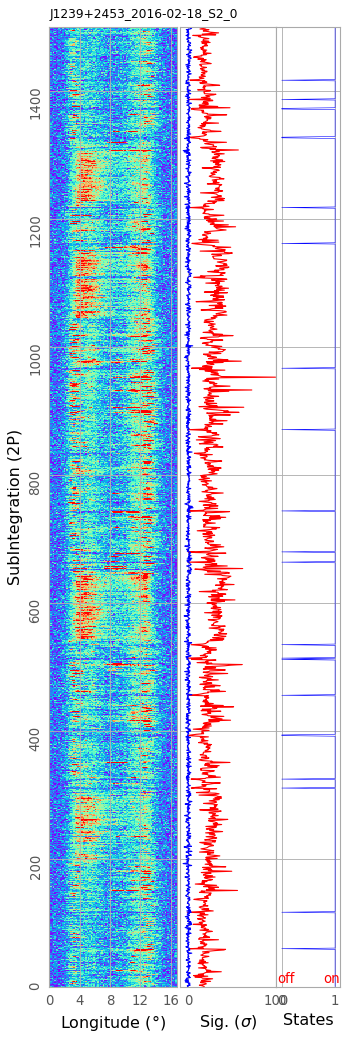}  
  \caption{J1239+2453 observed on 2015-12-19 and 2016-02-18 with every
  2 pulses folded. }
  \label{fig:J1239_20151219-20160218}
\end{figure}

%------------------------------J1509
\begin{figure}
  \centering
  \includegraphics[height=10.5cm,width=0.235\textwidth]{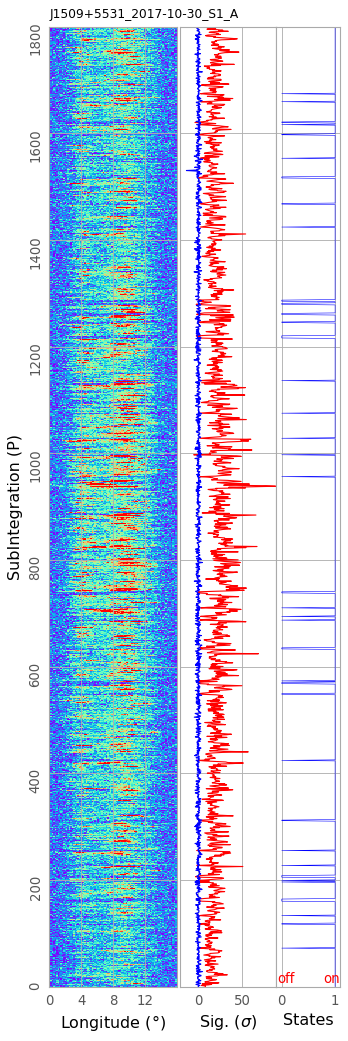}
  \includegraphics[height=10.5cm,width=0.235\textwidth]{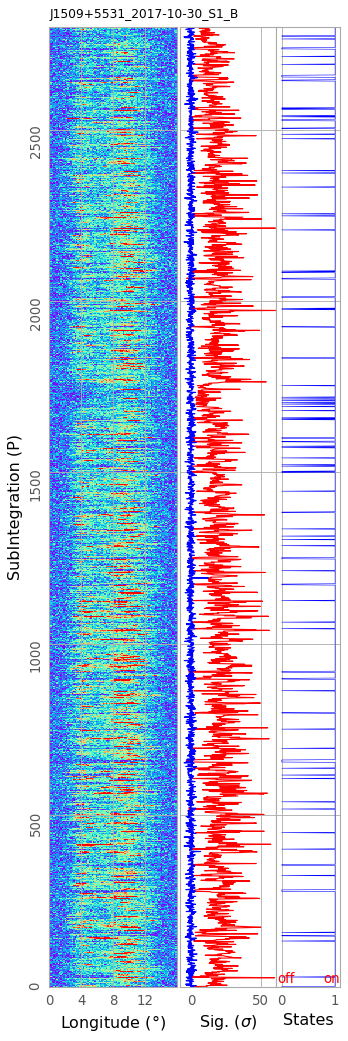}
  \caption{PSR J1509+5531 observed on 2017-10-30 (A and B) with single pulses.}
  \label{fig:J1509_20171030AB}
\end{figure}

%------------------------------J1709
\begin{figure}
  \centering
  \includegraphics[height=10.5cm,width=0.235\textwidth]{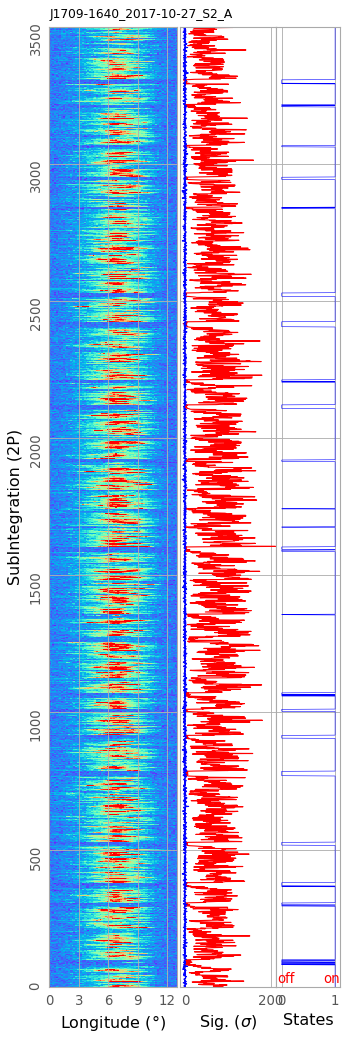}
  \caption{PSR J1709-1640 observed on 2017-10-27 with every 2 pulses folded.}
  \label{fig:J1709_20171027-20171103}
\end{figure}

%------------------------------J1844
\begin{figure}
  \centering
  \includegraphics[height=10.5cm,width=0.235\textwidth]{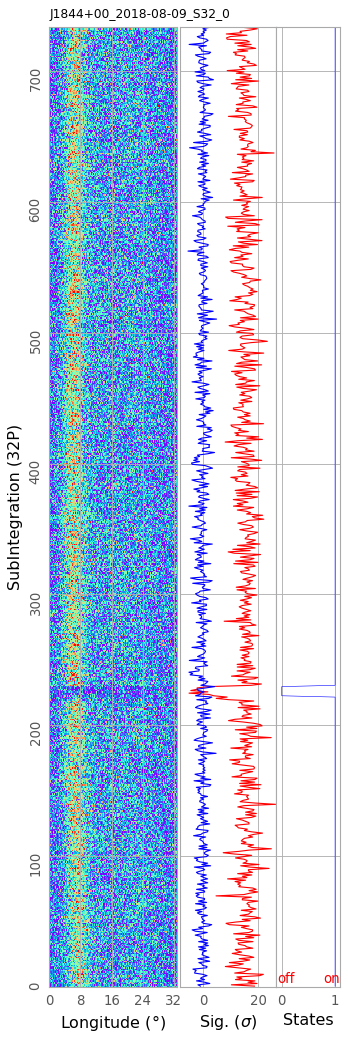}
  \caption{PSR J1844+00 observed on 2018-08-09 with 32 pulses folded for each subintegration.}
  \label{fig:J1844_20180809}
\end{figure}

%------------------------------J1932
\begin{figure}
  \centering
  \includegraphics[height=10.5cm,width=0.235\textwidth]{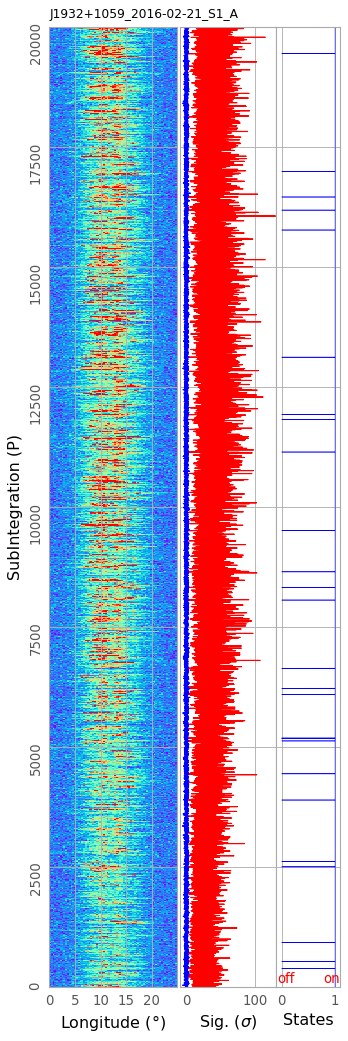}
  \includegraphics[height=10.5cm,width=0.235\textwidth]{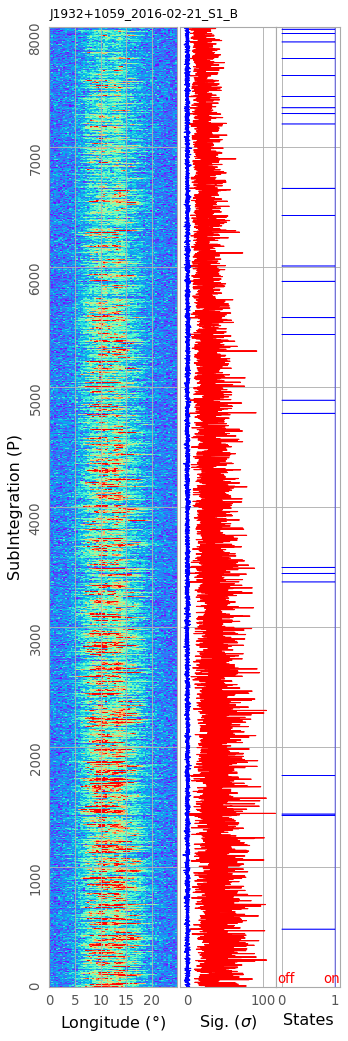}
  \caption{PSR J1932+1059 observed on 2016-02-21 (A and B) with single pulses.}
  \label{fig:J1932_20160221AB}
\end{figure}

%------------------------------
\begin{figure}
  \centering
  \includegraphics[height=10.5cm,width=0.235\textwidth]{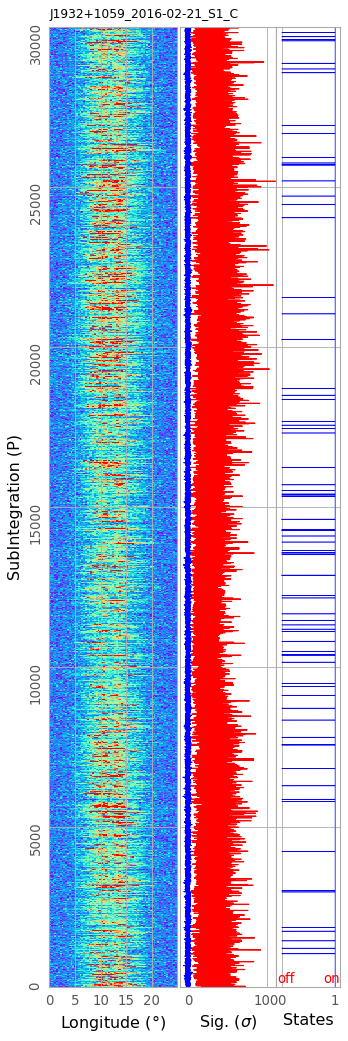}
  \includegraphics[height=10.5cm,width=0.235\textwidth]{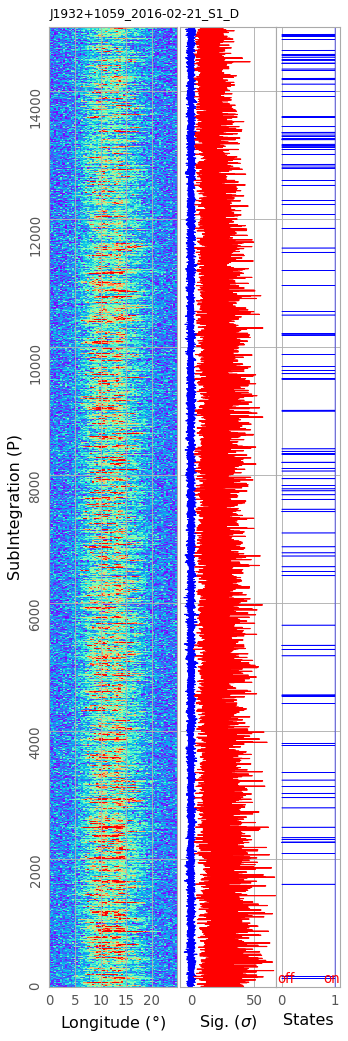}
  \caption{PSR J1932+1059 observed on 2016-02-21 (C and D) with single pulses.}
  \label{fig:J1932_20160221CD}
\end{figure}

%------------------------------J2022
\begin{figure}
  \centering
  \includegraphics[height=10.5cm,width=0.235\textwidth]{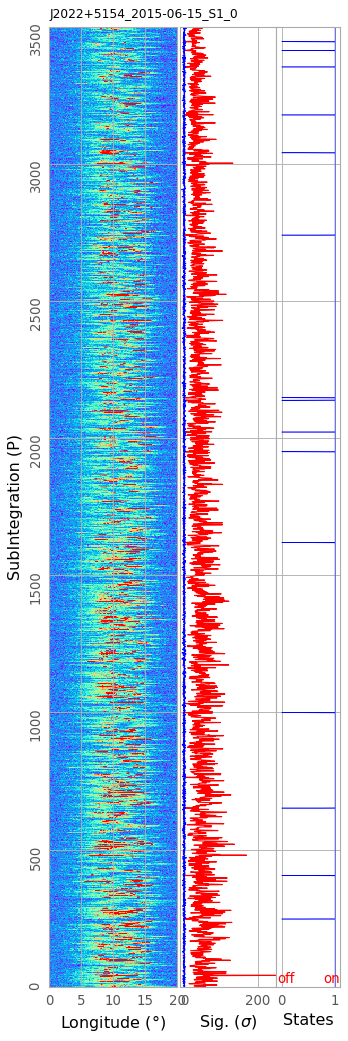}
  \includegraphics[height=10.5cm,width=0.235\textwidth]{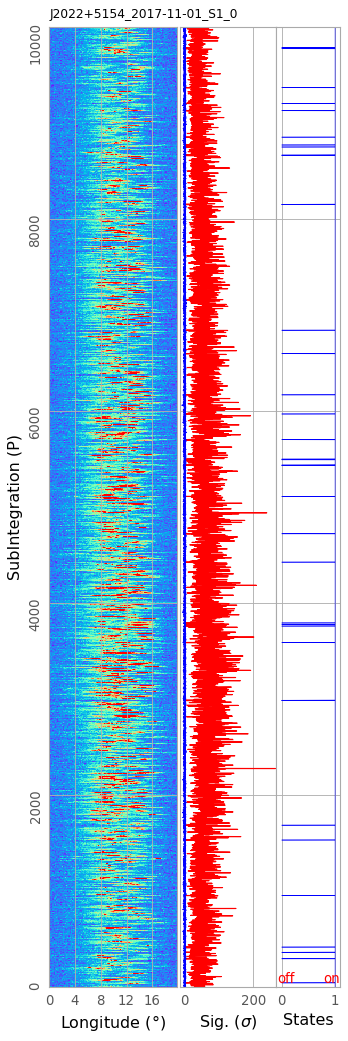}
  \caption{PSR J2022+5154 observed on 2015-06-15 and 2017-11-01 with single pulses.}
  \label{fig:J2022_20150615-20171101}
\end{figure}
%------------------------------
\begin{figure}
  \centering
  \includegraphics[height=10.5cm,width=0.235\textwidth]{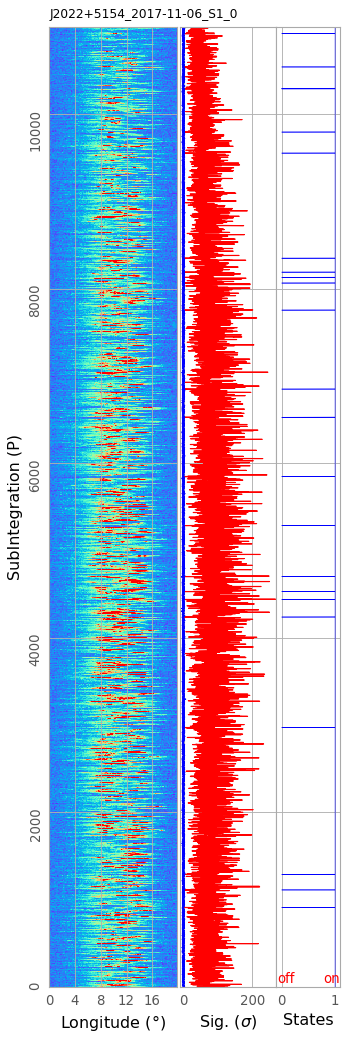}
  \caption{PSR J2022+5154 observed on 2017-11-06 with single pulses.}
  \label{fig:J2022_20171106}
\end{figure}

%------------------------------J2048
\begin{figure}
  \centering
  \includegraphics[height=10.5cm,width=0.235\textwidth]{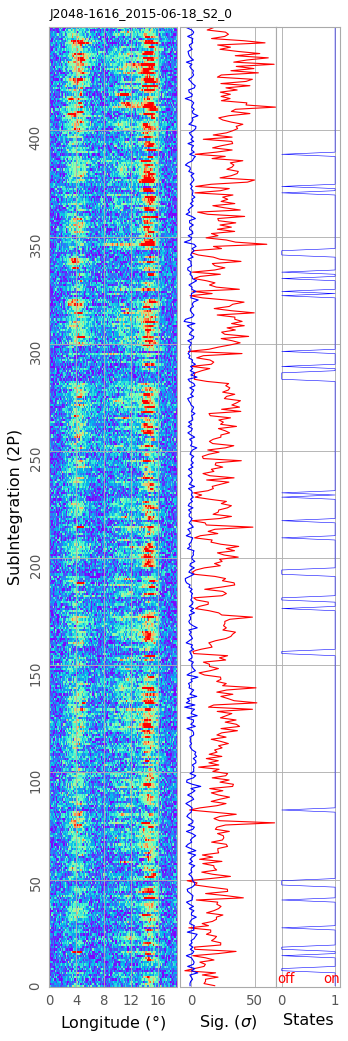}
  \includegraphics[height=10.5cm,width=0.235\textwidth]{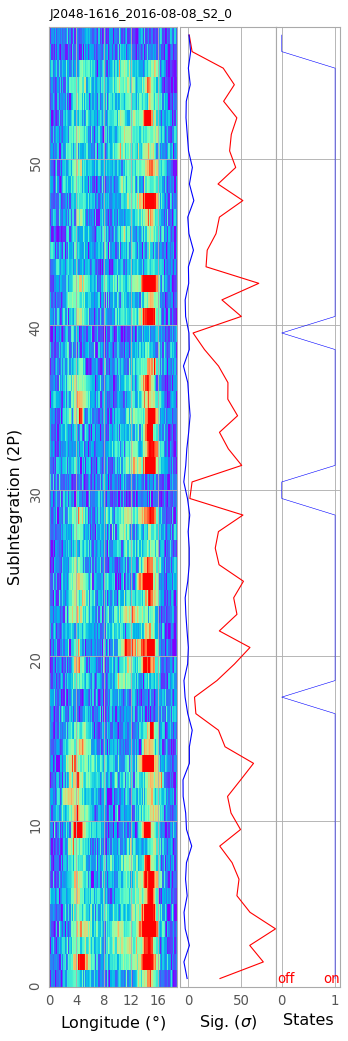}
  \caption{PSR J2048-1616 observed on 2015-06-18 and 2016-08-08 with every 2 pulses foleded.}
  \label{fig:J2048_20150618-20160808}
\end{figure}

%------------------------------J2313
\begin{figure}
  \centering
  \includegraphics[height=10.5cm,width=0.235\textwidth]{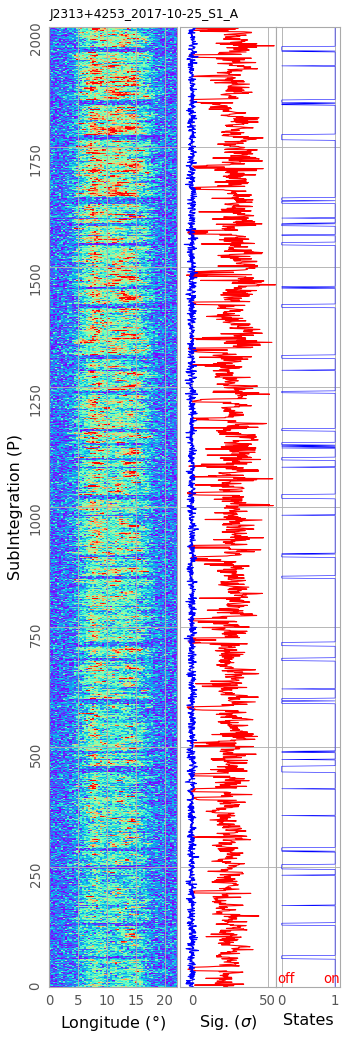}
  \includegraphics[height=10.5cm,width=0.235\textwidth]{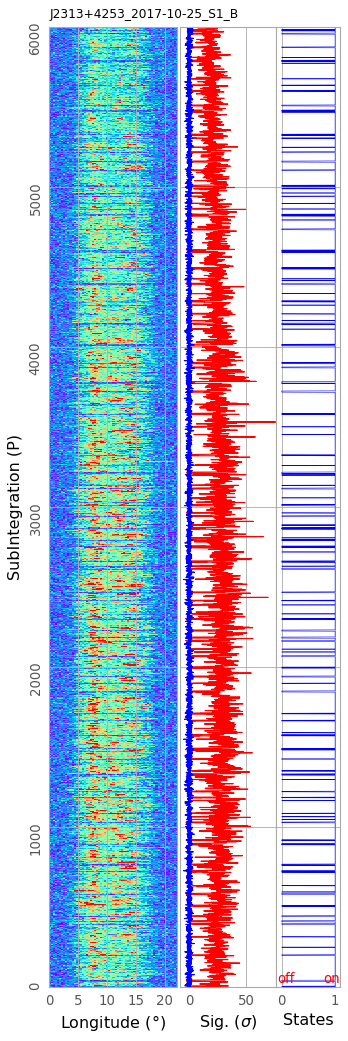}
  \caption{PSR J2313+4253 observed on 2017-10-25 (A and B) with single pulses.}
  \label{fig:J2313_20171025AB}
\end{figure}

%------------------------------J2321
\begin{figure}
  \centering
  \includegraphics[height=10.5cm,width=0.235\textwidth]{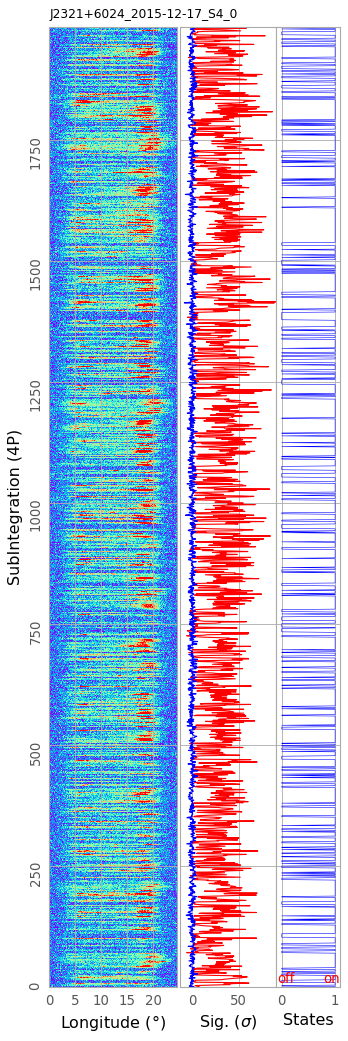}
  \caption{PSR J2321+6024 observed on 2015-12-17 with every 4 pulses foleded.}
  \label{fig:J2321_20151217}
\end{figure}

%------------------------------
\begin{figure}
  \centering
  \includegraphics[height=10.5cm,width=0.235\textwidth]{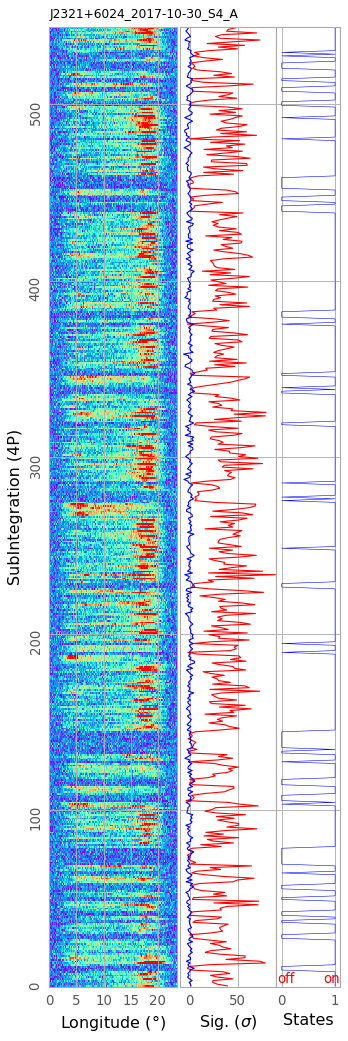}
  \includegraphics[height=10.5cm,width=0.235\textwidth]{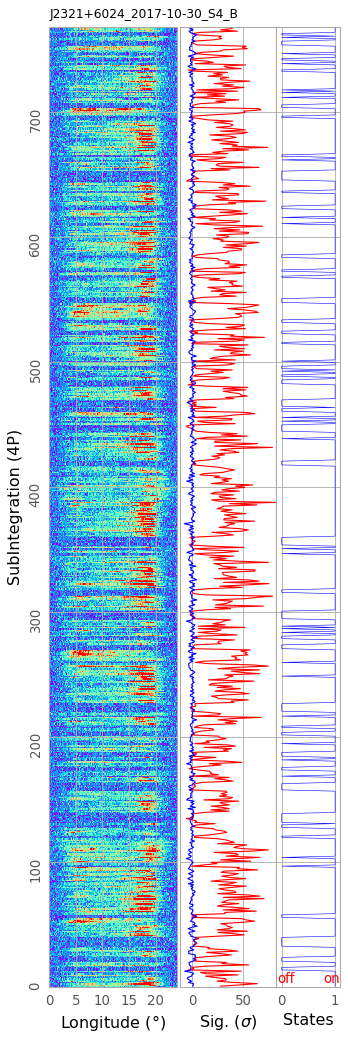}
  \caption{PSR J2321+6024 observed on 2017-10-30 (A and B) with every 4 pulses folded.}
  \label{fig:J2321_20171030AB}
\end{figure}

%------------------------------
\begin{figure}
  \centering
  \includegraphics[height=10.5cm,width=0.235\textwidth]{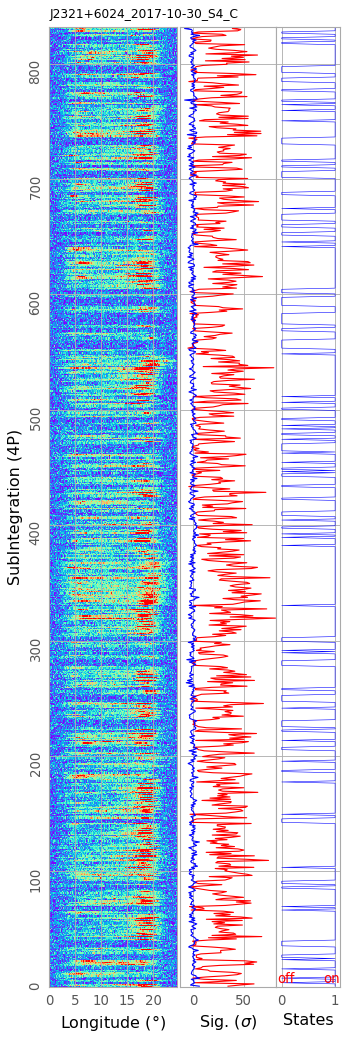}
  \includegraphics[height=10.5cm,width=0.235\textwidth]{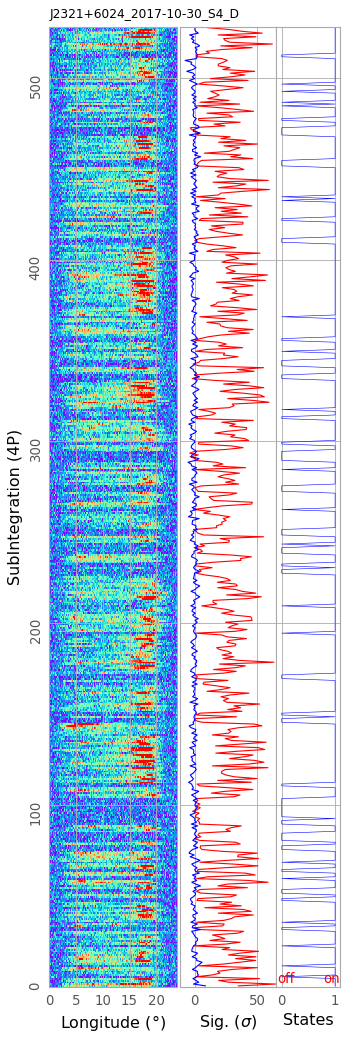}
  \caption{PSR J2321+6024 observed on 2017-10-30 (C and D) with every 4 pulses folded.}
  \label{fig:J2321_20171030CD}
\end{figure}

\clearpage

%\onecolumn
\section{Statistical analysis of nulling behaviours of the 20 pulsars}
\label{appendixB}

\begin{figure*}
  \centering
  \tabcolsep 2.0mm
  \begin{tabular}{cccc}
  \includegraphics[height=4.2cm,width=0.234\textwidth]{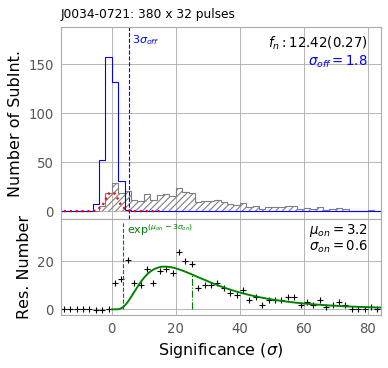} &
  \includegraphics[height=4.2cm,width=0.234\textwidth]{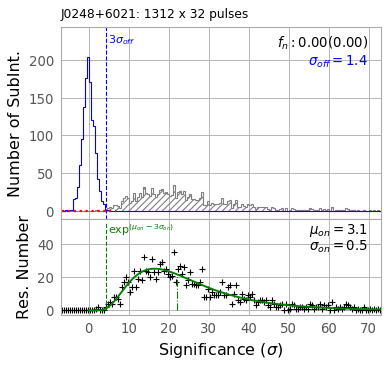} &
  \includegraphics[height=4.2cm,width=0.234\textwidth]{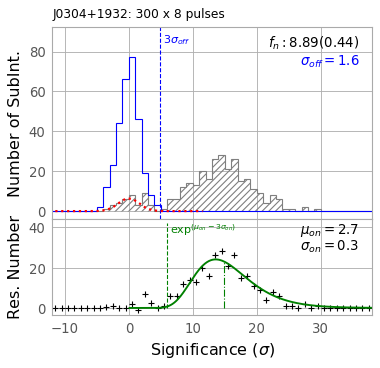} &
  \includegraphics[height=4.2cm,width=0.234\textwidth]{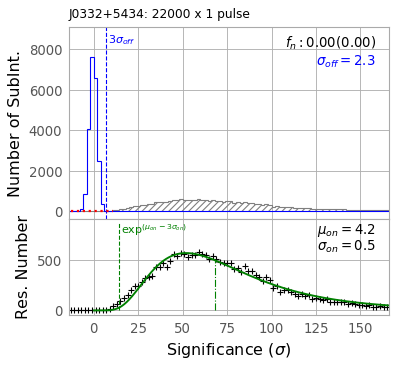}\\
  (1) J0034-0721 & (2) J0248+6021 & (3) J0304+1932 & (4) J0332+5434 \\
  \includegraphics[height=4.2cm,width=0.234\textwidth]{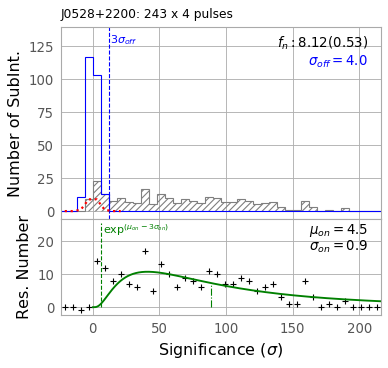} &
  \includegraphics[height=4.2cm,width=0.234\textwidth]{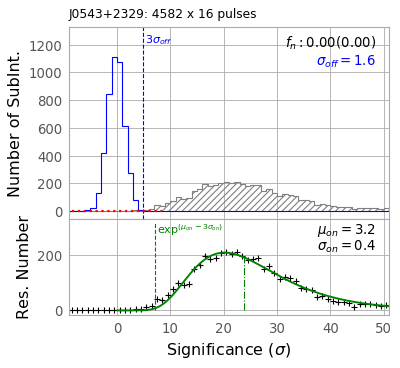} &
  \includegraphics[height=4.2cm,width=0.234\textwidth]{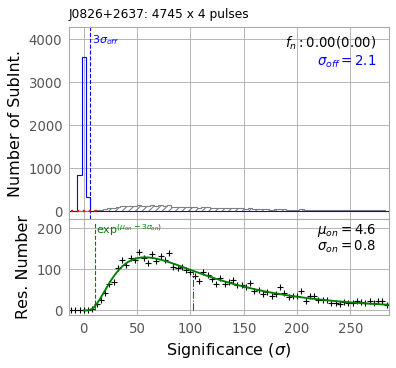} &
  \includegraphics[height=4.2cm,width=0.234\textwidth]{J0908-1739_NF_dis.png}\\
  (5) J0528+2200 & (6) J0543+2329 & (7) J0826+2637 & (8) J0908-1739 \\
  \includegraphics[height=4.2cm,width=0.234\textwidth]{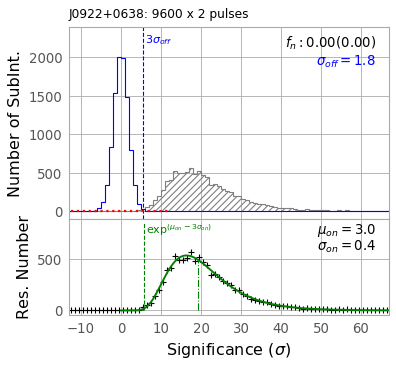} &
  \includegraphics[height=4.2cm,width=0.234\textwidth]{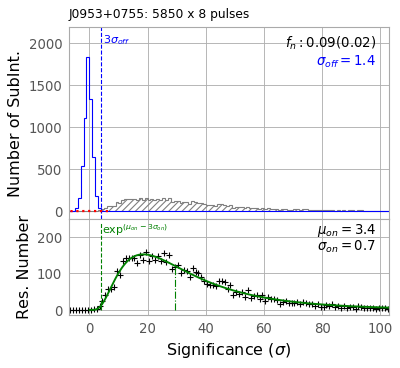} &
  \includegraphics[height=4.2cm,width=0.234\textwidth]{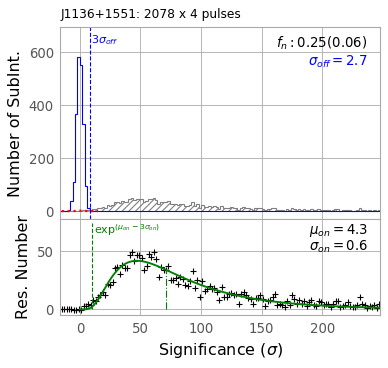} &
  \includegraphics[height=4.2cm,width=0.234\textwidth]{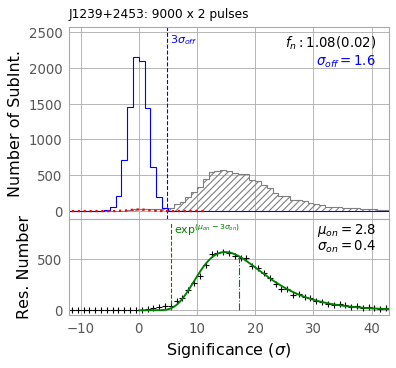}\\
  (9) J0922+0638 & (10) J0953+0755 & (11) J1136+1551 & (12) J1239+2453 \\
  \includegraphics[height=4.2cm,width=0.234\textwidth]{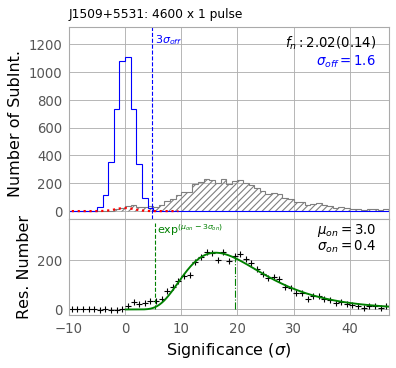} &
  \includegraphics[height=4.2cm,width=0.234\textwidth]{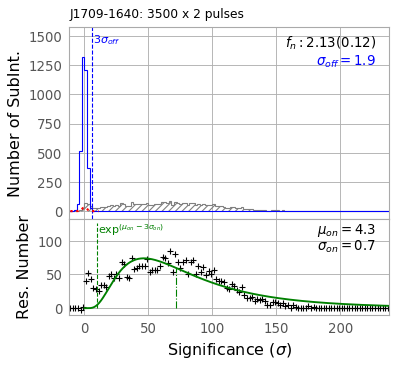} &
  \includegraphics[height=4.2cm,width=0.234\textwidth]{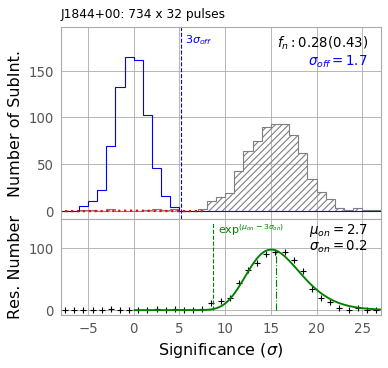}   &
  \includegraphics[height=4.2cm,width=0.234\textwidth]{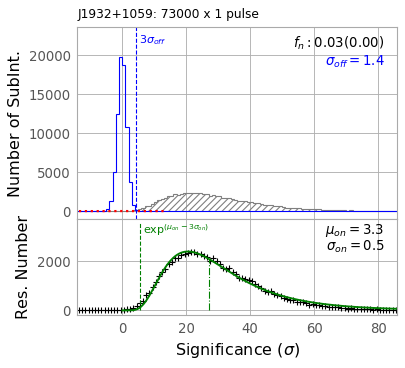}\\
  (13) J1509+5531 & (14) J1709-1640 & (15) J1844+00   & (16) J1932+1059 \\
  \includegraphics[height=4.2cm,width=0.234\textwidth]{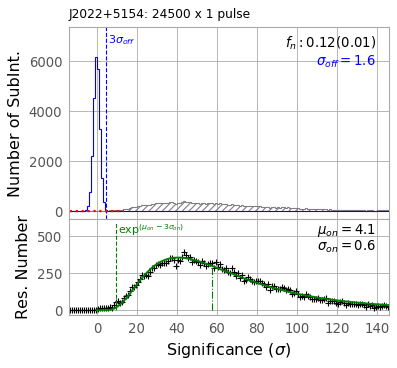} &
  \includegraphics[height=4.2cm,width=0.234\textwidth]{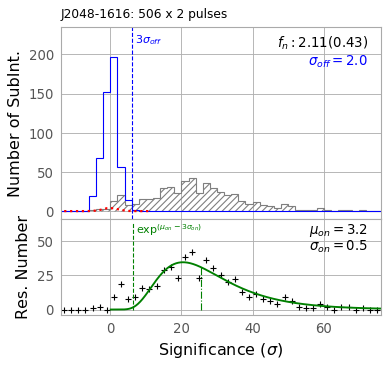} &
  \includegraphics[height=4.2cm,width=0.234\textwidth]{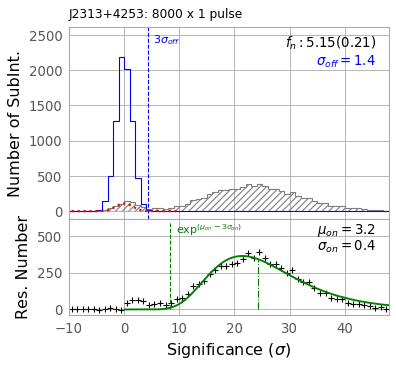} &
  \includegraphics[height=4.2cm,width=0.234\textwidth]{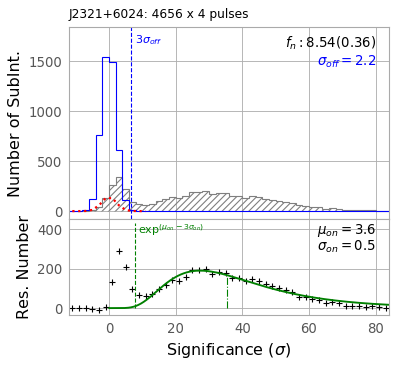}\\
  (17) J2022+5154 & (18) J2048-1616 & (19) J2313+4253 & (20) J2321+6024 \\
  \end{tabular}
  \caption{Histograms of the significances of on-pulses and off-pulses
    for the 20 pulsars. Normal functions with mean zero and standard
    derivation $\sigma_{\rm off}$ are employed to model the
    distribution of off-pulse significances, it is scaled with $f_{\rm
    n}$ and represented by red dotted line. The residual distribution
    of on-pulse significances is ploted below and modeled by lognormal
    functions with mean $\mu_{\rm on}$ and standard derivation
    $\sigma_{on}$.}  \label{fig:nfs}
\end{figure*}

\begin{figure*}
  \centering
  \tabcolsep 5.0mm
  \begin{tabular}{cc}
  \includegraphics[height=4.0cm,width=0.25\textwidth]{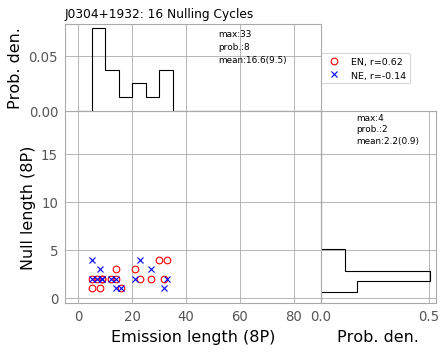}
  \includegraphics[height=3.5cm,width=0.21\textwidth]{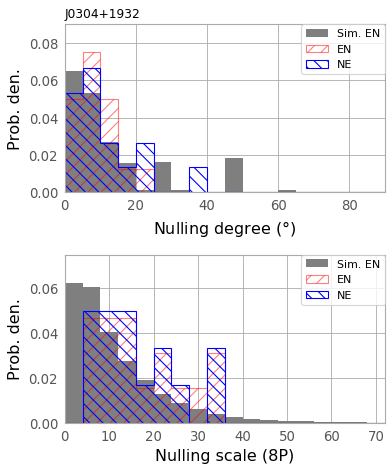} &
  \includegraphics[height=4.0cm,width=0.25\textwidth]{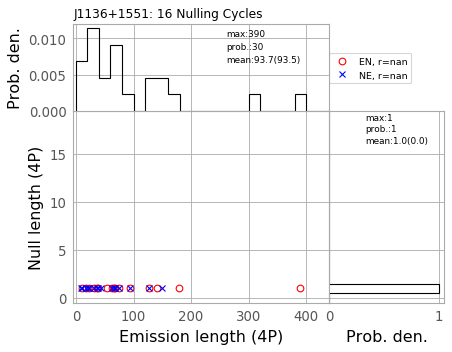}
  \includegraphics[height=3.5cm,width=0.21\textwidth]{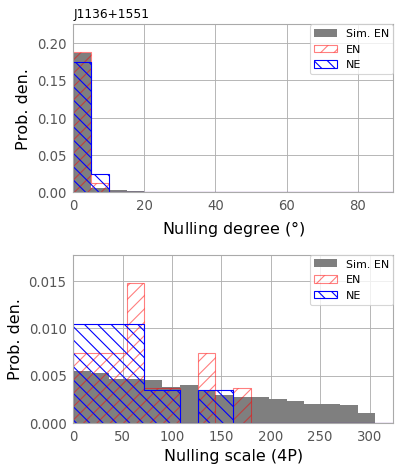}\\
  (1) J0304+1932  &  (2) J1136+1551 \\
  \includegraphics[height=4.0cm,width=0.25\textwidth]{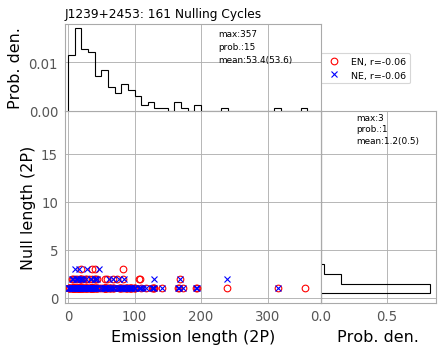}
  \includegraphics[height=3.5cm,width=0.21\textwidth]{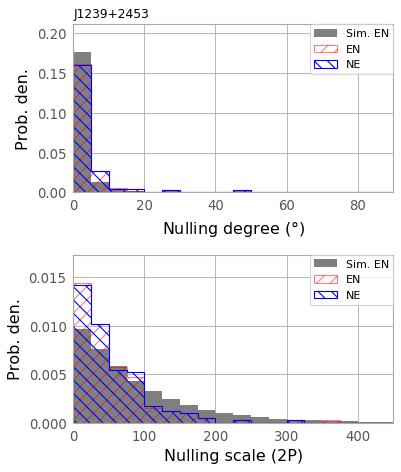} &
  \includegraphics[height=4.0cm,width=0.25\textwidth]{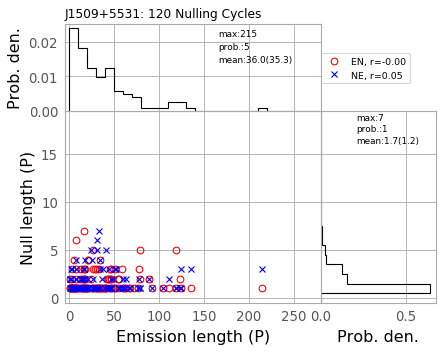}
  \includegraphics[height=3.5cm,width=0.21\textwidth]{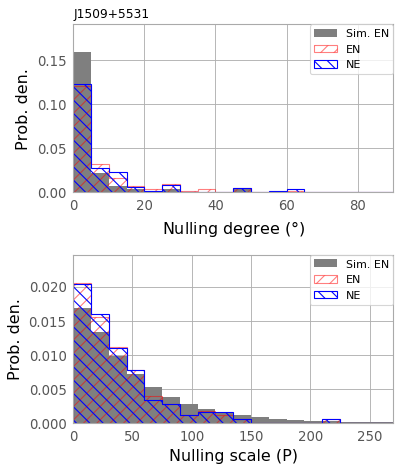}\\
  (3) J1239+2453  &  (4) J1509+5531 \\
  \includegraphics[height=4.0cm,width=0.25\textwidth]{J1709-1640_BN_dis.png}
  \includegraphics[height=3.5cm,width=0.21\textwidth]{J1709-1640_BN_cor.png} &
  \includegraphics[height=4.0cm,width=0.25\textwidth]{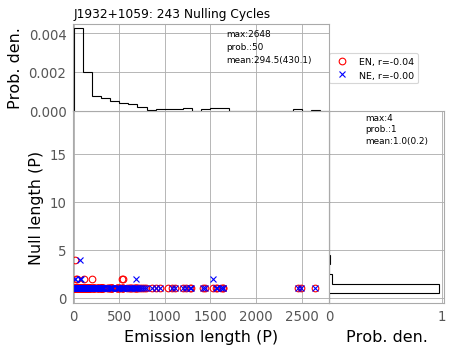}
  \includegraphics[height=3.5cm,width=0.21\textwidth]{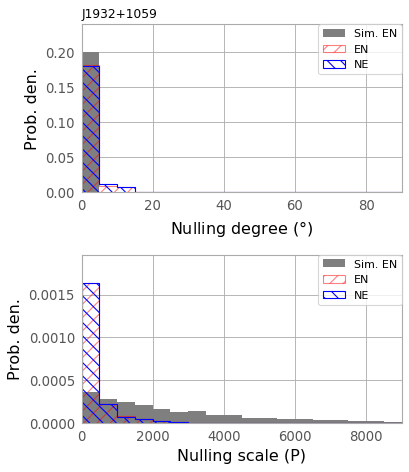}\\
  (5) J1709-1640  &  (6) J1932+1059 \\
  \includegraphics[height=4.0cm,width=0.25\textwidth]{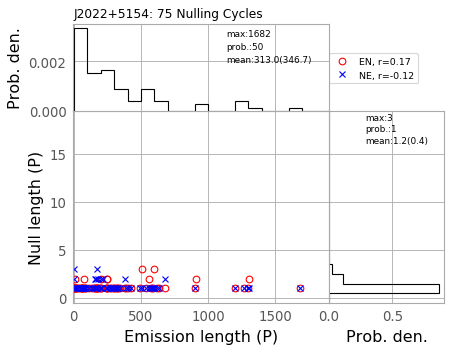}
  \includegraphics[height=3.5cm,width=0.21\textwidth]{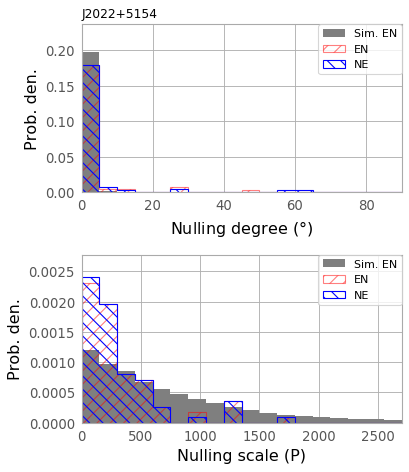} &
  \includegraphics[height=4.0cm,width=0.25\textwidth]{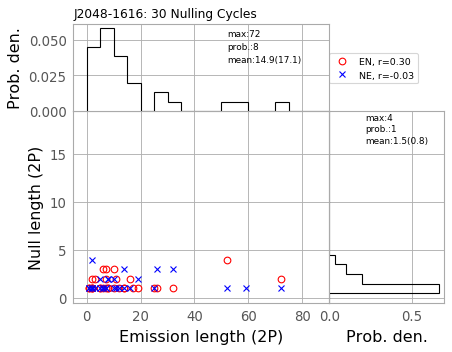}
  \includegraphics[height=3.5cm,width=0.21\textwidth]{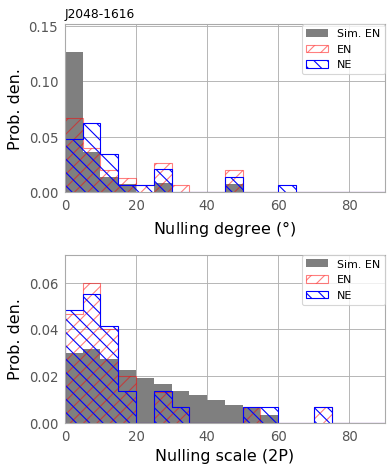}\\
  (7) J2022+5154  &  (8) J2048-1616 \\
  \includegraphics[height=4.0cm,width=0.25\textwidth]{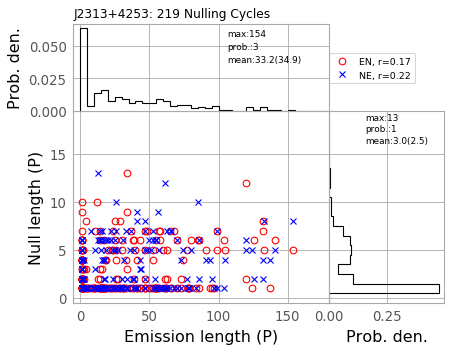}
  \includegraphics[height=3.5cm,width=0.21\textwidth]{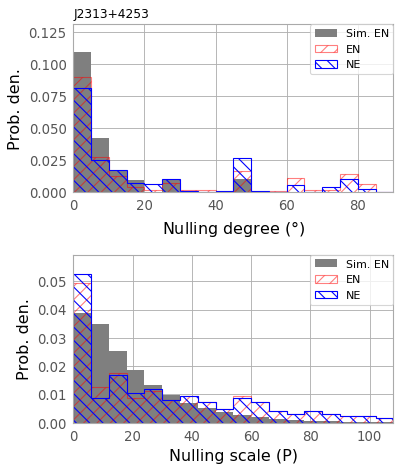} &
  \includegraphics[height=4.0cm,width=0.25\textwidth]{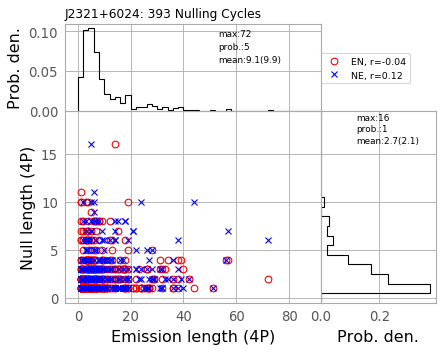}
  \includegraphics[height=3.5cm,width=0.21\textwidth]{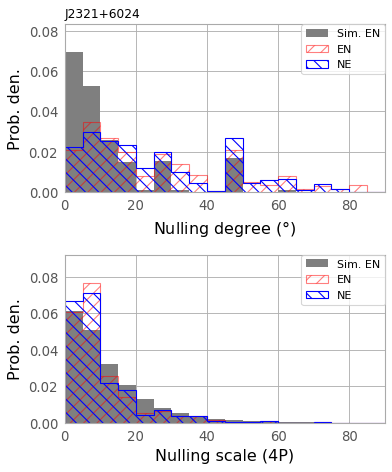}\\
  (9) J2313+4253  &  (10) J2321+6024 \\
  \end{tabular}
  \caption{Emission and null lengths and their correlations for 10
  pulsars.  {\it In the left panels}, distributions of emission and
  null lengths. Length pairs for the emission and the next null are
  denoted as $EN$, and length pairs for the emission and its pre null
  are denoted as $NE$. Correlation coefficients of emission
  lengths with respect to the null lengths are calculated and
  indicated in the top right corner for $EN$ and $NE$. Histograms for
  the emission and null lengths are drawn in the top and right
  panels. {\it In the right panels}, histograms of nulling degrees and
  scales of emission-null length pairs. The left hatched red step and
  right hatched blue step represent distributions for the $EN$ and
  $NE$ length pairs, respectively. The grey bars are for the simulated
  $EN$ length pairs of randomly distributed emission-null sequence.
  } \label{fig:dis-cor}
\end{figure*}

\begin{figure}
  \centering
  \includegraphics[height=3.5cm,width=0.242\textwidth]{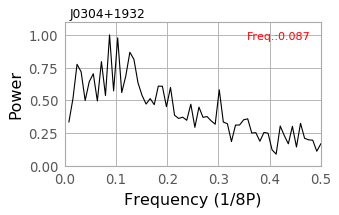}
  \includegraphics[height=3.5cm,width=0.242\textwidth]{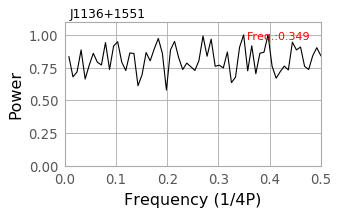}\\
  \includegraphics[height=3.5cm,width=0.242\textwidth]{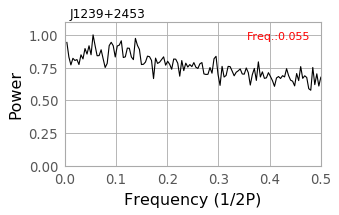}
  \includegraphics[height=3.5cm,width=0.242\textwidth]{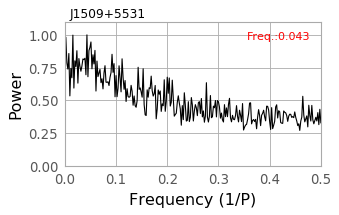}\\
  \includegraphics[height=3.5cm,width=0.242\textwidth]{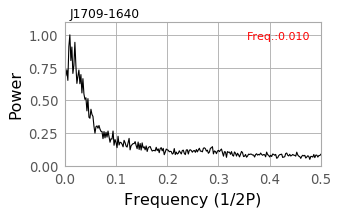}
  \includegraphics[height=3.5cm,width=0.242\textwidth]{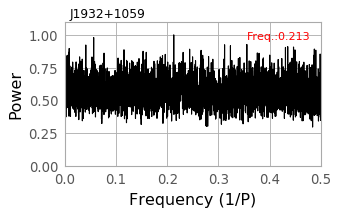}\\
  \includegraphics[height=3.5cm,width=0.242\textwidth]{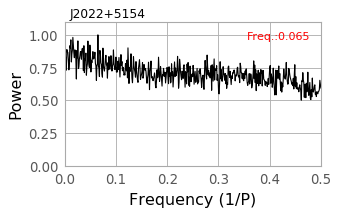}
  \includegraphics[height=3.5cm,width=0.242\textwidth]{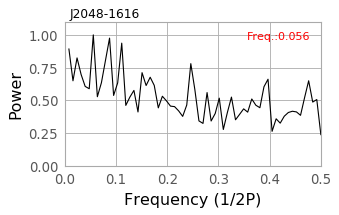}\\
  \includegraphics[height=3.5cm,width=0.242\textwidth]{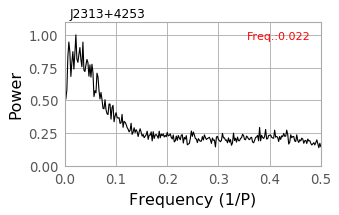}
  \includegraphics[height=3.5cm,width=0.242\textwidth]{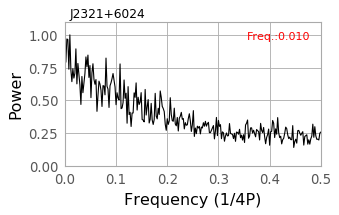}\\
  \caption{Average power spectrum for the 10 pulsars. It is calculated from sliding
    Fourier transform of the emission-null (1-0) sequence in
    Appendix~\ref{appendixA}.  }
    \label{fig:period}
\end{figure}

\begin{figure}
  \centering
  \includegraphics[height=3.7cm,width=0.242\textwidth]{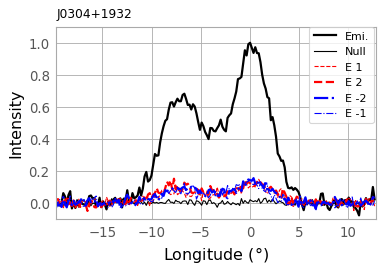}
  \includegraphics[height=3.7cm,width=0.242\textwidth]{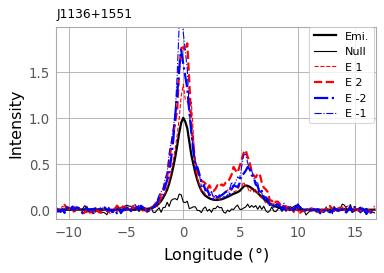}\\
  \includegraphics[height=3.7cm,width=0.242\textwidth]{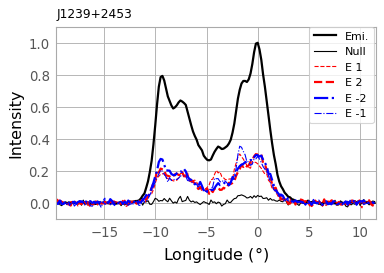}
  \includegraphics[height=3.7cm,width=0.242\textwidth]{J1509+5531_B_profs.png}\\
  \includegraphics[height=3.7cm,width=0.242\textwidth]{J1709-1640_B_profs.png}
  \includegraphics[height=3.7cm,width=0.242\textwidth]{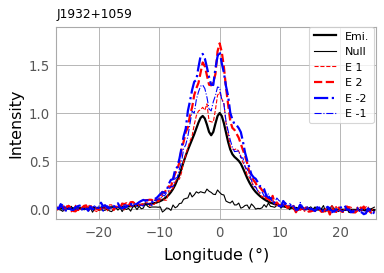}\\
  \includegraphics[height=3.7cm,width=0.242\textwidth]{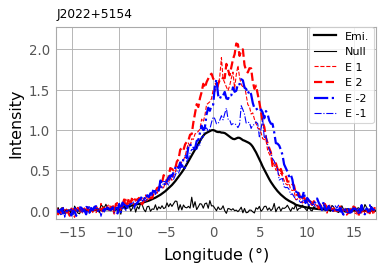}
  \includegraphics[height=3.7cm,width=0.242\textwidth]{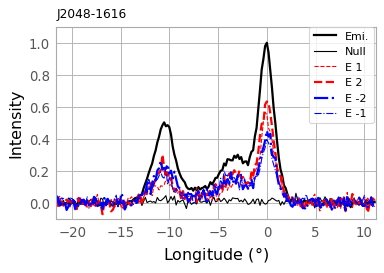}\\
  \includegraphics[height=3.7cm,width=0.242\textwidth]{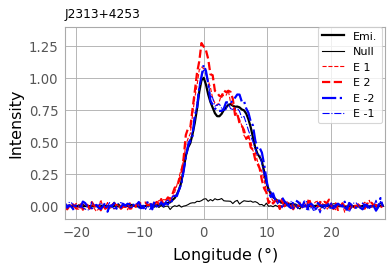}
  \includegraphics[height=3.7cm,width=0.242\textwidth]{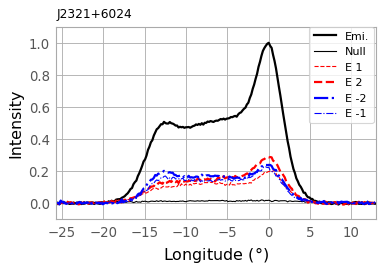}\\

\caption{Integrated pulse profiles for 10 pulsars. Profiles for the emission and
    nulls are represented by thick and thin black solid lines. For
    emission on states lasting for more than 5 pulses or
    subintegrations, profiles for the first and the second
    subintegration are represented by the thin and thick red dashed
    lines, and profiles for the last two subintegrations are
    represented by the thin and thick blue dash-dotted lines.}
    
\label{fig:profs}
\end{figure}

Statistical analysis on the nulling fractions are shown in
Figure~\ref{fig:nfs}, emission and null lengths and their correlations
in Figure~\ref{fig:dis-cor}, nulling periodicity in
Figure~\ref{fig:period} and intentensity variations in
Figure~\ref{fig:profs}.

%\clearpage
%----------------------------
%\twocolumn
%\section{Emission and null interaction of 12 pulsars}
%\label{appendixC}

\clearpage
\onecolumn      
    {
    %%%%%%%%%%%%%%%%%%%%%%%%%%%%%%%%%
    \section{Nulling pulsars reported hitherto}
    \label{appendixC}

    By now, 214 pulsars have been reported to null at various
    frequencies, including those collected from literatures and the
    ones presented in this work. They are classified into, those with
    nulling fraction measurements, those without nulling fraction
    measurements and intermittent pulsars. For each pulsar, the
    available nulling fractions are grouped into 5 frequency bands,
    303-333, 408-430, 607-645, 1308-1518 and 2000-4850~MHz, as listed
    in columns (5) to (9). Column (10) presents the average nulling
    fractions, which are taken as half the upper limit if one pulsar
    has only an upper limit reported, or as 1.5 times the lower limit
    if one has only a lower limit.
    
      %\small
      \scriptsize
      %\tiny
      \setlength{\tabcolsep}{0.5pt}
      \begin{longtable}{cccrcccccll}
        \caption{Nulling fractions for 214 pulsars.}
        \label{table:null} \\    
        \hline  
        \hline
%\begin{table*}
%        \centering     
%        \caption{Parameters for previous researches on nulling pulsars.}
%        \label{table:prev}
%        \tabcolsep 0.5mm
%        \tiny
        %\small
%        \begin{tabular}{cccrcccccll}
%        \hline  
%        \hline
JName & BName & Period & DM & \multicolumn{5}{c}{$f_{\rm n}$ ($\%$)} & <$f_{\rm n}$> & Reference  \\
\cline{5-9} 
      &       & (s) & (pc/cm$^3$)        &  303--333~MHz    &  408--430~MHz  &  607--645~MHz & 1308--1518~MHz& 2000--4850~MHz & ($\%$) &      \\
  (1) & (2)   & (3) & (4)                &  (5)             &   (6)          &  (7)          &  (8)          &  (9)           & (10)   &  (11) \\     
\hline
\endhead
\hline
\endfoot 
J0034$-$0721 & B0031$-$07 & 0.943 & 11.3 &44.6(1.3), 43(2), 31.3(2.3) &      &$\leqslant$16, 44(2), 22.8(1.8) & 43(2) & 24.9(13.3) & 36.2  & Big92, Viv95, GJK+14, BMM17, our \\
J0048$+$3412 & B0045$+$33 & 1.217 & 39.9 &21(3), 21          &               &               &               &            & 21   & HR09, RR09                  \\
J0151$-$0635 & B0148$-$06 & 1.464 & 25.6 &                   &               & $\leqslant$5  &               &            & 2.5  & Big92                       \\
J0152$-$1637 & B0149$-$16 & 0.832 & 11.9 & $\leqslant$2.5    &               &               &               &            & 1.3  & Viv95                       \\
J0248$+$6021 &            & 0.217 &370.0 &                   &               &               &               & 17.3(17.9) & 17.3 & our                         \\
J0304$+$1932 & B0301$+$19 & 1.387 & 15.7 &14(4), 13, 8.7(1.2)&10             & 6.1(0.6)      &               & 13.2(4.9)  & 10.8 & Ran86, HR09, RR09, BMM17, our  \\
J0332$+$5434 & B0329$+$54 & 0.714 & 26.7 &                   &$\leqslant$0.25&               &               &   0.0      & 0.0  & Rit76, our                  \\
J0452$-$1759 & B0450$-$18 & 0.548 & 39.9 &                   &$\leqslant$0.5 &$\leqslant$0.9 &               &            & 0.4  & Rit76, Big92                \\
J0458$-$0505 &            & 1.883 & 47.8 & 69                &               & 63            &               &            & 66   & LBR+13, WSE07               \\
J0525$+$1115 & B0523$+$11 & 0.354 & 79.4 &                   & <0.06         &               &               &            & 0.03 & WAB+86                      \\
J0528$+$2200 & B0525$+$21 & 3.745 & 50.9 &28(2), 25, 14.4(1.2)&              & 25(5)         &               &  15.2(7.7) & 21.5 & Rit76, HR09, RR09, BMM17, our   \\
J0529$-$6652 & B0529$-$66 & 0.975 &103.2 &                   &               &               & 83.3(1.5)     &            & 83.3 & CALL13                      \\
J0536$-$7543 & B0538$-$75 & 1.245 & 17.5 &                   &               &               & 32.5(6.5)     &            & 32.5 & CALL13                      \\
J0540$+$3207 &            & 0.524 & 61.9 & 54(1), 53         &               &               &               &            & 53.5 & HR09, RR09                  \\
J0543$+$2329 & B0540$+$23 & 0.245 & 77.7 &                   &               &               &               &  8.3(8.5)  & 8.3  & our                         \\
J0629$+$2415 & B0626$+$24 & 0.476 & 84.1 &                   & <0.02         &               &               &            & 0.01 & WAB+86                      \\
J0630$-$2834 & B0628$-$28 & 1.244 & 34.4 & 13.6(1.9)         & $\leqslant$2  &$\leqslant$0.3 &               &            & 13.6 & Rit76, Big92, BMM17         \\
J0659$+$1414 & B0656$+$14 & 0.384 & 14.0 &                   & 12(4)         &               &               &            & 12   & WAB+86                      \\
J0738$-$4042 & B0736$-$40 & 0.374 &160.8 &                   &               &$\leqslant$0.4 &               &            & 0.2  & Big92                       \\
J0742$-$2822 & B0740$-$28 & 0.166 & 73.7 &                   &$\leqslant$0.25&$\leqslant$0.2 &               &            & 0.13 & Rit76, Big92                \\
J0754$+$3231 & B0751$+$32 & 1.442 & 39.9 &38(6), 39          & 34(0.5)       &               &               &            & 37   & WAB+86, HR09, RR09          \\
J0814$+$7429 & B0809$+$74 & 1.292 &  5.7 &1.4(0.3)           &$\leqslant$5,1.42(0.02) &1.0(0.4), 1.6(0.4) &1.2(0.2) & >1  & 1.3  & Rit76, LA83, GJK12, GJK+14  \\
J0820$-$1350 & B0818$-$13 & 1.238 & 40.9 &9.8(1.2)           & 1.5(0.25),1.01 &1(0.5),0.9(1.8), 0.8 &     &        & 2.5  & Rit76, LA83, Big92, GJK12, BMM17   \\
J0820$-$4114 & B0818$-$41 & 0.545 &113.4 &30                 &               &               &               &            & 30   & BGG10                       \\
J0823$+$0159 & B0820$+$02 & 0.864 & 23.7 &                   & <0.06         &               &               &            & 0.03 & WAB+86                      \\
J0826$+$2637 & B0823$+$26 & 0.530 & 19.4 &7(2), 7            &$\leqslant$5   &               &               &   1.7(1.7) & 5.2  & Rit76, HR09, RR09, our      \\
J0828$-$3417 & B0826$-$34 & 1.848 & 52.2 &                   & >70           &$\leqslant$12  & -             &            & -    & DLL+79, Big92, BJB+12    \\
J0837$+$0610 & B0834$+$06 & 1.273 & 12.8 &7.0(2.0), 9.3, 9(1), 9, 3.9(0.3) &7.1(1)  & 4.7(0.7) &             &            & 7.1  & Rit76, Viv95, RW07, HR09, RR09, BMM17 \\
J0837$-$4135 & B0835$-$41 & 0.751 &147.2 &                   &               &$\leqslant$1.2, 1.7(1.2) &          &       & 1.7  & Big92, GJK12                \\
J0908$-$1739 & B0906$-$17 & 0.401 & 15.8 &26.8(1.7)          &               & 25.7(1.3)     &               &  24.0(6.1) & 25.5 & BMM17, our                  \\
J0922$+$0638 & B0919$+$06 & 0.430 & 27.2 &                   &<0.05          &               &               & 0.03(0.03) & 0.03 & WAB+86, our                 \\
J0934$-$5249 & B0932$-$52 & 1.444 & 100  & 5(3)              &               &               & -             &            & 5    & BJB+12, NJMK17              \\
J0942$-$5552 & B0940$-$55 & 0.664 &180.2 &                   &               &$\leqslant$12.5&               &            & 6.3  & Big92                       \\
J0943$+$1631 & B0940$+$16 & 1.087 & 20.3 &                   & 8(3)          &               &               &            & 8    & WAB+86                      \\
J0944$-$1354 & B0942$-$13 & 0.570 & 12.5 &$\leqslant$7, 14.4(0.9) &          &               &               &            & 14.4 & Viv95, BMM17                \\
J0953$+$0755 & B0950$+$08 & 0.253 &  2.9 &$\leqslant$5       &$\leqslant$5   &               &               &  5.9(6.1)  & 5.9  & Rit76, Viv95, our           \\
J1049$-$5833 &            & 2.202 &446.8 &                   &               &               &47(3), -, 33(35)&           & 40   & WMJ07, BJB+12, CALL13       \\
J1057$-$5226 & B1055$-$52 & 0.197 & 30.1 &                   &               &$\leqslant$11  &               &            & 5.5  & Big92                       \\
J1115$+$5030 & B1112$+$50 & 1.656 &  9.1 & -                 & 60(5)         & 64(6)         &               &            & 62   & Rit76, WSE07, GJK12         \\
J1116$-$4122 & B1114$-$41 & 0.943 & 40.5 &3.3(0.5)           &               &               &               &            & 3.3  & BMM17                       \\
J1136$+$1551 & B1133$+$16 & 1.187 &  4.8 &19.8(4.5)$^\sharp$,20.6, 20, 13.7(2.1) &15(2.5) &18.3(5.8)$^\sharp$,11.9(2.3) &  & 5.9(6.2)& 15.7 & Rit76, BGK+07, HR07, RR09, BMM17, our \\
J1239$+$2453 & B1237$+$25 & 1.382 &  9.2 &6(1), 2.0(0.1), 7(3) & 6(2.5)      & 3.1(0.4), 4(1)&               &   2.0(0.9) & 4.3  & Rit76, HR09, BMM17, NJMK17, our\\
J1243$-$6423 & B1240$-$64 & 0.388 &297.2 &                   &               &$\leqslant$4   & -             &            & 2    & Big92, BJB+12               \\
J1326$-$6700 & B1322$-$66 & 0.543 &209.6 &                   &               &               & 9.1           &            & 9.1  & WMJ07                       \\
J1328$-$4921 & B1325$-$49 & 1.478 &118   & 4.0               &               & 4.4           &               &            & 4.2  & BMM17                       \\
J1359$-$6038 & B1356$-$60 & 0.127 &293.7 &                   &               &               & 0.1(2.3)      &            & 0.1  & CALL13                      \\
J1401$-$6357 & B1358$-$63 & 0.842 & 98.0 &                   &               &               & 1.6           &            & 1.6  & WMJ07                       \\
J1430$-$6623 & B1426$-$66 & 0.785 & 65.3 &                   &               &$\leqslant$0.05&               &            & 0.03 & Big92                       \\
J1456$-$6843 & B1451$-$68 & 0.263 &  8.6 &                   &               &$\leqslant$3.3 &               &            & 1.7  & Big92                       \\
J1502$-$5653 &            & 0.535 &194.0 &                   &               &               &93(4), 93.6, -, 70(9) &     & 85.5 & WMJ07, LEM+12, BJB+12, CALL13 \\
J1509$+$5531 & B1508$+$55 & 0.739 & 19.6 &                   &               & 7(2)          &               &   2.0(0.1) & 4.5  & NJMK17, our                 \\
J1525$-$5417 &            & 1.011 & 235  &                   &               &               &16(5), 26(5)   &            & 21   & WMJ07, CALL13               \\
J1527$-$3931 & B1524$-$39 & 2.417 & 49   & 5.1(1.3)          &               &               & -             &            & 5.1  & BJB+12, BMM17               \\
J1532$+$2745 & B1530$+$27 & 1.124 & 14.6 &                   & 6(2)          &               &               &            & 6.0  & WAB+86                      \\
J1534$-$5334 & B1530$-$53 & 1.368 & 24.8 &                   &               &$\leqslant$0.25&               &            & 0.13 & Big92                       \\
J1543$-$0620 & B1540$-$06 & 0.709 & 18.3 & 2(1)              &               & 4(2)          &               &            & 3.0  & NJMK17                      \\ 
J1559$-$4438 & B1556$-$44 & 0.257 & 56.1 &                   &               &$\leqslant$0.01, 0.24&         &            & 0.24 & Big92, BMM17                \\
J1607$-$0032 & B1604$-$00 & 0.421 & 10.6 &                   &$\leqslant$1   &$\leqslant$0.1 &               &            & 0.05 & Rit76, Big92                \\
J1614$+$0737 & B1612$+$07 & 1.206 & 21.3 &10                 &<5             &               &               &            & 10.0 & WAB+86, RR09                \\
J1634$-$5107 &            & 0.507 &372.8 &                   &               &               & 90(5)         &            & 90.0 & YWS+15                      \\
J1639$-$4359 &            & 0.587 &258.9 &                   &               &$\leqslant$0.1 &               &            & 0.05 & GJK12                       \\
J1644$-$4559 & B1641-45 & 0.455 &478.8 &                     &               &$\leqslant$0.4 &               &            & 0.2  & Big92                       \\
J1645$-$0317 & B1642$-$03 & 0.387 & 35.7 &$\leqslant$0.5     &$\leqslant$0.25&               &               &            & 0.13 & Rit76, Viv95                \\
J1648$-$4458 &            & 0.629 &925   &                   &               &               & 1.4           &            & 1.4  & WMJ07                       \\
J1649$+$2533 &            & 1.015 & 34.4 &25(5), 20          & 30            &               &               &            & 25   & LWF+04, HR09, RR09          \\
J1701$-$3726 & B1658$-$37 & 2.454 &303.4 &                   &               & 19(6)         & 14(2), -      &            & 16.5 & WMJ07, GJK12,  BJB+12       \\
J1702$-$4428 &            & 2.123 &395   &                   &               &               & 26(3)         &            & 26   & WMJ07                       \\
J1703$-$3241 & B1700$-$32 & 1.211 &110.3 &1.6                &               & 0.4           & -             &            & 1.0  & BJB+12, BMM17               \\
J1703$-$4851 &            & 1.396 &150.2 &                   &               &               & 1.1           &            & 1.1  & WMJ07                       \\
J1709$-$1640 & B1706$-$16 & 0.653 & 24.8 &3.7(1.3), 31(2)    &               & 4.9(0.3)      &               &  2.7(0.8)  & 10.6 & BMM17, NJMK18, our          \\
J1715$-$4034 &            & 2.072 &254   &                   &               & $\geq6$       &               &            & 9.0  & GJK12                       \\
J1717$-$4054 & B1713$-$40 & 0.887 &306.9 &                   &               &               &>95, 80(15), 77(5)&         & 78.5 & WMJ07, KHS+14, YWS+15       \\
J1722$-$3207 & B1718$-$32 & 0.477 &126.0 &                   &               & 1(1)          &               &            & 1    & NJMK17                      \\
J1725$-$4043 &            & 1.465 &203   &                   &               &$\leqslant$70  & -             &            & 35.0 & GJK12, BJB+12               \\
J1727$-$2739 &            & 1.293 &147   &57.0(2.3)          &               &48.3(1.8)      &52(3), -, 68.2(1.1) &       & 56.4 & WMJ07, BJB+12, WWY+16, BMM17\\
J1731$-$4744 & B1727$-$47 & 0.829 &123.3 &                   &               &$\leqslant$0.1 &               &            & 0.05 & Big92                       \\
J1733$-$3716 & B1730$-$37 & 0.337 &153.5 &                   &               &52.4(3.5)      &               &            & 52.4 & BMM17                       \\
J1738$-$2330 &            & 1.978 & 99.3 &$\geq69$, 85.1(2.3)&               &               & -             &            & 85.1 & GJK12, BJB+12, GJW14        \\
J1740$+$1311 & B1737$+$13 & 0.803 & 48.6 &                   &$\leqslant$0.02&               & 0.04          &            & 0.04 & WAB+86, FR10                \\
J1741$-$0840 & B1738$-$08 & 2.043 & 74.9 & 15.7(1.7)         &               &15.8(1.4), 30(5)& -            &            & 20.5 & BJB+12, BMM17, GYY+17       \\
J1744$-$3922 &            & 0.172 &148.1 &                   &               &               & 75, -         &            & 75   & FSK+04, KEK+13              \\
J1745$-$3040 & B1742$-$30 & 0.367 & 88.3 & 40.2(1.9)         &               &$\leqslant$17.5, 24.8(1.0) & - &            & 32.5 & Big92, BJB+12, BMM17        \\
J1751$-$4657 & B1747$-$46 & 0.742 & 20.4 & 2.4(0.5)          &               & 2.4           &               &            &  2.4 & BMM17                       \\
J1752$+$2359 &            & 0.409 & 36.2 & 81, <89           & 75(5)$^\sharp$ &               &               &            & 78   & LWF+04, RR09, GJW14         \\
J1752$-$2806 & B1749$-$28 & 0.562 & 50.3 & 0.2               &$\leqslant$0.75& $\leqslant$1.4, 1.7(0.4)&       &          & 0.9  & Rit76, Big92, BMM17 \\
J1801$-$0357 & B1758$-$03 & 0.921 &120.3 & -, 27.7(1.3)      &               &26.1(2.6)      &               &            & 26.9 & WSE07, BMM17                \\
J1808$-$0813 &            & 0.876 &151.2 & -, 12.8(1.3)      &               &8.2(1.0)       & -             &            & 10.5 & WSE07, BJB+12, BMM17        \\
J1812$-$1718 & B1809$-$173& 1.205 &255.1 &                   &               &               & 5.8           &            & 5.8  & WMJ07                       \\
J1817$-$3618 & B1813$-$36 & 0.387 & 94.3 &16.7(0.7)          &               &               & -             &            & 16.7 & BJB+12, BMM17               \\
J1819$+$1305 &            & 1.060 & 64.8 &36.74, 41(6)       &               &               &               &            & 38.9 & RW08, HR09                  \\
J1820$-$0427 & B1818$-$04 & 0.598 & 84.4 &                   &$\leqslant$0.75&$\leqslant$0.25&               &            & 0.13 & Rit76, Big92                \\
J1820$-$0509 &            & 0.337 &104   &                   &               &               &67(3), -       &            & 67   & WMJ07, BJB+12               \\
J1822$-$2256 & B1819$-$22 & 1.874 &121.2 &4.7(0.9), 5.5(0.2) &               &5.5(0.7), 10(2)& -             &            & 6.4  & BJB+12, BMM17, NJMK17, BM18 \\
J1823$+$0550 & B1821$+$05 & 0.752 & 66.7 &                   &$\leqslant$0.4 &               &               &            & 0.2  & WAB+86                      \\
J1831$-$1223 &            & 2.857 &342   &                   &               &               & 4(1), -       &            & 4.0  & WMJ07, BJB+12               \\
J1833$-$1055 &            & 0.633 &543   &                   &               &               & 7(2)          &            & 7    & WMJ07                       \\
J1834$-$0010 & B1831$-$00 & 0.520 & 88.6 & <2                &               &               &               &            & 1    & HR09                        \\
J1840$-$0840 &            & 5.309 &272   &                   &               & 50(6)         & -             &            & 50   & BJB+12, GYY+17              \\
J1841$+$0912 & B1839$+$09 & 0.381 & 49.1 & <2                & <5            &               &               &            & 1    & WAB+86, HR09                \\
J1843$-$0211 &            & 2.027 &441.7 &                   &               &               & 6(2)          &            & 6    & WMJ07                       \\
J1844$+$00   &            & 0.460 &345.5 &                   &               &               &               &  17.2(17.5)& -    & our                         \\
J1844$+$1454 & B1842$+$14 & 0.375 & 41.4 &                   & <0.15         &               &               &            & 0.08 & WAB+86                      \\
J1847$-$0402 & B1844$-$04 & 0.597 &141.9 &                   &               & 3(1)          &               &            & 3    & NJMK17                      \\
J1848$-$1952 & B1845$-$19 & 4.308 & 18.2 & 27(6)             &               & 19(4)         & -             &            & 23   & BJB+12, NJMK17              \\
J1851$+$1259 & B1848$+$12 & 1.205 & 70.6 & 51(2), 54         &               &               &               &            & 52.5 & HR09, RR09                  \\
J1853$+$0505 &            & 0.905 &279   &                   &               &               & 67(8)         &            & 67   & YWS+15                      \\
J1900$-$2600 & B1857$-$26 & 0.612 & 37.9 & 20(3), 5.1(0.8)   & 10(2.5)       & 8.1(0.5)      &               &            & 10.8 & Rit76, MR08, BMM17          \\
J1901$+$0413 &            & 2.663 &352   &                   &               &$\leqslant$6   &               &            & 3    & GJK12                       \\
J1901$-$0906 &            & 1.781 & 72.6 & 2.9, 29(4)        &               &5.6(0.7), 30(1)&               &            & 16.9 & BMM17, NJMK17               \\
J1910$+$0358 & B1907$+$03 & 2.330 & 82.9 &                   & 4(0.2)        &               &               &            & 4    & WAB+86                      \\
J1913$-$0440 & B1911$-$04 & 0.825 & 89.3 &                   &$\leqslant$0.5 &               &               &            & 0.3  & Rit76                       \\
J1916$+$1023 &            & 1.618 &329.8 &                   &               &               & 47(4)         &            & 47   & WMJ07                       \\
J1919$+$0021 & B1917$+$00 & 1.272 & 90.3 &                   &$\leqslant$0.1 &               &               &            & 0.05 & Ran86                       \\
J1920$+$1040 &            & 2.215 &304   &                   &               &               & 50(4)         &            & 50   & WMJ07                       \\
J1921$+$1948 & B1918$+$19 & 0.821 &153.8 & 9(2), 8.9, 2.0    &               &               &               &            & 6.6  & HR09, RWB13, BMM17          \\
J1921$+$2153 & B1919$+$21 & 1.337 & 12.4 &$\leqslant$1.2     &$\leqslant$0.25&               &               &            & 0.13 & Rit76, Viv95                \\
J1926$-$0652 &            & 1.608 & 84.7 &                   & 75            &               &               &            & 75   & ZLH+19                      \\
J1926$-$1314 &            & 4.864 & 40.8 & 72.7(2.5)         &               &75.7(1.9)$^\ddag$&              &            & 74.2 & RSM+13                      \\
J1926$+$0431 & B1923$+$04 & 1.074 &102.2 &                   & <5            &               &               &            & 2.5  & Rit76                       \\
J1932$+$1059 & B1929$+$10 & 0.226 &  3.1 &                   & $\leqslant$1  &               &               & 0.03(0.003)& 0.03 & Rit76, our                  \\
J1935$+$1616 & B1933$+$16 & 0.358 &158.5 &                   &$\leqslant$0.25&$\leqslant$0.06&               &            & 0.03 & Rit76, Big92                \\
J1944$+$1755 & B1942$+$17 & 1.996 &175   &                   & $\geq60$      &               &               &            & 90   & LCX02                       \\
J1945$-$0040 & B1942$-$00 & 1.045 & 59.7 &28                 & 21(1)         &               &               &            & 24.5 & WAB+86, RR09                \\
J1946$+$1805 & B1944$+$17 & 0.440 & 16.1 &50(7),66.67, 29.7(1.4) & 64        &55(5),37.9(2.3)& 66.67         &            & 52.8 & Rit76, DCHR86, Viv95, KR10, BMM17  \\
J1948$+$3540 & B1946$+$35 & 0.717 &129.3 &                   &$\leqslant$0.75&               &               &            & 0.4  & Rit76                       \\
J2006$-$0807 & B2003$-$08 & 0.580 & 32.3 & 15.6(1.0)         &               & 24.2(1.5)     &               &            & 19.9 & BMM17                       \\
J2018$+$2839 & B2016$+$28 & 0.557 & 14.2 &$\leqslant$1.0, 1(2) &$\leqslant$0.25 & 2(2)       & 1(3)          &            & 1.0  & Rit76, Viv95, NJMK17        \\
J2022$+$2854 & B2020$+$28 & 0.343 & 24.6 &                   &$\leqslant$3   & 0.2(1.6)      &               &            & 0.2  & LA83, GJK12                 \\
J2022$+$5154 & B2021$+$51 & 0.529 & 22.6 &                   & $\leqslant$5  & 1.4(0.7)      &               &  0.12(0.01)& 0.76 & Rit76, GJK12, our           \\
J2033$+$0042 &            & 5.013 & 37.8 &48                 &               & 56$^\ddag$     &               &            & 52   & LBR+13                      \\
J2037$+$1942 & B2034$+$19 & 2.074 & 36.8 &44(4)              &               & $\geq26$      &               &            & 44   & HR09, GJK12                 \\
J2046$+$1540 & B2044$+$15 & 1.138 & 39.8 &                   & <0.04         &               &               &            & 0.02 & WAB+86                      \\
J2048$-$1616 & B2045$-$16 & 1.961 & 11.4 &10(2.5),8.3(1.4), 14(3)& 10(2.5)   &9.0(0.5), 17(6)& 22(5)         &  4.3(2.7)  & 11.8 & Rit76, Viv95, BMM17, NJMK17, our \\
J2055$+$3630 & B2053$+$36 & 0.221 & 97.4 &                   & <0.7          &               &               &            & 0.35 & WAB+86                      \\
J2113$+$2754 & B2110$+$27 & 1.202 & 25.1 & 30, -             &               &               &               &            & 30.0 & WSE07, RR09                 \\
J2113$+$4644 & B2111$+$46 & 1.014 &141.2 &                   & 12.5(2.5)     & 21(4)         &               &            & 16.8 & Rit76, GJK12                \\
J2116$+$1414 & B2113$+$14 & 0.440 & 56.2 &                   & <1            &               &               &            &  0.5 & WAB+86                      \\
J2124$+$1407 & B2122$+$13 & 0.694 & 30.2 &21(6), 22          &               &               &               &            & 21.5 & HR09, RR09                  \\
J2157$+$4017 & B2154$+$40 & 1.525 & 71.1 &                   & 7.5(2.5)      &               &               &            &  7.5 & Rit76                       \\
J2208$+$5500 &            & 0.933 &101.0 &                   &               & >7.5          &               &            & 11.3 & JML+09                      \\
J2219$+$4754 & B2217$+$47 & 0.538 & 43.5 &                   &$\leqslant$2   &               &               &            &  1   & Rit76                       \\
J2253$+$1516 &            & 0.792 & 29.2 &49                 &               &               &               &            & 49   & RR09                        \\
J2305$+$3100 & B2303$+$30 & 1.575 & 49.5 & 11(2), 11, 5.3(0.5)  & 1          &               &               &            &  7.1 & Ran86, HR09, RR09, BMM17    \\
J2313$+$4253 & B2310$+$42 & 0.349 & 17.2 &3.7(0.5)           &               &               &               &  5.2(0.2)  &  4.5  & BMM17, our                  \\
J2317$+$2149 & B2315$+$21 & 1.444 & 20.8 & 2.3(0.5), 3       & 3(0.5)        &               &               &            &  2.8 & WAB+86, HR09, RR09          \\
J2321$+$6024 & B2319$+$60 & 2.256 & 94.5 &35(5)              & 25(5)         & 29(1), 33(3)  & 31(2)         &>30,13.5(5.4)& 27.8& Rit76, GJK12, GJK+14, our  \\
J2330$-$2005 & B2327$-$20 & 1.643 &  8.4 & 9.6(0.9)          &               &12(1),13.1(1.5)&               &            & 11.6 & Big92, BMM17                \\
J2346$-$0609 &            & 1.181 & 22.5 &42.5(3.8)          &               &28.7(1.8)      &               &            & 35.6 & BMM17                       \\
\hline
J0014$+$4746 & B0011$+$47 & 1.240 & 30.4 & -                 &               &               &               &            & & WSE07  \\
J0726$-$2612 &            & 3.442 & 69.4 &                   &               &               & -             &            & & BJB+12 \\
J0855$-$3331 & B0853$-$33 & 1.267 & 86.6 &                   &               &               & -             &            & & BJB+12 \\
J0941$-$39   &            & 0.586 & 78.2 &                   &               &               & -             &            & & BB10   \\
J0943$+$2253 &            & 0.532 & 27.2 & -                 &               &               &               &            & & BFRS18 \\
J1012$-$5830 &            & 2.133 & 294  &                   &               &               & -             &            & & BJB+12 \\
J1055$-$6905 &            & 2.919 &142.8 &                   &               &               & -             &            & & BJB+12 \\
J1059$-$5742 & B1056$-$57 & 1.184 &108.7 &                   &               &               & -             &            & & BJB+12 \\
J1129$-$53   &            & 1.062 & 77.0 &                   &               &               & -             &            & & BJB+12 \\
J1133$-$6250 & B1131$-$62 & 1.022 &567.8 &                   &               &               & -             &            & & BJB+12 \\
J1157$-$6224 & B1154$-$62 & 0.400 &325.2 &                   &               &               & -             &            & & BJB+12 \\
J1225$-$6035 &            & 0.626 &176.1 &                   &               &               & -             &            & & BJB+12 \\
J1255$-$6131 &            & 0.657 &206.5 &                   &               &               & -             &            & & BJB+12 \\
J1307$-$6318 &            & 4.962 &374   &                   &               &               & -             &            & & BJB+12 \\
J1322$-$62   &            & 1.044 &733.6 &                   &               &               & -             &            & & KEK+13 \\
J1326$-$5859 & B1323$-$58 & 0.477 &287.3 &                   &               &               & -             &            & & BJB+12 \\
J1326$-$6408 & B1323$-$63 & 0.792 &502.7 &                   &               &               & -             &            & & BJB+12 \\
J1406$-$5806 &            & 0.288 &229   &                   &               &               & -             &            & & BJB+12 \\
J1423$-$6953 &            & 0.333 &123.9 &                   &               &               & -             &            & & BJB+12 \\
J1428$-$5530 & B1424$-$55 & 0.570 & 82.4 &                   &               &               & -             &            & & BJB+12 \\
J1453$-$6413 & B1449$-$64 & 0.179 & 71.0 &                   &               &               & -             &            & & BJB+12 \\
J1457$-$5122 & B1454$-$51 & 1.748 & 37   &                   &               &               & -             &            & & BJB+12 \\
J1514$-$4834 & B1510$-$48 & 0.454 & 51.5 &                   &               &               & -             &            & & BJB+12 \\
J1514$-$5925 &            & 0.148 &194.0 &                   &               &               & -             &            & & BJB+12 \\
J1534$-$46   &            & 0.364 & 64.4 &                   &               &               & -             &            & & BB10   \\
J1559$-$5545 & B1555$-$55 & 0.957 &212.9 &                   &               &               & -             &            & & BJB+12 \\
J1624$-$4613 &            & 0.871 &224.2 &                   &               &               & -             &            & & BJB+12 \\
J1633$-$4453 & B1630$-$44 & 0.436 &474.1 &                   &               &               & -             &            & & BJB+12 \\
J1646$-$6831 & B1641$-$68 & 1.785 & 43   &                   &               &               & -             &            & & BJB+12 \\
J1647$-$3607 &            & 0.212 &224   &                   &               &               & -             &            & & BJB+12 \\
J1649$-$4349 &            & 0.870 &398.6 &                   &               &               & -             &            & & BJB+12 \\
J1653$-$3838 & B1650$-$38 & 0.305 &207.2 &                   &               &               & -             &            & & BJB+12 \\
J1707$-$4729 &            & 0.266 &268.3 &                   &               &               & -             &            & & BJB+12 \\
J1726$-$31   &            & 0.123 &264.4 &                   &               &               & -             &            & & KEK+13 \\
J1736$-$2457 &            & 2.642 &170   &                   &               &               & -             &            & & BJB+12 \\
J1738$-$3211 & B1735$-$32 & 0.768 & 49.5 &                   &               &               & -             &            & & BJB+12 \\
J1741$-$3016 &            & 1.893 & 382  &                   &               &               & -             &            & & BJB+12 \\
J1742$-$4616 &            & 0.412 &115.9 &                   &               &               & -             &            & & BJB+12 \\
J1749$+$16   &            & 2.311 & 59.6 & -                 &               &               &               &            & & DSM+16 \\
J1750$+$07   &            & 1.908 & 55.4 & -                 &               &               &               &            & & DSM+16 \\
J1750$-$3157 & B1747$-$31 & 0.910 &206.3 &                   &               &               & -             &            & & BJB+12 \\
J1757$-$2223 &            & 0.185 &239.3 &                   &               &               & -             &            & & BJB+12 \\
J1758$-$2540 &            & 2.107 &218.2 &                   &               &               & -             &            & & BJB+12 \\
J1809$-$2109 & B1806$-$21 & 0.702 &381.9 &                   &               &               & -             &            & & BJB+12 \\
J1819$-$1458 &            & 4.263 &196   &                   &               &               & -             &            & & BJB+12 \\
J1823$-$1126 &            & 1.846 &607   &                   &               &               & -             &            & & BJB+12 \\
J1825$-$1446 & B1822$-$14 & 0.279 &357   &                   &               &               & -             &            & & BJB+12 \\
J1825$-$33   &            & 1.271 & 43.2 &                   &               &               & -             &            & & BB10   \\
J1827$-$0750 &            & 0.270 &381   &                   &               &               & -             &            & & BJB+12 \\
J1830$-$1135 &            & 6.221 &257   &                   &               &               & -             &            & & BJB+12 \\
J1837$-$0653 & B1834$-$06 & 1.905 &316.1 &                   &               &               & -             &            & & BJB+12 \\
J1837$-$1243 &            & 1.876 &300   &                   &               &               & -             &            & & BJB+12 \\
J1840$-$1419 &            & 6.597 & 19.4 &                   &               &               & -             &            & & BJB+12 \\
J1841$-$0310 &            & 1.657 &216   &                   &               &               & -             &            & & BJB+12 \\
J1852$-$0635 &            & 0.524 &171   &                   &               &               & -             &            & & BJB+12 \\
J1854$-$1557 &            & 3.453 &160   &                   &               &               & -             &            & & BBJ+11 \\
J1857$-$1027 &            & 3.687 &108.9 &                   &               &               & -             &            & & BJB+12 \\
J1904$+$1011 & B1901$+$10 & 1.856 &135   &                   & -             &               &               &            & & LCX02  \\
J1935$+$1159 &            & 1.939 &188.7 & -                 &               &               &               &            & & BFRS18 \\
J2050$+$1259 &            & 1.221 & 52.4 & -                 &               &               &               &            & & BFRS18 \\
\hline
J1107$-$5907$\dag$&            & 0.252 & 40.2 &                   &               & 93(4)         & 48(5),91.5(0.5),88.9(0.3) & 89(5)  & & BJB+12, YWS+14 \\
J1832$+$0029$\dag$&            & 0.533 & 28.3 &                   &               &               & 50            &            && LLM+12   \\
J1841$-$0500$\dag$&            & 0.912 &532   &                   &               &               &               & 30         && CRC+12   \\
J1910$+$0517$\dag$&            & 0.308 &300   &                   &               &               & 70(4)         &            && LSF+17   \\
J1929$+$1357$\dag$&            & 0.866 &150.7 &                   &               &               & 94.5(0.7)     &            && LSF+17   \\
J1933$+$2421$\dag$& B1931$+$24 & 0.813 &106.0 &                   &               &               & 80            &            && KLO+06   \\
\hline
\multicolumn{11}{l}{Note. $^\dag$ intermittent pulsars,
   $^\ddag$ the observation were taken at 820~MHz,
   $^\sharp$ $f_{\rm n}$ and uncertainties were taken as the mean and half
   the difference between the upper and lower limits. }                            \\
\multicolumn{11}{l}{References. Rit76: \citet{rit76}, DLL+79: \citet{dll+79},
  LA83: \citet{la83}, WAB+86: \citet{wab+86}, Ran86: \citet{ran86},
  DCHR86: \citet{dchr86},  }                                 \\
\multicolumn{11}{l}{Big92: \citet{big92}, Viv95: \citet{viv95}, LCX02: \citet{lcx02}, LWF+04: \citet{lwf+04},
  FSK+04:\citet{fsk+04}, KLO+06: \citet{klo+06},  }  \\
\multicolumn{11}{l}{BGK+07: \citet{bgk+07}, WMJ07: \citet{wmj07}, RW07: \citet{rw07},  HR07: \citet{hr07},
  WSE07:\citet{wse07}, MR08: \citet{mr08},   } \\
\multicolumn{11}{l}{RW08: \citet{rw08}, HR09: \citet{hr09}, RR09: \citet{rr09}, JML+09: \citet{jml+09}, KR10: \citet{kr10},
   } \\
\multicolumn{11}{l}{BGG10: \citet{bgg10}, FR10: \citet{fr10}, BB10: \citet{bb10}, BBJ+11:\citet{bbj+11}, GJK12: \citet{gjk12}, } \\
\multicolumn{11}{l}{CRC+12: \citet{crc+12}, LEM+12:\citet{lem+12}, BJB+12: \citet{bjb+12}, LLM+12: \citet{llm+12}, KEK+13:\citet{kek+13}, LBR+13: \citet{lbr+13},  } \\
\multicolumn{11}{l}{RWB13: \citet{rwb13}, CALL13: \citet{call13}, RSM+13: \citet{rsm+13}, GJW14: \citet{gjw14}, GJK+14: \citet{gjk+14}, KHS+14: \citet{khs+14}, } \\
\multicolumn{11}{l}{YWS+14: \citet{yws+14}, YWS+15: \citet{yws+15}, 
  DSM+16: \citet{dsm+16}, WWY+16: \citet{wwy+16}, LSF+17: \citet{lsf+17},
  BMM17: \citet{bmm17},   }  \\
\multicolumn{11}{l}{NJMK17: \citet{njmk17}, GYY+17:\citet{gyy+17}, NJMK18:\citet{njmk18},
  BM18:\citet{bm18}, BFRS18:\citet{bfrs18}, ZLH+19:\citet{zlh+19}. }\\
%
%        \end{tabular}
%\end{table*}
      \end{longtable}
    }
%\clearpage
%\twocolumn

%\label{lastpage}

\end{document}